\documentclass[a4paper,11pt]{article}
\pdfoutput=1 
\usepackage{jheppub} 
\usepackage[T1]{fontenc} 

\usepackage{amsmath}
\usepackage{lscape}
\DeclareMathOperator{\Tr}{Tr}
\DeclareMathOperator{\PE}{PE}

\title{The Giant Graviton Expansion}

\author[a]{Davide Gaiotto}
\author[a]{Ji Hoon Lee}

\affiliation[a]{Perimeter Institute for Theoretical Physics, Waterloo, Ontario, Canada N2L 2Y5}

\abstract{We propose and test a novel conjectural relation satisfied by the superconformal index of maximally supersymmetric $U(N)$ gauge theory in four dimensions. Analogous relations appear to be also valid for the superconformal indices of a large collection of other gauge theories, as well as for a broad class of index-like generating functions. The relation expresses the finite $N$ index as a systematic series of corrections to a large $N$ answer. Individual corrections have an holographic interpretation as the analytic continuation of contributions from ``giant graviton'' branes fixed by a specific symmetry generator.}

\begin{document} 
\maketitle

\section{Introduction and Conclusions}
The superconformal index \cite{Romelsberger:2005eg,Kinney:2005ej,Bhattacharya:2008zy} is a powerful observable of supersymmetric conformal field theories (SCFTs).
The superconformal index is a generating function for the Witten indices $c_\gamma$ of the spaces of local operators 
of fixed charge $\gamma$ under an appropriate collection of $U(1)$ generators.\footnote{It is also possible to make the index into an actual function of the fugacities by defining it as a partition function on a twisted $S^3 \times S^1$ geometry, but we will mostly focus on the generating function perspective.}  Schematically, we can write
\begin{equation}
Z(y) = \sum_\gamma c_\gamma y^\gamma \qquad \qquad\qquad y^\gamma \equiv \prod_i y_i^{\gamma_i}
\end{equation}
The calculation is set up so that the Witten indices receive contributions only from operators which are annihilated by a certain nilpotent supercharge $Q$ as well as an hermitean conjugate 
superconformal charge $S$. Their scaling dimension saturates a BPS bound. The superconformal index is invariant under continuous deformations and can be readily computed from an appropriate UV Lagrangian description.\footnote{In an SCFT the index can be computed or defined equivalently in terms of local operators or in terms of states on a sphere. The latter definition can be extended to situations where conformal invariance is lost with the help of supersymmetric sphere compactifications \cite{Festuccia:2011ws}. This helps insure invariance along the RG flow. The former definition may also be extended by considering the cohomology of $Q$.}  

The calculation is perturbative and combinatorial in four dimensions: one can turn off continuous couplings and compute the index of local operators which are BPS at tree level. 
These are polynomials in a certain collection of BPS (derivatives of) elementary fields, aka ``letters''. 
In other dimensions one has to account for non-trivial topological sectors of the theory: vortices in 2d, monopoles in 3d, instantons in 5d. 

The large $N$ behaviour of the superconformal index 
\begin{equation}
Z_N(y) = \sum_\gamma c_{N;\gamma} y^\gamma
\end{equation}
of four-dimensional $U(N)$ ${\cal N}=4$ Supersymmetric Yang-Mills theory can be compared with holographic calculations. There are various regimes of interest:
\begin{itemize}
\item If we keep $\gamma$ fixed as $N$ is increased, the corresponding index $c_{N;\gamma}$ stabilizes to some limiting value $c_{\infty;\gamma}$ which agrees 
the index $Z_\infty$ of supergravity/string modes in AdS \cite{Kinney:2005ej,Bhattacharya:2008zy}. 
\item If we allow $\gamma$ to grow linearly in $N$, we enter a regime where one needs to include the contribution of supersymmetric configurations of D-branes in AdS, usually dubbed ``giant gravitons'' \cite{McGreevy:2000cw,Grisaru:2000zn,Hashimoto:2000zp,Balasubramanian:2001nh,Mikhailov:2000ya}. There have been multiple efforts to  find the correct prescription to count the states associated to these configurations \cite{Biswas:2006tj,Mandal:2006tk,Kim:2006he}. For simple configurations a match is possible \cite{Arai:2019xmp,Arai:2019wgv,Arai:2019aou,Arai:2020qaj,Bourdier:2015wda} but a general story is missing.
\item If we allow $\gamma$ to grow quadratically in $N$, the indices grows exponentially in a manner consistent with the entropy of a black hole in the dual geometry \cite{Cabo-Bizet:2018ehj,Choi:2018hmj,Benini:2018ywd,Honda:2019cio,ArabiArdehali:2019tdm,Kim:2019yrz,Cabo-Bizet:2019osg,Amariti:2019mgp,GonzalezLezcano:2019nca,Choi:2019zpz,ArabiArdehali:2019orz,Cabo-Bizet:2019eaf,Murthy:2020rbd,Agarwal:2020zwm,Cabo-Bizet:2020nkr}.
\end{itemize}

Our initial motivation was to study a special case of the linear regime. In order to formulate our objective and results, it is useful to refine our notations a bit. We pick a specific letter, a bosonic scalar $X$ transforming in the adjoint representation of $U(N)$. We denote as $x$ the fugacity for a global symmetry $U(1)_x$ under which the field $X$ has charge $1$ and denote as $y$ the rest of the fugacities. We are interested in the large $N$ behaviour of $c_{N;\gamma}$ for arbitrarily large $U(1)_x$ charge, keeping all other charges finite. In practice, this means looking at indices counting gauge-invariant operators built from any number of $X$'s but only a finite number of other letters. 

Running some computational experiments, we find an interesting pattern to the large $N$ corrections to $Z_\infty$ in this regime. Corrections first occur
for $U(1)_x$ charge of order $N$. The form of the corrections only depends on $N$ through an overall factor of $x^N$, up to $U(1)_x$ charge of order $2N$. This means we can find an auxiliary generating function $\hat Z_1$ such that $Z_N$ 
agrees with an improved approximation 
\begin{equation}
Z_\infty + x^N \hat Z_1 Z_\infty
\end{equation}
up to $U(1)_x$ charges of order $2N$. The overall factor of $Z_\infty$ is included for future convenience:
it makes the analytic properties of $\hat Z_1$ particularly nice.

We then observe that corrections to the improved formula occur for $U(1)_x$ charge of order $2N$ and also depend on $N$ only 
through an overall factor of $x^{2N}$, up to $U(1)_x$ charge of order $3N$. This means we can find a second auxiliary generating function $\hat Z_1$ 
such that $Z_N$ agrees with an improved approximation 
\begin{equation}
Z_\infty + x^N \hat Z_1 Z_\infty+ x^{2N} \hat Z_2 Z_\infty
\end{equation}
up to $U(1)_x$ charges of order $3N$. The pattern continues on and allows us to derive a sequence of improved approximations
\begin{equation}
\left(1+ x^N \hat Z_1 + x^{2N} \hat Z_2 + \cdots + x^{k N} \hat Z_k \right)Z_\infty 
\end{equation}

The specific $U(1)_x$ charge at which corrections to these formulae first occur decreases progressively as we increase the power of $y$ fugacities we are focussing on.  This has two consequences:
\begin{itemize}
\item Increasingly large values of $N$ are required to derive a given coefficient of the auxiliary $\hat Z_k$. 
\item For a fixed power of $y$ and sufficiently low $N$ some of the corrections will involve negative powers of $x$
\end{itemize}
Our initial expectation was that the sequence of improved approximations would simply break down when the power of the $y$ fugacities becomes of order $N$.

To our surprise, this does not appear to be the case. As we increase the order $k$ of our approximation, the spurious terms with negative powers of $x$ cancel out and the sequence of approximants eventually stabilizes to the correct value of $c_{N;\gamma}$, even for powers of $y$ which are greater than $N$ and for small $N$. 
We thus find an exact relation
\begin{equation}  \label{intro:full}
\frac{Z_N(x;y)}{Z_\infty(x;y)} = 1+ \sum_{k=1}^\infty x^{k N} \hat Z_k(x;y)
\end{equation} 
valid for all $N$, involving a new infinite sequence of auxiliary series $\hat Z_k(x;y)$. The sum on the right hand side even makes sense for negative $N$:
the coefficient of every given monomial vanishes as soon as we include enough terms. 

It is instructive to look at the simplest specialization of the superconformal index, which counts half-BPS operators. These are gauge-invariant polynomials 
built from a single letter $X$. We have finite $N$ indices
\begin{equation}
Z_N(x) = \frac{1}{\prod_{n=1}^N (1-x^n)}
\end{equation}
and a systematic expansion 
\begin{equation} \label{intro:single}
\frac{Z_N(x)}{Z_\infty(x)} = 1+ \sum_{k=1}^\infty x^{k N} \hat Z_k(x)
\end{equation} 
with an exactly computable
\begin{equation}
\hat Z_k(x) = \frac{(-1)^k x^{\frac{k(k+1)}{2}}}{\prod_{n=1}^N (1-x^n)}
\end{equation}
Both the original and the auxiliary indices are resummed here to rational functions of $x$. There is a remarkable relation
\begin{equation}
\hat Z_k(x) = Z_k(x^{-1})
\end{equation}
We find that this relation is a special example of a more general result valid for the full index. 
Indeed, we find experimentally that the coefficient of any given power of $y$ in the auxiliary indices $\hat Z_k(x;y)$
can be re-summed to a rational function of $x$ with a denominator which depends on the power of $y$ but has zeroes only 
at $n$-th roots of unity, with $n\leq k$. 

It thus makes sense to define yet another sequence of auxiliary indices $\tilde Z_k(\tilde x;y)$
obtained from $\hat Z_k(\tilde x^{-1};y)$ by re-expanding these rational functions into a power series in $\tilde x = x^{-1}$.
Strikingly, $\tilde Z_k(\tilde x;y)$ is related to the original indices $Z_k(\tilde x;\tilde y)$ by a simple reorganization $y \to \tilde y$ of the 
fugacities. 

We can look at holography to gain some insight on this result. The best understood example of half-BPS giant gravitons are D3 branes 
which wrap the maximal S$^3 \subset$ S$^5$ fixed by $U(1)_x$. They have have $U(1)_x$ charge $N$. These are only special examples of 
a continuous one-parameter family which wraps smaller S$^3$'s and carry $U(1)_x$ charge smaller than $N$. The corresponding radial 
fluctuation mode in the world-volume theory of the maximal D3 branes carries $U(1)_x$ charge $-1$. 

If we consider a stack of $k$ maximal giants, the radial fluctuation mode becomes a $k \times k$ matrix of scalars and the world-volume theory has $U(k)$ gauge invariance. The gauge-invariant half-BPS excitations of branes will be counted by an auxiliary $U(k)$ index which is precisely $Z_k(x^{-1})$. 
Thus $\hat Z_k(x)$ is a re-summed form of the index counting infinitesimal fluctuations of a maximal giant graviton. It may be possible to formulate 
some equivariant localization argument to explain (\ref{intro:single}) holographically as a sum over fixed points of the $U(1)_x$ action consisting of $k$ maximal giant gravitons, weighed by the index for local fluctuations. Re-summations of rational fluctuation determinants are not uncommon in localization calculations. 

We will not attempt to formalize such an argument. Instead, we simply observe that the full relation (\ref{intro:full}) appears to have a similar holographic 
interpretation. The infinitesimal fluctuations of $k$ maximal giant gravitons are controlled by a twisted version of four-dimensional $U(k)$ ${\cal N}=4$ Supersymmetric Yang-Mills theory. The identification between the symmetry generators of the $U(k)$ auxiliary theory and of the original $U(N)$ theory 
is precisely the $y \to \tilde y$ reorganization relating $Z_k(\tilde x;\tilde y)$ to $\tilde Z_k(\tilde x;y)$. Thus $\tilde Z_k(\tilde x;y)$ is the index for 
infinitesimal fluctuations of $k$ giant gravitons and individual terms in (\ref{intro:full}) would be reproduced by re-summation of a contribution from $k$ maximal giants. 

We can repeat the analysis leading to (\ref{intro:full}) for the indices of other four-dimensional supersymmetric gauge theories with a nice large $N$ limit. 
Indeed, we can apply it to a large class of index-like quantities $Z_N(x;y)$ which have analogous combinatorial properties but are {\it not} the indices 
of actual SCFTs. For example, we could take $Z_N(x;y)$ to count $U(N)$ gauge-invariant polynomials built from any reasonable collection of letters 
including $X$, counted by some single-letter index $f(x;y)$. 

We find that the formula (\ref{intro:full}) holds in great generality, together with the re-summation relation $\hat Z_k(x;y) \leftrightarrow \tilde Z_k(\tilde x;y)$
involving a second collection of index-like quantities built from a single-letter index $\tilde f(\tilde x;y)$ with $x \tilde x =1$ and
\begin{equation}
(1-f)(1-\tilde f) = (1-x)(1-\tilde x)
\end{equation}
The relation is symmetric: the new $\tilde Z_k(\tilde x;y)$ indices satisfy an analogue of (\ref{intro:full}) involving on the right hand side the analytic continuation of the original $Z_N(x;y)$ indices.
 
Generalizations of this relation are available if we add (anti)-fundamental letters or consider quiver gauge theories with $X$ being an adjoint or bi-fundamental field. In all cases we checked where the $Z_N(x;y)$ have an holographic interpretation we find that the re-summed auxiliary indices $\tilde Z_k(\tilde x;y)$ 
coincide with the indices of appropriate giant graviton worldvolume theories. 

The exceedingly broad applicability of these relations is a bit confusing at first. It is rather clear that (\ref{intro:full}) and its generalizations should have a purely combinatorial explanation. 
In all examples we considered where the formula works, the $x$ dependence of the index is essentially controlled by the number of ways an operator in some generic representation of $U(N)$ can be combined in a gauge-invariant way with an arbitrary number of $X$'s. If that restricted counting problem satisfies an appropriate analogue of (\ref{intro:full}), the full indices will do so as well. 

Such a combinatorial explanation may appear in tension with the apparently physical relation between the right hand side of (\ref{intro:full}) and the indices of world-volume theories for 
giant gravitons. The tension is ameliorated if we recall that {\it any} $U(N)$ gauge theory is dual to some kind of string theory,
simply because of 't Hooft large $N$ combinatorics. Operators of finite charge are associated to string excitations and operators of size $N$ can be associated to  D-branes \cite{Witten:1979kh,Witten:1998xy}. From that perspective, it is perhaps natural that the world-volume index for such D-branes 
could be determined in a purely combinatorial fashion, irrespectively of the existence of a weakly-curved holographic dual. 

In order to probe the combinatorial nature of the dual $\tilde Z_k(\tilde x;y)$ indices, we make contact with the known gauge theory description of giant gravitons. 
The maximal giant gravitons are dual to operators of the form $\det X$. Powers $(\det X)^k$ represent a stack of $k$ coincident D3 branes. There is a well-defined strategy \cite{Berenstein:2002ke,Balasubramanian:2002sa} to construct finite modifications of a $(\det X)^k$ operator which may be matched to small fluctuations of the stack of D3 branes, governed by an appropriate world-volume $U(k)$ gauge theory. 

Following this strategy, we will solve an associated combinatorial problem: compute (in a theory-agnostic fashion) the index counting the operators defined at large $N$ as finite modifications of a given product of determinants, in a manner analogous to the way one can count single trace operators built from a finite number of fields. The counting problem is slightly ill-defined at tree level. If we assume a certain non-linear correction to the tree-level supercharge $Q$ we can regularize the problem and find that the finite modifications of $(\det X)^k$ are counted precisely by $\tilde Z_k(\tilde x;y)$. Without that assumption we can still recover $\tilde Z_k(\tilde x;y)$ by a formal manipulation whose meaning is somewhat unclear. 

Another potential puzzle associated to (\ref{intro:full}) is the absence of further corrections of order $N^2$ which could be associated to non-trivial gravitational saddles 
such as black holes. In the initial regime of interest this is not a complete surprise: SUSY black holes involve multiple large charges, rather than just $U(1)_x$. 
Still, there are rich families of half-BPS supergravity solutions \cite{Lin:2004nb} which could in principle shown up in a large $N$ expansion for large $U(1)_x$ charges. 
The simplest potential explanation is that (\ref{intro:full}) is a purely stringy result, associated to a regime of small 't Hooft coupling where non-trivial geometries 
and even black holes may be fully described as large stacks of D-brane and thus do not need to be separately accounted for. 

We will also discuss in some detail the 3d version of the story, looking at the superconformal index for the SCFTs associated to stacks of M2 branes. 
The contribution of monopole sectors make the analysis more intricate, but the payoff is a new insight on the superconformal index of stacks of M5 branes, 
which appear as giant gravitons in the setup. 

While this paper was concluded another work appeared \cite{Imamura:2021ytr} which has some overlap with our analysis. In the language of the current paper, the work computes the indices for the $U(n_1) \times U(n_2) \times U(n_3)$ world-volume theory of three stacks of maximal giant gravitons of different orientations, which roughly speaking would be associated to modifications of a product of determinants $(\det X)^{n_1} \times (\det Y)^{n_2} \times (\det Z)^{n_3}$. The objective is to write a formula $(3)$ which is 
somewhat analogous to our (\ref{intro:full}), but involves a triple summation over analytically-continued indices for 
all values of the three $n_i$.\footnote{A technical difference is that the analytic continuation is done within the integral over $U(n_1) \times U(n_2) \times U(n_3)$ fugacities with the help of a pole-selection rule which appears to require the combination of different terms with the same $n = n_1 + n_2 + n_3$.}

The reader may well wonder how could our (\ref{intro:full}) and the formula $(3)$ from the reference \cite{Imamura:2021ytr} be simultaneously true, as our formula 
is essentially a sum over a subset of the terms in $(3)$. In lieu of an explanation, we provide a simplified and tractable example of this phenomenon which appears in Section 
\ref{sec:M2}, involving a certain simplified index $Z_N(x,y)$ of the 3d SCFTs associated to M2 branes. We have a simple expression 
\begin{equation}
\sum_{N=0}^\infty \zeta^N Z_N(x,y) = \frac{1}{\prod_{a\geq 0,b\geq 0}(1- \zeta x^a y^b)}
\end{equation}
We can evaluate $Z_N(x,y)$ as a contour integral in the $\zeta$ plane, giving rise to an exact analogue of $(3)$ 
\begin{equation}
Z_N(x,y) = \sum_{k,k' \geq 0} Z_N^{k,k'}(x,y)
\end{equation}
valid for finite values of $x$ and $y$ within in the unit circle, with 
\begin{equation}
Z_N^{k,k'}(x,y) = x^{k N} y^{k' N} \frac{1}{\prod_{a\geq -k,b\geq -k'|(a,b)\neq (0,0)}(1- x^{a} y^{b})}
\end{equation}
These individual terms have denominators such as $(x-y)$ which cannot be expanded unambiguously into a power series in $x$ and $y$. 
The ambiguous denominators must go away when multiple terms, perhaps with the same value of $k+k'$, are added up. 

On the other hand, we can also find experimentally that the analogue of (\ref{intro:full})
\begin{equation}
Z_N(x,y) = \sum_{k \geq 0} Z_N^{k,0}(x,y)
\end{equation}
holds as well. Here the right hand side is expanded first into a power series in $y$ and then into a power series in $x$. We leave a further investigation of these wall-crossing-like mathematical phenomena to future work. 

\subsection{Future directions}
We conclude with a brief list of open questions which are left open by our work:
\begin{enumerate}
\item It should be possible to test (\ref{intro:full}) and the physical identification of $\tilde Z_k$ in a systematic manner in all 4d SCFTs theories with holographic duals,
possibly in the presence of defects of various co-dimension. 
\item The contribution of monopole sectors in 3d SCFTs is well understood but offers some extra computational challenges which prevent us from deriving an explicit prediction for the corresponding $\tilde Z_k$. Overcoming this obstruction would give novel explicit expressions for the indices of the corresponding giant graviton theories, which are often strongly-coupled 6d SCFTs.
\item We have not explored the analogues of (\ref{intro:full}) in 2d SCFTs or in dimension greater than four.  
\item The expansion (\ref{intro:full}) can be interpreted as a non-perturbative string theory result. It would be interesting to explore this perspective further.  
\item A localization argument for (\ref{intro:full}) could help explain the coexistence of many different expansion formulae which take into account distinct collections of 
giant graviton configurations. 
\item It may be possible to motivate (\ref{intro:full}) by a saddle expansion of the superconformal index in a regime where $x$ is finite and other $y$'s are infinitesimal. 
\end{enumerate}
\subsection{Structure of the paper}
We devote Section \ref{sec:infty} to a review of the ``naive'' large $N$ limit $Z_N \to Z_\infty$. We then proceed in an acausal manner and devote Section \ref{sec:det} to the counting problem for small fluctuations of a product of determinant operators. This allows us to derive immediately our conjectural formula for the auxiliary 
$\tilde Z_k(\tilde x;y)$ and $\hat Z_k(x;y)$ indices. Prior knowledge of the $\hat Z_k(x;y)$ greatly simplifies our work in Section \ref{sec:expansion}, as an explicit calculation of $Z_N$ for very large $N$ is challenging, but (\ref{intro:full}) can be tested at small $N$ as long as candidate $\hat Z_k(x;y)$ are available. 
We devote Section \ref{sec:M2} to the case of 3d gauge theories.

In Appendix \ref{app:reflex} we count the number of ways an operator in the adjoint representation of $U(N)$ can be combined in a gauge-invariant way with an arbitrary number of $X$'s. In Appendix \ref{section: schur}, we write a generating function for the Schur index that greatly facilitates their computation. We present tests of the giant graviton expansion for the Schur index in Appendix \ref{appendix: check of schur proposal}. We review a useful formula for $U(N)$ gauge theory indices in terms of $S_n$ characters in Appendix \ref{app: character formula for U(N)}. We present tests of expansion for a specialized $\mathcal{N}=4$ index and for general $U(N)$ indices in Appendix \ref{appendix: check of reduced proposal} and \ref{appendix: check of general proposal}.

\section{Single trace operators in the large $N$ limit}\label{sec:infty}
We begin by reviewing the large $N$ limit of the Witten indices $c_{N,\gamma}$ at fixed $\gamma$. 

\subsection{$U(N)$ gauge theory with adjoint matter}
The superconformal indices of an $U(N)$ gauge theory with adjoint matter has a simple combinatorial interpretation: it counts with signs 
the gauge-invariant operators built from a collection of adjoint BPS letters $L_a$ with Grassmann parity $|L_a|$. 
The letters consist of all the elementary fields and their derivatives
which are in the tree-level cohomology of the selected supercharge $Q$. 

The letters $L_a$ have some specific charge $\gamma^a$ under the $U(1)$ global symmetries which organize the index. 
We denote the single-letter index as 
\begin{equation}
f(y) = \sum_a (-1)^{|L_a|} y^{\gamma^a} 
\end{equation}
In a physical theory the charges $\gamma^a$ satisfy some unitarity bounds and only a finite collection of monomials in the $L_a$ can 
contribute to the space of BPS local operators with any given charge. Whenever we consider more general counting problems of this form which do not arise from an SCFT
we should impose a similar constraint by hand. For example, $f(x) = x + x^{-1}$ would not give a good counting problem: any power of the product of the two letters 
would contribute to the charge $0$ index.  

The index can be expressed as an integral over the unitary group. In terms of eigenvalues $\mu_a$ of an $U(N)$ group element, we can write 
\begin{equation}
Z_N(y) = \frac{1}{N!} \int \frac{d\mu_a}{2 \pi i \mu_a} \prod_{a \neq b} (1-\mu_a/\mu_b) \PE\left[(\sum_a \mu_a)(\sum_b \mu_b^{-1})f(y)\right]
\end{equation}
where $PE$ denotes the Plethystic Exponential:
\begin{equation}
\PE\left[\sum_a n_a \rho_a\right] =  \frac{1}{\prod_a (1-\rho_a)^{n_a}}\,.
\end{equation}
Here $\rho_a$ are monomials in the fugacities and $n_a$ are integers. The right hand side is expanded into a formal power series 
in the $\rho_a$. The plethystic exponential in the integrand counts polynomials in the adjoint letters.

We write
\begin{equation}
Z_N(y) = \sum_{\gamma} c_{N;\gamma} y^\gamma
\end{equation}
Gauge-invariant operators built from a number of letters which is kept finite as $N$ is increased can be written uniquely as a product of traces 
\begin{equation}
\mathrm{Tr} \,L_{a_1} L_{a_2} \cdots L_{a_n}
\end{equation}
of cyclic words built from the letters in the alphabet: any relations between trace involve a product of at least $N$ letters and can thus be disregarded. 

This implies that the coefficients $c_{N;\gamma}$ will stabilize to some limit $c_{\infty;\gamma}$ for any given $\gamma$
as $N$ increases. The stabilization occurs as soon as $N$ is larger than the maximum number of letters which can appear in an operator of charge $\gamma$. 
We will denote it as $Z_{\infty}(y)$ the generating function of the $c_{\infty;\gamma}$, i.e. the index counting polynomials in single trace operators. It can be written explicitly as an infinite product:
\begin{equation}
Z_{\infty}(y) = \prod_{n=1}^\infty \frac{1}{1-f(y^n)}\,.
\end{equation}
This is also the result of a large $N$ saddle-point evaluation of $Z_N(y)$, around a saddle for which the $\mu_a$ are distributed uniformly on the unit circle \cite{Kinney:2005ej}. 

The single letter partition function for the superconformal index of ${\cal N}=4$ SYM is
\begin{equation}
f(x,y,z,p,q)=1-\frac{(1-x)(1-y)(1-z)}{(1-p)(1-q)} 
\end{equation}
The fugacities are restricted by the relation $p q = x y z$. 

This expression for $f$ is a shorthand for 
\begin{equation}
f(x,y,z,p,q)=1-(1-x)(1-y)(1-z)\sum_{n=0}^\infty \sum_{m=0}^\infty p^n q^m 
\end{equation}
Concretely, the three complex scalars $X$, $Y$, $Z$ have fugacities $x$, $y$, $z$, while the holomorphic derivatives $\partial_{\alpha +}$ have fugacities $p$ and $q$. 

The factorization of $1-f$ leads to a particularly simple large $N$ index:
\begin{equation}
Z_{\infty}(x,y,z,p,q) = \prod_{n=1}^\infty \frac{(1-p^n)(1-q^n)}{(1-x^n)(1-y^n)(1-z^n)}
\end{equation}

There are a variety of simplifying limits which keep the letter $X$:
\begin{itemize}
\item We can send all fugacities except $x$ to $0$ to get the half-BPS index, counting local operators built from $X$ alone: $f = x$. The large $N$ answer 
\begin{equation}
Z_\infty = \prod_{n=1}^\infty \frac{1}{1-x^n}
\end{equation}
counts polynomials in the $\mathrm{Tr} X^n$ single trace operators.
\item We can send $p$,$q$ and $z$ to zero to get a $\frac14$-BPS index, with $f = x+y- x y$. The individual letters are now two scalar fields $X$ and $Y$, 
as well as a fermion $\zeta$ with the same charge as the commutator $[X,Y]$. The supercharge indeed maps $\zeta \to [X,Y]$, so that the large $N$ 
Q-cohomology includes symmetrized traces $\mathrm{STr} X^n Y^m$.
The large $N$ answer 
\begin{equation}
Z_\infty = \prod_{n=1}^\infty \frac{1}{(1-x^n)(1-y^n)} =   \frac{\prod_{n=0}^\infty \prod_{m=0}^n (1-x^{n-m+1} y^{m+1})}{\prod_{n=1}^\infty \prod_{m=0}^n (1-x^{n-m} y^m)} 
\end{equation}
can be interpreted as counting polynomials built from symmetrized traces as well as the fermionic operators $\mathrm{STr} \zeta X^n Y^m$.
\item We can send $p$, $y$ and $z$ to zero to get 
\begin{equation}
f = \frac{x-q}{1-q}
\end{equation}
\item We can set $p=z$ to get the Schur index, with $q = x y$:
\begin{equation}
f = \frac{x+y-2 q}{1-q}
\end{equation}
\end{itemize}
We could also consider some purely combinatorial problems which do not occur as an index, such as counting operators built from two letters $X$ and $Y$ only, i.e. $f = x+y$. 
Finally, we could consider problems where $X$ is not present, such the limit where $x$,$y$,$z$ and $p$ are sent to $0$, leaving a single fermionic letter. 

\subsection{$U(N)$ gauge theory with adjoint and (anti)fundamental matter}

There is a simple generalization of the large $N$ formula which is applicable in the situation where (anti)fundamental letters are present. The integral is now 
\begin{align}
Z_N(y) &= \frac{1}{N!} \int \frac{d\mu_a}{2 \pi i \mu_a} \prod_{a \neq b} (1-\mu_a/\mu_b) \PE\left[(\sum_a \mu_a)(\sum_b \mu_b^{-1})f(y)\right] \cdot \cr &\cdot
\PE\left[(\sum_a \mu_a)v(y)\right]\PE\left[(\sum_b \mu_b^{-1}) \bar v(y)\right]
\end{align}
with $v(y)$ and $\bar v(y)$ the indices counting the fundamental and anti-fundamental letters. 

Gauge-invariant operators built from a number of letters which is kept finite as $N$ is increased are built from both traces and ``mesons'', 
i.e. sequences of adjoint letters sandwiched between a fundamental and an anti-fundamental letters. These operators are counted by 
\begin{equation}\label{eq:largemesons}
Z_{\infty}(y) =  \frac{1}{\prod_{n=1}^\infty \left(1-f(y^n) \right)} \PE\left[ \frac{\bar v(y) v(y)}{1-f(y)}\right]
\end{equation}
which is also the saddle-point evaluation of $Z_N(y)$ around a saddle for which the $\mu_a$ are distributed uniformly on the unit circle.

As a physical example, consider the index for an half-BPS (aka $(4,4)$) 2d surface defect in 4d SYM, defined by coupling the theory to a 
2d hypermultiplet transforming in the fundamental representation of the gauge group. This defects breaks some supersymmetry, but preserves the one employed in the definition of the superconformal index. The choice of geometric orientation and unbroken R-symmetry 
breaks the symmetry between $p$ and $q$ and part of the symmetry between $x$, $y$ and $z$. 

With a specific choice of orientation 
we find the contribution of the 2d hypermultiplets:
\begin{equation}
v =  p^{-1}z \frac{ (1-x)(1-y)}{1-q}s^{-1} \qquad \qquad \qquad \bar v =  \frac{ (1-x)(1-y)}{1-q} s
\end{equation}
with $s$ being the fugacity for the $U(1)_s$ flavour symmetry rotating the 2d hypermultiplets. Here $q$ is associated to rotations in the plane of the 2d defect. 

We obtain 
\begin{equation}
Z_{\infty}(x,y,z,p,q) =  \prod_{n=1}^\infty \frac{(1-p^n)(1-q^n)}{(1-x^n)(1-y^n)(1-z^n)} \PE\left[ z p^{-1} \frac{(1-p)(1-x)(1-y)}{(1-z)(1-q)}\right]
\end{equation}
The surface defect is expected to be holographically dual to a D3 brane wrapping $AdS_3 \times S^1$. It would be nice to 
reproduce the right hand side in that description. As a rational function, the argument of the plethystic exponential on the right hand side is the usual $f$ with $p$ and $z$ replaced respectively by $z^{-1}$ and $p^{-1}$. 
\footnote{Another interesting observation is that the plethystic exponential on the right hand side contains a $\prod_n (1-z^n)$ factor 
which cancels the same factor on the right hand side. This likely means that operators of the form $\mathrm{Tr} Z^n$ 
are not BPS anymore in the presence of the defect. This is compatible with the fact that
the coupling to the 2d hypermultiplets mixes $Z$ with the moment map operators for the hypermultiplets. }

\subsection{Quiver gauge theories}

We can employ this formula recursively to study quiver gauge theories, with some $\prod_r U(N_r)$ gauge group and adjoint, bifundamental and 
(anti)fundamental letters. 

For example, consider an $U(N_1) \times U(N_2)$ gauge theory with adjoint letters $f_{11}(y_i)$ and $f_{22}(y_i)$, bifundamental $f_{12}(y_i)$ and $f_{21}(y_i)$. Taking $N_2$ large first we compute $Z_{N_1,\infty}(y_i)$, which is the product of $\prod_{n=1}^\infty \frac{1}{1-f_{22}(y_i^n)}$ and of an integral over $U(N_1)$ fugacities, taking the form of the index of a $U(N_1)$ gauge theory with adjoint index $f(y_i) = f_{11}(y_i) + \frac{f_{12}(y_i) f_{21}(y_i)}{1-f_{22}(y_i)}$. Taking $N_1$ large as well, we get a neat final expression 
\begin{equation}
Z_{\infty,\infty}(y_i) = \prod_{n=1}^\infty \frac{1}{1-f_{11}(y_i^n)-f_{22}(y_i^n)+f_{11}(y_i^n)f_{22}(y_i^n)-f_{12}(y_i^n) f_{21}(y_i^n)}
\end{equation}
which counts operators built from a finite number of adjoint and bifundamental letters, with no trace relations. 

We can write this as 
\begin{equation}
Z_{\infty,\infty}(y_i) = \prod_{n=1}^\infty \frac{1}{\det_{2 \times 2} \left[1 - \hat f(y_i^n)\right]} 
\end{equation}
where $\hat f$ is the $ 2 \times 2$ matrix with entries $f_{rs}$. This matches a direct saddle point evaluation of the 
full integral \cite{Gadde:2010en}.

If we also had fundamental letters counted by $v_1$ and $v_2$ as well as anti-fundamental counted by $\bar v_1$ and $\bar v_2$, 
the effective $U(N_1)$ index which appears after the large $N_2$ limit has $v = v_1(y_i) + \frac{f_{12}(y_i) v_2(y_i)}{1-f_{22}(y_i)}$ and
$\bar v = \bar v_1(y_i) + \frac{\bar v_2(y_i) f_{21}(y_i) }{1-f_{22}(y_i)}$. The final answer neatly simplifies to 
\begin{align}
Z_{\infty,\infty}(y_i) &= \prod_{n=1}^\infty \frac{1}{1-f_{11}(y_i^n)-f_{22}(y_i^n)+f_{11}(y_i^n)f_{22}(y_i^n)-f_{12}(y_i^n) f_{21}(y_i^n)} \cdot \cr
&\cdot \PE\left[ \frac{(1-f_{22})\bar v_1 v_1+(1-f_{11})\bar v_2 v_2+f_{12} \bar v_1 v_2+f_{21}\bar v_2 v_1}{1-f_{11}-f_{22}+f_{11}f_{22}-f_{12} f_{21}}  \right]
\end{align}
We can write this as 
\begin{equation}
Z_{\infty,\infty}(y_i) =  \frac{1}{\prod_{n=1}^\infty \det_{2 \times 2} \left[1 - \hat f(y_i^n)\right]} \PE\left[ \sum_{r,s} \bar v_r \left[1 - \hat f(y_i)\right]^{-1}_{rs}v_s \right]
\end{equation}
The same expressions, with a $n \times n$ matrix $\hat f$, can be derived for quivers with $n$ nodes. It matches a direct saddle point evaluation of the 
full integral \cite{Gadde:2010en}.

\section{Modifications of a determinant} \label{sec:det}
The objective of this section is to introduce the auxiliary indices $\tilde Z_k$ which in later section will provide the 
analytic continuation of the $\hat Z_k$. In this section we introduce the $\tilde Z_k$ as indices of open string fluctuations of 
a collection of determinants. The definition and computation seem to be valid under somewhat restrictive sets of assumptions, even though 
the expression we derive appears to provide the analytic continuation of the $\hat Z_k$ in a much broader context. 

\subsection{$U(N)$ gauge theory with adjoint matter}
Next, we pick a specific bosonic letter $X$ with fugacity $x$. We now use $L_a$ to denote the remaining letters and $y_i$ to denote the remaining fugacities. We will thus refer to the single-letter partition function as $f(x;y_i)$ and to the partition function as $Z_N(x;y_i)$. We denote as $Z_{\infty}(x;y_i)$ the count of operators of finite size from the previous section.

We want to count ``finite'' gauge-invariant modification of the operator $\det X$. Loosely speaking, the modifications are obtained from $\det X$ by replacing a finite collection of $X$ symbols with some other finite strings of letters. For example, replacing a single $X$ with $L_1 L_2 L_1$ gives an operator 
\begin{equation}
\epsilon^{i_1 i_2 i_3 \cdots}\epsilon_{j_1 j_2 j_3 \cdots} (L_1 L_2 L_1)^{j_1}_{i_1} X^{j_2}_{i_2}X^{j_3}_{i_3} \cdots
\end{equation}
etcetera. In the holographic/large $N$ perspective, these modifications represent open strings attached to a giant graviton. 

As we will illustrate later on with the help of a fermionization trick, the finite modifications of a determinant in the large $N$ limit
are obtained by a finite collection of replacements $X \to \cdots$ as well as multiplication by some overall polynomial of the traces. 
These contribution of the closed traces will give rise to an overall power of $Z_\infty$, which we will strip off in this section 
to focus on $X \to \cdots$ modifications only. 

This description involves some overcounting, due to a class of relations which survive in the large $N$ limit. Indeed, if we replace $X$ by $X W$ for some word $W$, 
antisymmetry allows one to rewrite the answer as $\Tr W \det X$. The same occurs for a $W X$ modification. These appear to be the only 
non-trivial relations surviving at large $N$. We will momentarily devise a strategy to avoid over-counting the possible $X \to \cdots$ modifications.

Before we move on, we need to discuss a potential obstruction to our counting objective. A determinant modification which replaces $X$ with ``$1$'' has effectively the opposite charge to $X$ and changes the overall fugacity of the operator by $x^{-1}$. On the other hand, a determinant modification containing $X$ fields in the middle of the word, 
such as $X \to L_1 X^n L_2$, will change the fugacity of the operator schematically by $y_1 y_2 x^n$. This creates a problem: for every integer $n$ we can combine the $X\to L_1 X^n L_2$ modification with $n$ other $X \to 1$ modifications to get a distinct modified determinant operator with total fugacity $x^N y_1 y_2$. This gives an infinite class of operators 
which are distinct in the large $N$ limit and have the same fugacity, ruining the counting problem. 

Concretely, we can define a ``single modification'' index which counts individual modifications of the determinant, but the computation will break down when we try to count multiple modifications by a Plethystic Exponential. The problem is somewhat similar to what would have happened in the usual index calculation if we had included by mistake letters which were not BPS at tree level by dropping some terms in the tree-level supercharge: the Witten index would not, strictly speaking, change, but we may have added infinitely many 
pairs of states which would have cancelled out in the correct calculation. Our interpretation of the situation here is that there must be some non-linear part of the 
supercharge $Q$ which enforces the cancellation of these infinitely many determinant modifications to recover the count of open strings attached to a giant graviton. 

As this phenomenon occurs in a situation where we have $N$ copies of $X$ in the background, it is natural to look at some non-linear piece of $Q$ which 
can give an effective linear contribution in a background of $X$'s. Suppose that all the letters come in pairs $L_a$, $L'_a$ with opposite Grassmann parity,
such that $L'_a$ has the same charge as $X L_a$. Suppose also that the non-linear part of the supercharge $Q$ includes terms which map $L'_a \to [X,L_a]$. 
This is indeed the case in important physical examples. This term in the supercharge would allow one to commute $X$ along a chain of symbols, up to $Q$-exact operators.
Then we could bring $X$ to the very left of a modification and eliminate it by the relation discussed above. Modifications of the form $L_1 X^n L_2$  would then drop out of the cohomology. 

In this situation, we would have 
\begin{equation}
f = x + h(x,y) - x h(x,y)
\end{equation}
where $h$ counts the $L_a$ letters. Equivalently, we would be able to factorize 
\begin{equation}
1-f = (1-x)(1-h)
\end{equation}
We will indeed see that this factorization leads to important simplifications below and makes it possible to exponentiate the single-modification index. 

We are ready to introduce a powerful fermionization trick. Write
\begin{equation} 
\det X = \int d \psi d \bar \psi e^{\bar \psi X \psi} 
\end{equation}
for some (anti)fundamental fermions $\bar \psi$ and $\psi$. Modifications of the determinant can be written as expressions of the form 
\begin{equation}
\int d \psi d \bar \psi e^{\bar \psi X \psi} (\bar \psi W_1 \psi) (\bar \psi W_2 \psi) \cdots
\end{equation}
This gives a quick explanation for the large-$N$ combinatorics of determinants: inserting a determinant in a correlation function is the same as 
introducing auxiliary (anti)fundamental variables $\bar \psi$ and $\psi$, which add boundaries to the 't Hooft ribbon graphs and thus add a D-brane 
to the dual string theory. Fermion bilinears of the form $\bar \psi L_1 \cdots L_s \psi$ represent open strings attached to the D-brane, in the same way as $\mathrm{Tr}L_1 \cdots L_s$
represents closed strings. 

The relations we mentioned above are just the Ward identities for the fermions:  
\begin{equation}
\int d \psi d \bar \psi e^{\bar \psi X \psi} (\bar \psi X W_1 \psi) (\bar \psi W_2 \psi) \cdots = \int d \psi d \bar \psi e^{\bar \psi X \psi} \left(\frac{d}{d\psi} W_1 \psi \right) (\bar \psi W_2 \psi) \cdots
\end{equation}
where the derivative acts on the whole expression to the right of it. 

For counting purposes, we can attempt to impose these Ward identities by adding some bosonic antifields $\bar u$ and $u$, with a BRST differential 
mapping the antifields to the Ward identities. The BRST cohomology in ghost number $0$ will reproduce the operators counting we are interested in:
ghost number $0$ operators do not contain the antifields and are all BRST closed. The BRST exact operators are precisely these which will vanish in the 
fermion integral. 

The main problem with this approach is that there is cohomology in non-zero ghost number. This is closely related to the 
fact that the expression $\bar \psi X \psi$ is set to zero by two distinct Ward identities: it appears in the variation of $\bar \psi u$ as well as $\bar u \psi$. The combination $\bar \psi u + \bar u \psi$ is thus BRST close, but it is not BRST exact. It gives a fermionic zeromode with ghost number $-1$ and trivial fugacity. Other operators which are cancelled by two Ward identities, such as $\bar \psi X X \psi$, do not give this problem: this appears in the BRST image of both $\bar \psi X u$ and $\bar u X \psi$, but $\bar \psi X u + \bar u X \psi$ is itself in the image of $\bar u u$ and cancels out in cohomology.

We find that the large $N$ operator counting in the enlarged theory vanishes: every BRST-closed operator in ghost number $0$ we are interested in can be multiplied by the zeromode to give an operator in ghost number $-1$ with the same charges. This appears to exhaust the large $N$ BRST cohomology. 
We thus conjecture that this problem can be addressed in a simple manner: we remove the offending fermionic zeromode by hand. 

We thus arrive at the following conjectural statement: in the large $N$ limit, finite modifications of $\det X$ can be counted as
polynomials in single trace operators as well as mesons built from fermions and anti-fields, with the $\bar \psi u + \bar u \psi$ meson subtracted off. These auxiliary fundamental letters are counted by $v = (x-1) \lambda$ and the anti-fundamental by $\bar v = (1-x^{-1})\lambda^{-1}$. Here $\lambda$ denotes a fugacity for an extra symmetry $U(1)$ which only acts on these auxiliary variables. It will actually drop out of calculations, but will be useful later on.

The index counting individual mesons is thus 
\begin{equation}
\tilde f=1+\frac{\bar v v}{1-f} = 1-\frac{(1-x)(1-x^{-1})}{1-f}
\end{equation}
We eliminated the contribution of the spurious zeromode by adding $1$. 

The expression simplifies and becomes more transparent if we have the factorization $(1-f) = (1-x)(1-h)$ discussed above:
\begin{equation}
1-\tilde f=\frac{1-x^{-1}}{1-h} =1 - \frac{x^{-1}}{1-h} + \frac{h}{1-h}
\end{equation}
We tentatively recognize $\frac{x^{-1}}{1-h}$ as the contribution of $X\to L_1 \cdots L_s$ modifications which do not include any $X$'s. The extra term must is a bit more mysterious.  

In any case, we can introduce the $\tilde x = x^{-1}$ fugacity for the $X \to 1$ modifications and write the relation as
\begin{equation}
\frac{1-\tilde f}{1-\tilde x} = \frac{1-x}{1-f}
\end{equation}
Curiously, this is an involution. 

We can exponentiate $\tilde f$ to get the counting function for any number of modifications:
\begin{equation}
\tilde Z_1(\tilde x,y) = \mathrm{PE}[\tilde f]
\end{equation}

Consider the example of ${\cal N}=4$ SYM. We have
\begin{equation}
\tilde f(\tilde x,y,z,p,q) = \frac{(1-\tilde x)(1-p)(1-q)}{(1-y)(1-z)} = f(\tilde x,p,q,y,z)
\end{equation}
This is just the single-letter index of an $U(1)$ ${\cal N}=4$ SYM with fugacities transformed as $x \to x^{-1}$, $y \leftrightarrow p$, $z \leftrightarrow q$.

As we mentioned in the introduction, the $\det X$ operator is holographically dual to a giant graviton wrapping an $\mathbb{R} \times $S$^3$ submanifold in AdS$_5 \times$S$^5$. The low energy worldvolume theory on the giant graviton is thus precisely $U(1)$ ${\cal N}=4$ SYM compactified on S$^3$, just as one would employ to compute the index by a state-operator map. The fugacities $y$ and $z$ of the original theory, though, now control the angular momenta on $S^3$, while 
the fugacities $p$ and $q$ are associated to scalar fields describing fluctuations in AdS$_5$. This motivates the redefinition of the fugacities. Fluctuations in the size of the $S^3$ inside $S^5$ actually lower the $x$ charge and thus have fugacity $\tilde x = x^{-1}$. 

We should stress that even after making sense of the index for determinant modifications with the single trace contribution stripped out, 
we would not be able to actually combine it with $Z_{\infty}(x;y_i)$ to produce an index counting all modifications of the determinant: $Z_{\infty}(x;y_i)$ 
is a formal power series in $x$ but $\tilde Z_1(\tilde x,y)$ is a power series in $x^{-1}$, so the product is ill-defined. 

\subsection{$U(N)$ gauge theory with adjoint and (anti)fundamental matter}

Next, we can add (anti)fundamental letters in the original gauge theory. These will also contribute determinant modifications. Using the (anti)fundamental letters we could write gauge-invariant expressions which have different numbers of $\psi$ and $\bar \psi$ letters. These do not actually correspond to determinant modifications.
Indeed, they vanish when inserted into the auxiliary $\int d \psi d \bar \psi e^{\bar \psi X \psi}$ integral. We can take this constraint into account by projecting onto operators invariant under the extra $U(1)_\lambda$ symmetry acting on $\psi$ and $\bar \psi$ with opposite charge. 

Including both types of (anti)fundamental letters in the large $N$ formula \ref{eq:largemesons} by replacing $v \to v+ (x-1) \lambda$ and $\bar v \to \bar v+ (1-x^{-1})\lambda^{-1}$ there, we arrive to our conjectural fomula for the large $N$ index counting finite modifications of $\det X$: 
\begin{equation}
 \tilde Z_1(x;y_i) =  \PE[\tilde f] \oint \frac{d \lambda}{2 \pi i \lambda} \PE\left[\frac{x-1}{1-f}\bar v\lambda\right]PE\left[\frac{1-x^{-1}}{1-f}  v\lambda^{-1}\right]
\end{equation}
The corrections thus take the form of the index for an auxiliary $U(1)$ gauge theory. 

The formula can be written in a more compact form by a shift $\lambda \to x^{-\frac12} \lambda$:
\begin{equation}
 \tilde Z_1(x;y_i) =   \PE[\tilde f] \oint \frac{d \lambda}{2 \pi i \lambda} \PE\left[\frac{x^{\frac12}-x^{-\frac12}}{1-f}\left(v\lambda^{-1}+ \bar v\lambda\right)\right]
\end{equation}
This takes the form of the index for an auxiliary $U(1)$ gauge theory, with (anti)fundamental letters counted by indices
\begin{equation}
\tilde v = \frac{x^{\frac12}-x^{-\frac12}}{1-f} v \qquad \qquad \tilde {\bar v} = \frac{x^{\frac12}-x^{-\frac12}}{1-f} \bar v
\end{equation}
Note the simplification when we factorize $1-f=(1-x)(1-h)$:
\begin{equation}
\tilde v =- \frac{x^{-\frac12}}{1-h} v \qquad \qquad \tilde {\bar v} = -\frac{x^{-\frac12}}{1-h} \bar v
\end{equation}
 
 As an example, consider the index for an half-BPS (aka $(4,4)$) 2d surface defect in 4d SYM, defined by coupling the theory to a 
2d hypermultiplet transforming in the fundamental representation of the gauge group:
\begin{equation}
v = p^{-1} z \frac{ (1-x)(1-y)}{1-q}s \qquad \qquad \qquad \bar v =  \frac{ (1-x)(1-y)}{1-q} s^{-1}
\end{equation}
so that $\mathrm{Tr} X^n$ and presumably $\det X$ survive as protected operators on the defect 
and it makes sense to consider their fluctuations. We get a correction factor 
\begin{equation}
\PE\left[x p^{-1} \frac{(1-x^{-1})( 1-p)}{1-y} s^{-1}\lambda\right]PE\left[y \frac{(1-x^{-1})(1-p)}{1-y}s  \lambda^{-1}\right]
\end{equation}
A shift of $s$ gives 
\begin{equation}
\PE\left[q  \frac{(1-x^{-1})( 1-p)}{1-y} s^{-1}\lambda\right]PE\left[z^{-1} \frac{(1-x^{-1})(1-p)}{1-y}s  \lambda^{-1}\right]
\end{equation}

On the holographic dual side, the giant graviton brane and the $AdS_3 \times S^1$ brane intersect along a 
2d submanifold $\mathbb{R} \times S^1$. Hence we find the usual 2d bifundamental hypermultiplets 
coupled to the two brane worldvolume theories, precisely as our index calculation shows. 

\subsection{Bifundamental determinants}
It is straightforward to generalize these formulae to modifications of the determinant of an adjoint field in quiver gauge theories.

Another simple generalization is to consider modifications of the determinant of a bifundamental field. 
These can be accommodated simply by taking $\psi$ and $\bar \psi$ to be charged under distinct gauge groups. 
The rest of the discussion proceeds essentially as before. 

\subsection{Powers of a determinant}
Next, consider the modifications of $(\det X)^k$. For simplicity, we consider first a theory with adjoints only. 

We can remove some $X$'s from any determinants and add extra words attached to any pair of indices in the $\epsilon$ tensors. We fermionize the determinants by adding $k$ sets of auxiiary fermions, $\bar \psi^\alpha$ and $\psi_\alpha$ 
with $\alpha = 1,\cdots k$, allowing us to write operators of the form 
\begin{equation}
\int d \psi d \bar \psi e^{ \bar \psi^\alpha X \psi_\alpha} (\bar \psi^{\beta_1} W_1 \psi_{\gamma_1}) (\bar \psi^{\beta_1}W_2 \psi_{\gamma_2}) \cdots
\end{equation}
representing such general modifications.

Although we have a lot of dangling indices, because of the explicit $U(k)$ global symmetry in the exponent the integral will produce 
some linear combination of $U(k)$-invariant tensors. Any linear combination of operators which is not $U(k)$ invariant will vanish. 
At the end of our calculation, we will thus need to impose $U(k)$ invariance by hand. 

We add again antifields $u_\alpha$ and $\bar u^\alpha$ to impose the Ward identities for the fermions. There is a bit of redundancy, as $\bar \psi^\beta X \psi_\gamma$ is eliminated twice. As a result, we have $k^2$ spurious fermionic zeromodes, which we will remove by hand. Now $v = (x-1) \sum_a \lambda_a$ counts the fundamental letters and $\bar v = (1-x^{-1})\sum_a \lambda_a^{-1}$ counts the anti-fundamental letters and the partition function is
\begin{equation}
 \tilde Z_k(x;y_i) = \frac{1}{k!} \int \frac{d\lambda_a}{2 \pi i \lambda_a} \prod_{a \neq b} (1-\lambda_a/\lambda_b) PE\left[\tilde f(\sum_a \lambda_a)(\sum_b \lambda_b^{-1})\right]
\end{equation}
The integral over the $\lambda_a$ fugacities takes precisely the form of the index for an auxiliary $U(k)$ gauge theory built from $U(k)$ adjoint letters counted by $\tilde f$.

If we apply this to $U(N)$ ${\cal N}=4$ SYM, we find that, as expected, $\tilde Z_k(x;y_i)$ counts the worldvolume fluctuations of $k$ coincident giant gravitons, giving rise to an auxiliary $U(k)$ ${\cal N}=4$ SYM theory on $\mathbb{R} \times S^3$. 

If we include (anti)fundamentals, we get 
\begin{align}
 Z^{(\det X)^k}_\infty(x;y_i) =  \frac{1}{k!} \int \frac{d\lambda_a}{2 \pi i \lambda_a}&  \prod_{a \neq b} (1-\lambda_a/\lambda_b) PE\left[\tilde f(\sum_a \lambda_a)(\sum_b \lambda_b^{-1})\right] \cdot \cr
&\cdot \PE\left[\frac{x-1}{1-f}\bar v \sum_a \lambda_a \right]PE\left[\frac{1-x^{-1}}{1-f}  v\sum_a \lambda_a^{-1}\right]
\end{align}
i.e. we have an auxiliary $U(k)$ theory with (anti)fundamental letter indices
\begin{equation}
\tilde v = \frac{x^{\frac12}-x^{-\frac12}}{1-f} v \qquad \qquad \tilde {\bar v} = \frac{x^{\frac12}-x^{-\frac12}}{1-f} \bar v
\end{equation}

\subsection{Counting modifications of products of distinct determinants}
Next, we pick two bosonic letters $X_1$ and $X_2$ with fugacities $x_1$ and $x_2$. This situation will not be needed in the rest of the paper, but we include it for completeness. 
We can study modifications of 
$\det X_1 \det X_2$. We start from the case with no (anti)fundamental letters. 

Using the fermionic trick we have auxiliary $v = (x_1-1) \lambda_1+(x_2-1) \lambda_2$ and
 $\bar v = (1-x_1^{-1})\lambda_1^{-1}+(1-x_2^{-1})\lambda_2^{-1}$. We get an index
 \begin{align}
 &\PE\left[1+x_1^{-1}\frac{(1-x_1)^2}{1-f}\right]\PE\left[1+x_2^{-1}\frac{(1-x_2)^2}{1-f}\right] \cdot\cr
\cdot\oint \frac{d \lambda_1}{2 \pi i \lambda_1} \frac{d \lambda_2}{2 \pi i \lambda_2}& \PE\left[\frac{ (x_1-1)(1-x_2^{-1})}{1-f}\lambda_2^{-1}\lambda_1\right]\PE\left[\frac{ (x_2-1)(1-x_1^{-1})}{1-f}\lambda_1^{-1}\lambda_2\right]
\end{align}
The auxiliary theory is an $U(1) \times U(1)$ quiver gauge theory. 

It is easy to generalize this to include (anti)fundamental fields. It is also straightforward to extend this to modifications of 
$(\det X_1)^{k_1} (\det X_2)^{k_2}$ described by the index of an auxiliary $U(k_1) \times U(k_2)$ gauge theory, as well as 
to products of several types of determinants. 
 
Similarly, one can extend this to the case of quiver gauge theories, including both determinants of adjoint fields and determinants of 
bifundamental fields. All of these can be treated with the fermionization trick. 

As a concrete example, consider $\det X \det Z$ in ${\cal N}=4$ SYM:
 \begin{align}
 &\PE\left[1-\frac{(1-x^{-1})(1-p)(1-q)}{(1-y)(1-z)}\right]\PE\left[1-\frac{(1-z^{-1})(1-p)(1-q)}{(1-x)(1-y)}\right] \cdot\cr
\cdot\oint \frac{d \lambda_1}{2 \pi i \lambda_1} \frac{d \lambda_2}{2 \pi i \lambda_2}& \PE\left[z^{-1} \frac{ (1-p)(1-q)}{1-y}\lambda_2^{-1}\lambda_1\right]\PE\left[ x^{-1} \frac{ (1-p)(1-q)}{1-y}\lambda_1^{-1}\lambda_2\right]
\end{align}
In the holographic dual geometry, the two giant gravitons wrap distinct maximal $S^3$'s in $S^5$, which intersect along an $S^1$. 
At the intersection the open strings stretched between the giant graviton D3 branes give 2d hypermultiplets in a bifundamental
representation of the gauge groups. 

In the index we see precisely the contributions of these hypermultiplets:
\begin{equation}
\PE\left[z^{-1} \frac{ (1-p)(1-q)}{1-y}\lambda\right]\PE\left[ x^{-1} \frac{ (1-p)(1-q)}{1-y}\lambda^{-1}\right] 
\end{equation}

\section{Systematic corrections to $Z_\infty$} \label{sec:expansion}
As we discussed in Section \ref{sec:infty}, the coefficients of $Z_N$ approach limiting values from $Z_\infty$ 
as $N$ becomes sufficiently large. More precisely, the coefficient for some charge $\gamma$ stabilizes whenever the maximum number of letters contributing to 
it is smaller than $N$.  

In this section we pick a specific bosonic field $X$ with fugacity $x$ and focus on operators built with an arbitrary number of $X$ 
letters but a finite fixed number of other letters. For convenience, we will now denote the fugacities as $(x,y_i)$ and the charges as 
a pair $(n,\gamma)$, writing the index as 
\begin{equation}
Z_N(x;y) = \sum_{n,\gamma} c_{N;n,\gamma} x^n y^\gamma = \sum_\gamma Z_{N;\gamma}(x) y^\gamma
\end{equation}
Concretely, we will study the large $N$ behaviour of the $Z_{N;\gamma}(x)$ power series at fixed $\gamma$ and the relation to the naive large $N$ limit 
\begin{equation}
Z_\infty(x;y) = \sum_{n,\gamma} c_{\infty;n,\gamma} x^n y^\gamma = \sum_\gamma Z_{\infty;\gamma}(x) y^\gamma
\end{equation}

We will now describe some structure which appears to hold experimentally for the ${\cal N}=4$ SYM index and its simplifications, but also for 
index-like quantities built from a variety of unphysical $f$'s. 

The discrepancies between $Z_{N;\gamma}(x)$ and $Z_{\infty;\gamma}(x)$ will begin at some order $x^{N+n^{(1)}_\gamma}$ 
for some integer $n^{(1)}_\gamma$. We find that the difference in the coefficients of $x^{N+n}$ for finite $n$ is independent of $N$ at large $N$. 
This statement only fails when $n-N$ hits a certain finite value.  

\subsection{A systematic sequence of corrections}
For future convenience, we will work with the ratio $\frac{Z_N(x;y)}{Z_\infty(x;y)}$ and write
\begin{equation}
\frac{Z_N(x;y)}{Z_\infty(x;y)} \sim_x 1 + \sum_{\gamma, n\geq n^{(1)}_\gamma} d^{(1)}_{N;n,\gamma} x^{N+n} y^\gamma 
\end{equation}
The $\sim_x$ denotes the fact that the structure on the right hand side is only expected to hold 
at finite $\gamma$ and for sufficiently large $N$. 

Our observation means that $d^{(1)}_{N;n,\gamma}$ is $N$-independent until we hit some $n =  n^{(2)}_\gamma$ and thus 
we can define some auxiliary power series $\hat Z_1(x;y)$ such that 
\begin{equation}
\frac{Z_N(x;y)}{Z_\infty(x;y)} \sim_x 1 + x^N \hat Z_1(x;y) + \sum_{\gamma, n\geq n^{(2)}_\gamma} d^{(2)}_{N;n,\gamma} x^{2 N+n} y^\gamma 
\end{equation}
We then find that $d^{(2)}_{N;n,\gamma}$ is $N$-independent until we hit some $n =  n^{(3)}_\gamma$ and thus 
we can define some auxiliary power series $\hat Z_2(x;y)$ such that 
\begin{equation}
\frac{Z_N(x;y)}{Z_\infty(x;y)} \sim_x 1 + x^N \hat Z_1(x;y) + x^{2N} \hat Z_2(x;y) + \sum_{\gamma, n\geq n^{(2)}_\gamma} d^{(2)}_{N;n,\gamma} x^{2 N+n} y^\gamma 
\end{equation}
Assuming that the pattern continues, given sufficient computational powers one could identify a systematic sequence of auxiliary functions 
$\hat Z_k(x;y)$ and write 
\begin{equation}
\frac{Z_N(x;y)}{Z_\infty(x;y)} \sim_x 1 + \sum_{k=1}^\infty x^{k N} \hat Z_k(x;y) + \cdots
\end{equation}
The functions $\hat Z_k(x;y)$ are determined from the large $N$ behaviour of the indices for powers of $x$ which scale linearly in $N$. At this point, we do not try to account for 
powers of $x$ which scale faster than $N$. We indicate this omission by the final ellipsis. 

After we have computed the $\hat Z_k(x;y)$, we could try to compute both sides of this relation at finite $N$. This only makes sense for $N$ such that
the overall powers $n^{(k)}_\gamma+ k N$ grow with $k$ so that the sum is sensible for each $\gamma$. Experimentally, $n^{(k)}_\gamma$ grows quadratically in $k$, 
so this is the case for all $N$. We thus find the surprising result described in the introduction: an exact equality 
\begin{equation}
\frac{Z_N(x;y)}{Z_\infty(x;y)}= 1 + \sum_{k=1}^\infty x^{k N} \hat Z_k(x;y)
\end{equation}
valid for all $N$. This is our ``giant graviton expansion'' for the index. The right hand side is well-defined even for negative $N$, and we find that large cancellations between $\hat Z_k(x;y)$ force the right hand side to vanish at negative $N$.

We will now demonstrate this expansion for various specializations of the ${\cal N}=4$ index as well as for some index-like examples. We refer to Appendix
 \ref{appendix: check of general proposal} for extensive checks of the expansion formula for the full ${\cal N}=4$ index. 

\subsection{The single-matrix index}
As a basic example, consider the case with a single letter $X$. The index is just 
\begin{equation}
Z_N(x) = \frac{1}{\prod_{n=1}^N (1-x^n)}
\end{equation}
and correspondingly 
\begin{equation}
Z_\infty(x) = \frac{1}{\prod_{n=1}^\infty (1-x^n)}.
\end{equation}
We can write 
\begin{equation}
Z_N(x) = Z_\infty(x)\prod_{n=1}^N (1-x^{N+n})
\end{equation}
Expanding the product at the leading order in $x^N$, we have
\begin{equation}
Z_N(x) =Z_\infty(x)  -\frac{x^{N+1}}{1-x} Z_\infty(x) + O(x^{2N})
\end{equation}
We see that the differences begin at order $x^{N+1}$, where the first trace relations occur, and have a structure which is actually independent of $N$ up to order $x^{2N}$. 

In this simple example we can readily do a systematic expansion:
\begin{equation}
\frac{Z_N(x)}{Z_\infty(x)} = \prod_{n=1}^\infty (1-x^{N+n}) =\sum_{k=0}^\infty (-1)^k x^{kN} \frac{x^{\frac{k(k+1)}{2}}}{\prod_{n=1}^k (1-x^n)}
\end{equation} 

Now we make a crucial observation: as rational functions of $x$, we have 
\begin{equation}
(-1)^k  \frac{x^{\frac{k(k+1)}{2}}}{\prod_{n=1}^k (1-x^n)} = \frac{1}{\prod_{n=1}^k (1-x^{-n})} = Z_k(x^{-1}) = \tilde Z_k(\tilde x)
\end{equation} 
In other words, although the ``giant graviton'' index $\tilde Z_k(\tilde x)$ solves the problem of counting determinant modifications 
when expanded in powers of $\tilde x = x^{-1}$, it appears to also provide the systematic corrections $\hat  Z_k(x)$ to the large $N$ expansion when 
analytically continued as a meromorphic function to a power series in $x$. 

This pattern will persist in other examples.

\subsection{The $\frac14$-BPS index}
Next, consider 
\begin{equation}
f = x + y - x y
\end{equation}
so that $1-f = (1-x)(1-y)$ and 
\begin{equation}
Z_\infty(x;y) = \frac{1}{\prod_{n>0} (1-x^n)(1-y^n)}
\end{equation}

We can look first at the coefficient of $y$ in $Z_1(x;y), \cdots, Z_6(x;y)$:
\begin{align}
&1+O\left(x^{18}\right) \cr
&1+x+x^2+x^3+x^4+x^5+x^6+x^7+x^8+x^9+x^{10}+x^{11}+O\left(x^{12}\right) \cr 
   &1+x+2 x^2+2 x^3+3 x^4+3
   x^5+4 x^6+4 x^7+5 x^8+5 x^9+6 x^{10}+6 x^{11}+O\left(x^{12}\right)  \cr
   &1+x+2 x^2+3 x^3+4 x^4+5 x^5+7 x^6+8 x^7+10
   x^8+12 x^9+14 x^{10}+16 x^{11}+O\left(x^{12}\right) \cr
   &1+x+2 x^2+3 x^3+5 x^4+6 x^5+9 x^6+11 x^7+15 x^8+18
   x^9+23 x^{10}+27 x^{11}+O\left(x^{12}\right) \cr
   &1+x+2 x^2+3 x^3+5 x^4+7 x^5+10 x^6+13 x^7+18 x^8+23
   x^9+30 x^{10}+37 x^{11}+O\left(x^{12}\right) \cr
\end{align}
We see the stabilization clearly, with first corrections at order $x^N$. 

Subtracting off $Z_\infty$ we get the first sequence of corrections
\begin{align}
&-x-2 x^2-3 x^3-5 x^4-7 x^5-11 x^6-15 x^7-22 x^8-30 x^9-42 x^{10}-56 x^{11}+O\left(x^{12}\right) \cr
   &-x^2-2 x^3-4 x^4-6 x^5-10 x^6-14 x^7-21 x^8-29 x^9-41
   x^{10}-55 x^{11}-76 x^{12}+O\left(x^{13}\right) \cr
   &-x^3-2 x^4-4 x^5-7 x^6-11 x^7-17 x^8-25 x^9-36 x^{10}-50
   x^{11}-70 x^{12}-94 x^{13}+O\left(x^{14}\right)\cr
   &-x^4-2 x^5-4 x^6-7 x^7-12 x^8-18 x^9-28 x^{10}-40
   x^{11}-58 x^{12}-80 x^{13}-111 x^{14}+O\left(x^{15}\right)\cr
   &-x^5-2 x^6-4 x^7-7 x^8-12 x^9-19 x^{10}-29 x^{11}-43
   x^{12}-62 x^{13}-88 x^{14}-122 x^{15}+O\left(x^{16}\right)\cr
   &-x^6-2 x^7-4 x^8-7 x^9-12 x^{10}-19 x^{11}-30 x^{12}-44
   x^{13}-65 x^{14}-92 x^{15}-130 x^{16}+O\left(x^{17}\right)
\end{align} 
We see clearly the stabilization to a power series 
\begin{equation}
x^N\left(-1-2 x-4 x^2-7 x^3-12 x^4-19 x^5-30 x^6+O\left(x^{7}\right) \right)
\end{equation}
up to corrections which start at order $x^{2N+1}$:
\begin{align}
& x^3+2 x^4+5 x^5+8 x^6+15 x^7+23 x^8+37 x^9+55 x^{10}+83 x^{11}+118 x^{12}+171
   x^{13}+238 x^{14}+O\left(x^{15}\right)\cr
   &x^5+2 x^6+5
   x^7+9 x^8+16 x^9+26 x^{10}+42 x^{11}+63 x^{12}+95 x^{13}+138 x^{14}+198 x^{15}+O\left(x^{16}\right)\cr
   &x^7+2 x^8+5 x^9+9 x^{10}+17 x^{11}+27
   x^{12}+45 x^{13}+68 x^{14}+104 x^{15}+151 x^{16}+O\left(x^{17}\right)\cr
   &x^9+2 x^{10}+5 x^{11}+9 x^{12}+17 x^{13}+28 x^{14}+46
   x^{15}+71 x^{16}+109 x^{17}+O\left(x^{18}\right)\cr
   &x^{11}+2 x^{12}+5 x^{13}+9
   x^{14}+17 x^{15}+28 x^{16}+47 x^{17}+O\left(x^{18}\right)\cr
   &x^{13}+2 x^{14}+5 x^{15}+9
   x^{16}+17 x^{17}+O\left(x^{18}\right)
\end{align} 
We still see a stabilization to a power series 
\begin{equation}
x^{2N}\left(x+2 x^{2}+5 x^{3}+9 x^{4}+17 x^{5}+28 x^{6}+47 x^{7}+O\left(x^{8}\right) \right)
\end{equation}
up to corrections which start at order $x^{3N+3}$. 

We can repeat the exercise for the coefficients of higher powers $y^s$ of $y$. At each stage, the corrections occur earlier as $s$ increases, at order
\begin{equation}
k N + \frac{k(k+1)}{2}- k s 
\end{equation} 
so that we need to go to higher $N$ to see the coefficients stabilize. But if we blindly subtract off these corrections from the low $N$ indices, even at the cost to 
introduce negative powers of $x$, the initial order at which the order $k$ corrections kick in becomes increasingly positive as $k$ increase, 
so that the expansion 
\begin{equation}\label{eq:exp}
\frac{Z_N(x;y)}{Z_\infty(x;y)}= 1 + \sum_{k=1}^\infty x^{k N} \hat Z_k(x;y)
\end{equation}
holds exactly for any value of $N$. 

Next, we can compare the correction coefficients $\hat Z_k(x;y)$ with the dual indices $\tilde Z_k(\tilde x,y)$ computed from 
\begin{equation}
1 -\tilde f = \frac{1-\tilde x}{1-y}
\end{equation}
i.e. 
\begin{equation}
\tilde f = \tilde x - y + \tilde x y + \cdots
\end{equation}

The analytic continuation $\tilde x \to x^{-1}$ is made possible by the following observation: the coefficient $\tilde Z_{k;s}(\tilde x)$ of $y^s$ 
in $\tilde Z_k(\tilde x,y)$ is a rational function with denominator which divides $\prod_{i=1}^k (1-\tilde x^i)$. Thus 
\begin{equation}
\tilde P_{k;s}(\tilde x)\equiv  \tilde Z_{k;s}(\tilde x)\prod_{i=1}^k (1-\tilde x^i)
\end{equation}
is a Laurent polynomial in $\tilde x$ and we can define the analytic continuation of $\tilde Z_{k;s}(\tilde x)$ as the power series expansion in $x$ of 
\begin{equation}
(-1)^k \frac{x^{k(k+1)/2}}{\prod_{i=1}^k (1-x^i)} \tilde P_{k;s}(x^{-1})
\end{equation}
We find that this agrees with the coefficient $\hat Z_{k;s}(x)$ of $y^s$ in $\hat Z_k(x;y)$. Indeed, having an explicitly computable candidate 
expression for $\hat Z_{k;s}(x)$ allows us to check \eqref{eq:exp} to rather high powers of $x$ and $y$ even for small $N$. For negative $N$, the right hand side 
magically vanishes order by order in $x$ and $y$. 

\subsection{A fictitious two-letter index}
As the next example, we consider a problem which does not occur as a limit of 4d SYM: the index counting 
gauge-invariant operators built from two bosonic matrices $X$ and $Y$, with no extra relations: $f = x+y$. 

Now the naive large $N$ index is
\begin{equation}
Z_\infty(x;y) = \frac{1}{\prod_{n>0} (1-x^n -y^n)}
\end{equation}

We can look first at the coefficient of $y$ in $Z_1(x;y), \cdots, Z_6(x;y)$:
\begin{align}
&1+x+x^2+x^3+x^4+x^5+x^6+x^7+x^8+x^9+x^{10}+x^{11}+O\left(x^{12}\right)
   \cr 
   &   1+2 x+3 x^2+4 x^3+5 x^4+6 x^5+7 x^6+8 x^7+9 x^8+10 x^9+11
   x^{10}+12 x^{11}+O\left(x^{12}\right)\cr 
   &1+2 x+4
   x^2+6 x^3+9 x^4+12 x^5+16 x^6+20 x^7+25 x^8+30 x^9+36
   x^{10}+42 x^{11}+O\left(x^{12}\right)\cr 
   &1+2 x+4
   x^2+7 x^3+11 x^4+16 x^5+23 x^6+31 x^7+41 x^8+53 x^9+67
   x^{10}+83 x^{11}+O\left(x^{12}\right)\cr
   &1+2 x+4
   x^2+7 x^3+12 x^4+18 x^5+27 x^6+38 x^7+53 x^8+71 x^9+94
   x^{10}+121 x^{11}+O\left(x^{12}\right)\cr
   &1+2 x+4
   x^2+7 x^3+12 x^4+19 x^5+29 x^6+42 x^7+60 x^8+83 x^9+113
   x^{10}+150 x^{11}+O\left(x^{12}\right)\cr
\end{align}
We see the stabilization to the infinite $N$ value:
\begin{equation}
1+2 x+4x^2+7 x^3+12 x^4+19 x^5+30 x^6+O\left(x^{7}\right)
\end{equation}
up to corrections appearing at order $x^N$:
\begin{align}
&-x-3 x^2-6 x^3-11 x^4-18 x^5-29 x^6-44 x^7-66 x^8-96
   x^9-138 x^{10}-194 x^{11}+O\left(x^{12}\right)\cr
   &-x^2-3 x^3-7 x^4-13 x^5-23
   x^6-37 x^7-58 x^8-87 x^9-128 x^{10}-183 x^{11}-259
   x^{12}+O\left(x^{13}\right)\cr
   &-x^3-3 x^4-7 x^5-14 x^6-25
   x^7-42 x^8-67 x^9-103 x^{10}-153 x^{11}-223 x^{12}-317
   x^{13}+O\left(x^{14}\right)\cr
   &-x^4-3 x^5-7 x^6-14 x^7-26
   x^8-44 x^9-72 x^{10}-112 x^{11}-170 x^{12}-250 x^{13}-361
   x^{14}+O\left(x^{15}\right)\cr
   &-x^5-3 x^6-7 x^7-14 x^8-26
   x^9-45 x^{10}-74 x^{11}-117 x^{12}-179 x^{13}-267
   x^{14}-389 x^{15}+O\left(x^{16}\right)\cr
   &-x^6-3 x^7-7 x^8-14 x^9-26
   x^{10}-45 x^{11}-75 x^{12}-119 x^{13}-184 x^{14}-276
   x^{15}-406 x^{16}+O\left(x^{17}\right)
   \end{align}
This is stabilizing to 
\begin{equation}
x^N\left(-1-3 x-7 x^2-14 x^3-26
   x^{4}-45 x^{5}-75 x^{6}+O\left(x^{7}\right)\right)
\end{equation}
up to corrections which appear at order $x^{2N+1}$:
\begin{align}
&x^3+3 x^4+8 x^5+16 x^6+31 x^7+54 x^8+91 x^9+146
   x^{10}+229 x^{11}+347 x^{12}+518 x^{13}+O\left(x^{14}\right) \cr
   &x^5+3
   x^6+8 x^7+17 x^8+33 x^9+59 x^{10}+101 x^{11}+164
   x^{12}+259 x^{13}+397 x^{14}+595 x^{15}+O\left(x^{16}\right)\cr
   &x^7+3 x^8+8 x^9+17 x^{10}+34
   x^{11}+61 x^{12}+106 x^{13}+174 x^{14}+278 x^{15}+429
   x^{16}+649 x^{17}+O\left(x^{18}\right)\cr
   &x^9+3 x^{10}+8
   x^{11}+17 x^{12}+34 x^{13}+62 x^{14}+108 x^{15}+179
   x^{16}+288 x^{17}+O\left(x^{18}\right)\cr
   &x^{11}+3 x^{12}+8
   x^{13}+17 x^{14}+34 x^{15}+62 x^{16}+109
   x^{17}+O\left(x^{18}\right)\cr
   &x^{13}+3 x^{14}+8 x^{15}+17
   x^{16}+34 x^{17}+O\left(x^{18}\right)
\end{align}
This is stabilizing up to corrections at order $x^{3N+3}$, etc. 

We can analyze in a similar manner higher powers of $y$ and find the usual general pattern, with systematic corrections controlled by $\hat Z_{k;s}(x)$
starting at order 
\begin{equation}
k N + \frac{k(k+1)}{2}- k s 
\end{equation} 

The analytic continuation procedure also seems to work, with the $\hat Z_{k;s}(x)$ taking the form of nice rational functions. The main difference with the previous example is that 
the denominator is slightly more complicated: the $(1-x)$ factor is replaced by $(1-x)^s$ for $s>0$. The resulting $\tilde Z_k(\tilde x, y)$
take the form of fictitious $U(k)$ indices with 
\begin{equation}
\tilde f = \tilde x \frac{1-\tilde x + y}{1-\tilde x + \tilde x y}
\end{equation} 
to be expanded out as a power series in $\tilde x$ and $y$ in the plethystic exponential.

This function does satisfy the expected relation
\begin{equation}
(1-f)(1-\tilde f)= (1-x)(1-\tilde x)
\end{equation} 
as a rational function, but we should stress that $\tilde f$ is {\it not} expanded in a manner compatible with describing determinant modifications, 
which would involve a series expansion around $x=0$. 

Hence the $\tilde Z_k(\tilde x, y)$ here have a more mysterious meaning and cannot be directly related to determinant modifications. 
Indices related by analytic continuation of the argument of a plethystic exponent do occur in some circumstances, such as wallcrossing of holomorphic blocks 
in \cite{Beem:2012mb}. Perhaps something similar is happening here. 

\subsection{Schur index}

The Schur index is a specialization of the full $\mathcal{N}=4$ index, setting $p=z$. It is also the torus partition function of the $U(N)$-gauged $\beta\gamma$ system, the 2d chiral algebra associated to $\mathcal{N}=4$ super Yang-Mills. We consider Schur index in the Ramond sector for simplicity, to avoid powers of $q^{1/2}$ in our formulae. The index in Neveu-Schwarz sector can be obtained from the Ramond index by a shift of the $SU(2)_R$ fugacity $x \to \sqrt{q} x$.

The single letter index in this example is
\begin{equation}
    f = \frac{1}{1-q} \left( x + \frac{q}{x} - 2q \right) = 1 - \frac{(1 - x)(1 - q/x)}{(1 - q)},
\end{equation}
with the naive large $N$ index
\begin{equation}
    Z_\infty (x;q) = \prod_{n>0} \frac{(1-q^n)}{(1-x^n)(1- q^n/x^n)}.
\end{equation}
The Schur index reduces to the $\frac{1}{2}$-BPS single matrix index for $q \to 0$ and has a $\mathbb{Z}_2$ Weyl symmetry exchanging $x \leftrightarrow q/x$.
We can efficiently compute Schur indices with the help of a generating function derived in Appendix \ref{appendix: check of schur proposal}. 

Repeating the analysis in previous examples, we can look at the coefficients of $q$ in $Z_1(x;q), \cdots, Z_6(x;q)$:
\begin{align}
    &\frac{1}{x}-1+O\left(x^{14}\right) \nonumber \\
    &\frac{1}{x}+O\left(x^{14}\right) \nonumber \\
    &\frac{1}{x}+x+x^3+x^5+x^7+x^9+x^{11}+x^{13}+O\left(x^{14}\right) \nonumber \\
    &\frac{1}{x}+x+x^2+x^3+x^4+2 x^5+x^6+2 x^7+2 x^8+2 x^9+2 x^{10}+3 x^{11}+2 x^{12}+3 x^{13}+O\left(x^{14}\right) \nonumber \\
    &\frac{1}{x}+x+x^2+2 x^3+x^4+3 x^5+2 x^6+4 x^7+3 x^8+5 x^9+4 x^{10}+7 x^{11}+5 x^{12}+8 x^{13}+O\left(x^{14}\right) \nonumber \\
    &\frac{1}{x}+x+x^2+2 x^3+2 x^4+3 x^5+3 x^6+5 x^7+5 x^8+7 x^9+7 x^{10}+10 x^{11}+10 x^{12}+13 x^{13}+O\left(x^{14}\right) 
\end{align}
The coefficients stabilize to the infinite $N$ value
\begin{equation}
    \frac{1}{x}+x+x^2+2 x^3+2 x^4+4 x^5+4 x^6+7 x^7+8 x^8+12 x^9+14 x^{10}+21 x^{11}+24 x^{12}+34 x^{13}+O\left(x^{14}\right)
\end{equation}
up to corrections appearing at order $x^{N-1}$:
\begin{align}
    &-1-x-x^2-2 x^3-2 x^4-4 x^5-4 x^6-7 x^7-8 x^8-12 x^9-14 x^{10}-21 x^{11}-24 x^{12}-34 x^{13}+O\left(x^{14}\right) \nonumber \\
    &-x-x^2-2 x^3-2 x^4-4 x^5-4 x^6-7 x^7-8 x^8-12 x^9-14 x^{10}-21 x^{11}-24 x^{12}-34 x^{13}+O\left(x^{14}\right) \nonumber \\
    &-x^2-x^3-2 x^4-3 x^5-4 x^6-6 x^7-8 x^8-11 x^9-14 x^{10}-20 x^{11}-24 x^{12}-33 x^{13}+O\left(x^{14}\right) \nonumber \\
    &-x^3-x^4-2 x^5-3 x^6-5 x^7-6 x^8-10 x^9-12 x^{10}-18 x^{11}-22 x^{12}-31 x^{13}+O\left(x^{14}\right) \nonumber \\
    &-x^4-x^5-2 x^6-3 x^7-5 x^8-7 x^9-10 x^{10}-14 x^{11}-19 x^{12}-26 x^{13}+O\left(x^{14}\right) \nonumber \\
    &-x^5-x^6-2 x^7-3 x^8-5 x^9-7 x^{10}-11 x^{11}-14 x^{12}-21 x^{13}+O\left(x^{14}\right).
\end{align}
We see that the corrections are stabilizing to
\begin{equation}
    x^N \left( -\frac{1}{x}-1-2 x-3 x^2-5 x^3-7 x^4-11 x^5-15 x^6-22 x^7+O\left(x^8\right) \right),
\end{equation}
up to further corrections appearing at order $x^{2N}$:
\begin{align}
    &x^2+x^3+3 x^4+3 x^5+7 x^6+8 x^7+14 x^8+18 x^9+28 x^{10}+35 x^{11}+53 x^{12}+67 x^{13}+O\left(x^{14}\right) \nonumber \\
    &x^4+x^5+3 x^6+4 x^7+7 x^8+10 x^9+16 x^{10}+21 x^{11}+32 x^{12}+43 x^{13}+O\left(x^{14}\right) \nonumber \\
    &x^6+x^7+3 x^8+4 x^9+8 x^{10}+10 x^{11}+18 x^{12}+23 x^{13}+O\left(x^{14}\right) \nonumber \\
    &x^8+x^9+3 x^{10}+4 x^{11}+8 x^{12}+11 x^{13}+O\left(x^{14}\right) \nonumber \\
    &x^{10}+x^{11}+3 x^{12}+4 x^{13}+O\left(x^{14}\right) \nonumber \\
    &x^{12}+x^{13}+O\left(x^{14}\right).
\end{align}
This in turn stabilizes up to corrections at order $x^{3N+2}$, and so on.

Continuing the procedure for higher powers of $q^s$, we observe that the corrections controlled by $\hat{Z}_{k;s}(x)$ occur at order
\begin{equation} \label{eq:schur correction order}
    N(k+1) + \frac{k(k+3)}{2} - (k+2)s  + 1
\end{equation}
of $x$. In other words, at $O(q^s)$, the truncated formula
\begin{equation} \label{eq:schur truncated - new}
    \frac{Z_N (x;q)}{Z_\infty (x;q)} = 1 + x^N \hat{Z}_1 (x;q) + x^{2N} \hat{Z}_2(x;q) + \cdots + x^{kN} \hat{Z}_k(x;q)
\end{equation}
is valid as power series in $x$ up to order \eqref{eq:schur correction order} minus $1$. At large $s$ and small $N$, one may need many correction terms on the right hand side of \eqref{eq:schur truncated - new} to see the convergence of coefficients. Our claim, however, is that the right hand side of \eqref{eq:schur truncated - new} converges to the left to arbitrarily high orders of $x$ given sufficient number $k$ of corrections.

Upon analytic continuation $\tilde{x} \to x^{-1}$, the auxiliary indices $\hat{Z}_k(x;q)$ agree with the dual indices $\tilde{Z}_k(\tilde{x};q)$ defined from
\begin{equation}
    1 - \tilde{f} = \frac{(1-q)(1-\tilde{x})}{(1 - q \tilde{x})}.
\end{equation}
The above definition of $\tilde{f}$ coincides with taking $x \to \tilde{x}$, $q \to q \tilde{x}$ in $f$. This observation allows us to perform extensive checks of the formula
\begin{equation} \label{eq:schur proposal - new}
    \frac{Z_N (x;q)}{Z_\infty (x;q)} = 1 + \sum_{k=1}^\infty x^{k N} \hat{Z}_k (x;q)
\end{equation}
at finite $N$ for the Schur index in Appendix \ref{appendix: check of schur proposal}. In particular, we find that \eqref{eq:schur proposal - new} holds even for negative integer $N$ and for orders of $x$ well beyond $O(x^{N^2})$, at any order of $q$. These properties suggest that the above expansion is indeed exact rather than asymptotic in $N$.

Before moving on to $\mathcal{N}=4$ SYM, we note another striking feature of \eqref{eq:schur proposal - new}. The Schur index has a $\mathbb{Z}_2$ Weyl symmetry exchanging $x \leftrightarrow y=q/x$. It would be natural to seek an alternative formula which treats $x$ and $y$ symmetrically and involves a sum of correction terms controlled by some 
$x^{k N} y^{k' N}$. This was indeed done in \cite{Arai:2020qaj}. We strongly suspect that both formulae may be simultaneously true, as discussed at the end of the introduction. 

\subsection{A final specialization of the $\mathcal{N}=4$ SYM index} \label{subsection: specialized N=4 index}
For a final demonstration, we specialize the $\mathcal{N}=4$ index as
\begin{equation}
    p = q = r, \quad y = z = \frac{r}{\sqrt{x}}.
\end{equation}
This examples includes operators which carry both non-zero spin and R-charges. We refer the reader to Appendix \ref{appendix: check of reduced proposal} for more detailed checks for this specialization. 

Let us look at the coefficients of $r$ in $Z_1(x;r), \cdots, Z_5(x;r)$:
\begin{align}
    &\frac{2}{\sqrt{x}}-2+O\left(x^{11/2}\right) \nonumber \\
    &\frac{2}{\sqrt{x}}-2+2 \sqrt{x}-2 x+2 x^{3/2}-2 x^2+2 x^{5/2}-2 x^3+2 x^{7/2}-2 x^4+2 x^{9/2}-2 x^5+O\left(x^{11/2}\right) \nonumber \\
    &\frac{2}{\sqrt{x}}-2+2 \sqrt{x}-2 x+4 x^{3/2}-4 x^2+4 x^{5/2}-4 x^3+6 x^{7/2}-6 x^4+6 x^{9/2}-6 x^5+O\left(x^{11/2}\right) \nonumber \\
    &\frac{2}{\sqrt{x}}-2+2 \sqrt{x}-2 x+4 x^{3/2}-4 x^2+6 x^{5/2}-6 x^3+8 x^{7/2}-8 x^4+10 x^{9/2}-10 x^5+O\left(x^{11/2}\right) \nonumber \\
    &\frac{2}{\sqrt{x}}-2+2 \sqrt{x}-2 x+4 x^{3/2}-4 x^2+6 x^{5/2}-6 x^3+10 x^{7/2}-10 x^4+12 x^{9/2}-12 x^5+O\left(x^{11/2}\right)
    %&\frac{2}{\sqrt{x}}-2+2 \sqrt{x}-2 x+4 x^{3/2}-4 x^2+6 x^{5/2}-6 x^3+10 x^{7/2}-10 x^4+14 x^{9/2}-14 x^5+O\left(x^{11/2}\right).
\end{align}
The coefficients stabilize to the infinite $N$ value
\begin{equation}
    \frac{2}{\sqrt{x}}-2+2 \sqrt{x}-2 x+4 x^{3/2}-4 x^2+6 x^{5/2}-6 x^3+10 x^{7/2}-10 x^4+14 x^{9/2}-14 x^5+O\left(x^{11/2}\right)
\end{equation}
up to corrections that appear at order $x^{N - \frac{1}{2}}$:
\begin{align}
    &-2 \sqrt{x}+2 x-4 x^{3/2}+4 x^2-6 x^{5/2}+6 x^3-10 x^{7/2}+10 x^4-14 x^{9/2}+14 x^5+O\left(x^{11/2}\right) \nonumber \\
    &-2 x^{3/2}+2 x^2-4 x^{5/2}+4 x^3-8 x^{7/2}+8 x^4-12 x^{9/2}+12 x^5-20 x^{11/2}+20 x^6+O\left(x^{13/2}\right)\nonumber \\
    &-2 x^{5/2}+2 x^3-4 x^{7/2}+4 x^4-8 x^{9/2}+8 x^5-14 x^{11/2}+14 x^6-22 x^{13/2}+22 x^7+O\left(x^{15/2}\right)\nonumber \\
    &-2 x^{7/2}+2 x^4-4 x^{9/2}+4 x^5-8 x^{11/2}+8 x^6-14 x^{13/2}+14 x^7-24 x^{15/2}+24 x^8+O\left(x^{17/2}\right)\nonumber \\
    &-2 x^{9/2}+2 x^5-4 x^{11/2}+4 x^6-8 x^{13/2}+8 x^7-14 x^{15/2}+14 x^8-24 x^{17/2}+24 x^9+O\left(x^{19/2}\right).
    %&-2 x^{11/2}+2 x^6-4 x^{13/2}+4 x^7-8 x^{15/2}+8 x^8-14 x^{17/2}+14 x^9-24 x^{19/2}+24 x^{10}+O\left(x^{21/2}\right).
\end{align}
We see that the corrections are stabilizing to
\begin{equation}
    x^{N} \left( -\frac{2}{\sqrt{x}}+2-4 \sqrt{x}+4 x-8 x^{3/2}+8 x^2-14 x^{5/2}+14 x^3-24 x^{7/2}+24 x^4+O\left(x^{9/2}\right) \right)
\end{equation}
up to further corrections appearing at order $x^{2N +\frac{1}{2}}$
\begin{align}
    &2 x^{5/2}-2 x^3+4 x^{7/2}-4 x^4+10 x^{9/2}-10 x^5+16 x^{11/2}-16 x^6+30 x^{13/2}-30 x^7+O\left(x^{15/2}\right) \nonumber \\
    &2 x^{9/2}-2 x^5+4 x^{11/2}-4 x^6+10 x^{13/2}-10 x^7+18 x^{15/2}-18 x^8+32 x^{17/2}-32 x^9+O\left(x^{19/2}\right) \nonumber \\
    &2 x^{13/2}-2 x^7+4 x^{15/2}-4 x^8+10 x^{17/2}-10 x^9+18 x^{19/2}-18 x^{10}+34 x^{21/2}-34 x^{11}+O\left(x^{23/2}\right) \nonumber \\
    &2 x^{17/2}-2 x^9+4 x^{19/2}-4 x^{10}+10 x^{21/2}-10 x^{11}+18 x^{23/2}-18 x^{12}+34 x^{25/2}-34 x^{13}+O\left(x^{27/2}\right) \nonumber \\
    &2 x^{21/2}-2 x^{11}+4 x^{23/2}-4 x^{12}+10 x^{25/2}-10 x^{13}+18 x^{27/2}-18 x^{14}+34 x^{29/2}-34 x^{15}+O\left(x^{31/2}\right).
    %&2 x^{25/2}-2 x^{13}+4 x^{27/2}-4 x^{14}+10 x^{29/2}-10 x^{15}+18 x^{31/2}-18 x^{16}+34 x^{33/2}-34 x^{17}+O\left(x^{35/2}\right)
\end{align}
This in turn stabilizes up to corrections at order $x^{3N+\frac{5}{2}}$, and so on.

For general powers $r^s$, the corrections controlled by $\hat{Z}_{k;s}(x)$ occur at order
\begin{equation} \label{eq:reduced correction order}
    N(k+1) + \frac{1}{2}k(k+3) - \frac{1}{2}(2k+3)s + 1
\end{equation}
of $x$. That is, at $O(r^s)$, the truncated formula
\begin{equation} \label{eq:reduced truncated - new}
    \frac{Z_N (x;r)}{Z_\infty (x;r)} = 1 + x^N \hat{Z}_1 (x;r) + x^{2N} \hat{Z}_2(x;r) + \cdots + x^{kN} \hat{Z}_k(x;r)
\end{equation}
is valid as power series in $x$ up to order \eqref{eq:reduced correction order} minus $1$. We find that the right hand side of \eqref{eq:reduced truncated - new} converges to the left to arbitrarily high orders of $x$ given sufficient number $k$ of corrections, at any order of $r$.

\subsection{Testing the limits of the giant graviton expansion}

We conclude with a couple of examples of fictitious indices for which the giant graviton expansion does {\it not} work.
A natural question to ask at this point is whether our prescription for $\tilde{Z}_k$ and the associated giant graviton expansion are compatible with any index-like quantity. In other words, do $\tilde{Z}_k(\tilde{x})$ defined through
\begin{equation} \label{eq:tildef relation}
    \frac{1-\tilde{f}}{1-\tilde{x}} = \frac{1-x}{1-f}
\end{equation}
always yield the right corrections in the expansion formula? We expect the answer to be negative: the whole structure should apply only to situations where 
we have a single bosonic letters of fugacity $x$ and we can suppress other letters by positive powers of other fugacities. 

To demonstrate how our prescription breaks down, we study two more indices $f = - x$ and $f = 2 x$. The former is the index of a theory with a single fermionic matrix, while the latter is a specialization of the two-letter index studied earlier. 

The integral over gauge fugacities for the fermion matrix index $f = -x$ gives the exact expression\footnote{We inserted an extra $(-1)^n$ into the plethystic exponential sum as is standard for fermion counting. The result without $(-1)^n$ simply changes the sign in front of $x^{2n-1}$ in $Z_N(x)$.}
\begin{equation} \label{fermion index}
    Z_N(x) = \prod_{n=1}^N (1 + x^{2 n - 1 }).
\end{equation}
It is simple to repeat the exercise of finding corrections to $Z_\infty(x)$ based on stabilizing coefficients. Experimentally, we find that the corrections appear only at even powers of $x^N$:
\begin{equation}
    \frac{Z_N(x)}{Z_\infty(x)} = 1 - x^{2 N} \frac{x}{(1-x^2)} + x^{4 N} \frac{x^2}{(1-x^2)(1-x^4)} - x^{6N} \frac{x^3}{(1-x^2)(1-x^4)(1-x^6)} + \cdots,
\end{equation}
This is readily understood. A fermionic adjoint operator $\xi$ has vanishing determinant, but $\det \xi^2$ is a non-trivial gauge-invariant operator. 
We can tentatively attribute the sequence of corrections to modifications of $(\det \xi^2)^2$. It would be nice to develop this idea further. 

It is also instructive to study the case $f = 2 x$. The indices $Z_1(x), \cdots, Z_5(x)$ are
\begin{align}
    &1+2 x+3 x^2+4 x^3+5 x^4+6 x^5+7 x^6+8 x^7+9 x^8+10 x^9+11 x^{10}+O\left(x^{11}\right) \nonumber \\
    &1+2 x+6 x^2+10 x^3+20 x^4+30 x^5+50 x^6+70 x^7+105 x^8+140 x^9+196 x^{10}+O\left(x^{11}\right) \nonumber \\
    &1+2 x+6 x^2+14 x^3+29 x^4+56 x^5+107 x^6+186 x^7+320 x^8+530 x^9+851 x^{10}+O\left(x^{11}\right) \nonumber \\
    &1+2 x+6 x^2+14 x^3+34 x^4+68 x^5+144 x^6+276 x^7+534 x^8+974 x^9+1774 x^{10}+O\left(x^{11}\right) \nonumber \\
    &1+2 x+6 x^2+14 x^3+34 x^4+74 x^5+159 x^6+324 x^7+657 x^8+1286 x^9+2488 x^{10}+O\left(x^{11}\right).
\end{align}
They stabilize to the infinite $N$ value
\begin{equation}
    1+2 x+6 x^2+14 x^3+34 x^4+74 x^5+166 x^6+350 x^7+746 x^8+1546 x^9+3206 x^{10}+O\left(x^{11}\right),
\end{equation}
up to corrections that appear at order $x^{N+1}$:
\begin{align}
    &-3 x^2-10 x^3-29 x^4-68 x^5-159 x^6-342 x^7-737 x^8-1536 x^9-3195 x^{10}+O\left(x^{11}\right) \nonumber \\
    &-4 x^3-14 x^4-44 x^5-116 x^6-280 x^7-641 x^8-1406 x^9-3010 x^{10}+O\left(x^{11}\right) \nonumber \\
    &-5 x^4-18 x^5-59 x^6-164 x^7-426 x^8-1016 x^9-2355 x^{10}+O\left(x^{11}\right) \nonumber \\
    &-6 x^5-22 x^6-74 x^7-212 x^8-572 x^9-1432 x^{10}+O\left(x^{11}\right) \nonumber \\
    &-7 x^6-26 x^7-89 x^8-260 x^9-718 x^{10}+O\left(x^{11}\right).
\end{align}
The coefficients here \textit{do not} stabilize to definite values, but they do stabilize to $N$-dependent values:
\begin{multline}
    - x^N \Big((N+2) x + (4 N + 6)x^2 + (15 N + 14)x^3 + (48 N + 20) x^4 \\ + (146 N - 12) x^5 + (416 N - 232) x^6 + O(x^7) \cdots \Big).
\end{multline}
It may be possible to develop a systematic expansion in powers of $x^N$ and $N$. We will not attempt to do so.

\section{The M2 brane index}\label{sec:M2}
In this section we study the superconformal index of 3d SCFTs with holographic duals. As a representative example we look at the SCFTs which describe the low energy world-volume theory of $N$ coincident $M2$ branes. 

The world-volume theory of M2 branes is a strongly coupled 3d SCFT. It can be given a variety of distinct UV descriptions:
\begin{itemize}
\item The 3d $U(N)$ {\cal N}=8 SYM
\item The 3d ADHM {\cal N}=4 SYM with $N_f=1$
\item The ABJM theory with level $k=1$
\end{itemize}
The three descriptions preserve different symmetries of the IR theory. Only the latter two are suitable for index calculations, and 
only the second one has a known brane interpretation: a stack of D2 branes in the presence of a D6 brane, which gives a dual description of 
M2 branes at the center of a Taub-NUT geometry. We will focus on that description. 

The main challenge is that 3d gauge theories admit protected monopole operators. The index is thus written as an infinite sum of modified $U(N)$ fugacity integrals, 
one for each monopole charge \cite{Imamura:2011su}. The individual integrals can be treated as we did in the rest of the paper, but each monopole sector has to be treated separately. 

The UV gauge theory description breaks the IR $\mathrm{Spin}(8)_R$ symmetry to $\mathrm{Spin}(4)\times \mathrm{Spin}(4)$, where each subgroup contains 
an $SU(2)$ R-symmetry factor and an $SU(2)$ global symmetry. The permutation of the two $\mathrm{Spin}(4)$ is a self-mirror symmetry in 3d ${\cal N}=4$
language. It exchanges determinant operators and Abelian monopole operators. This gives us two independent ways to study operators dual to giant gravitons. 

In this setting, giant gravitons are M5 branes wrapping $S^5 \subset S^7$. Hence our analysis can shed light on the 
superconformal index of the 6d SCFT living on the worldvolume of M5 branes. 
Past efforts to compute that M5 index focussed on a KK circle reduction of the $S^5$ space geometry to a $\mathbb{C}P^2$ \cite{Kim:2013nva}. This involves a $\mathbb{C}P^2$ compactification of $U(k)$ 5d SYM. The main computational challenge is to properly account for the contribution of instanton configurations on $\mathbb{C}P^2$. 

In our analysis below we will find conjectural expressions for the M5 SCFT index which appear to represent the index of an auxiliary $U(k)$ gauge theory with instanton sectors, but does not appear to be 5d SYM on $\mathbb{C}P^2$. Given the Taub-NUT geometry behind the ADHM quiver description, we suspect that the formulae we find involve a different, singular KK compactification of the space $S^5$ down to an $S^4$, with a circle of Taub-NUT singularities. Equivalently, the M5 index should be identified with the index of 5d SYM in the presence of a codimension 3 defect involving coupling to 2d chiral fermions. We leave a detailed analysis to future work. 
It would also be interesting to determine which formulation of the M5 index may emerge from a large $N$ analysis of the index in the ABJM formulation. 

\subsection{The ADHM theory}
The ADHM theory is a 3d $U(N)$ ${\cal N}=4$ gauge theory with a single adjoint hypermultiplet as well as a single (anti)fundamental hypermultiplet. 
We refer to \cite{Okazaki:2019ony} and references therein for several examples of index calculations for 3d ${\cal N}=4$ gauge theories and of mirror symmetry relations. 

The hypermultiplet single letter index is usually written as 
\begin{equation}
f_H = \frac{q^{\frac14} t s+q^{\frac14} t s^{-1}-q^{\frac34} t^{-1} s-q^{\frac34} t^{-1} s^{-1} }{1-q}
\end{equation}
with $s$ being a flavour fugacity and the vectormultiplet as 
\begin{equation}
f_V = \frac{q^{\frac12} t^{-2} -q^{\frac12} t^{2}}{1-q}
\end{equation}

The vectormultiplet together with the adjoint hypermultiplet combine into 
\begin{equation}
f = \frac{q^{\frac14} t s+q^{\frac14} t s^{-1}+q^{\frac12} t^{-2} -q^{\frac12} t^{2} -q^{\frac34} t^{-1} s-q^{\frac34} t^{-1} s^{-1} }{1-q}
\end{equation}
i.e. in a factorized form
\begin{equation}
1-f =\frac{(1-q^{\frac14} t s)(1-q^{\frac14} t s^{-1})(1-q^{\frac12} t^{-2})}{1-q}
\end{equation}

 The extra (anti)fundamental fields give 
\begin{equation}
v = \frac{q^{\frac14} t-q^{\frac34} t^{-1}}{1-q} \qquad \qquad \bar v = \frac{q^{\frac14} t -q^{\frac34} t^{-1} }{1-q}
\end{equation}
Notice that the single-letter index for the adjoint fields is symmetric under permutation of the three complex scalars, with fugacities $q^{\frac14} t s^{\pm 1}$
and $q^{\frac12} t^{-2}$. The symmetry is broken by the (anti)fundamentals. 

We will also introduce a fugacity $r$ associated to the monopole charge. Monopoles are weighed by powers of $q^{\frac14}t^{-1} r^{\pm 1}$. Self-mirror symmetry exchanges $r$ and $s$ and inverts $t$.

We can simplify the notation by defining
\begin{align}
x &= q^{\frac14} t s \cr
y &= q^{\frac14} t s^{-1} \cr
z &= q^{\frac14} t^{-1} r \cr
w &= q^{\frac14} t^{-1} r^{-1} 
\end{align}
and keep $q$ as well, with the constraint $q=x y z w$. 

Then we can write:
\begin{align}
1-f &=\frac{(1-x)(1-y)(1-z w)}{1-q} \cr
v &= \sqrt{x y} \frac{1-z w}{1-q} \cr
 \bar v &= \sqrt{x y} \frac{1 -z w}{1-q}
\end{align}

The full index will be invariant under the exchange of the pairs of fugacities $(x,y)$ and $(z,w)$, because of the self-mirror property of the ADHM theory. The 
low energy SUSY enhancement to ${\cal N}=8$ actually implies that the index should be invariant under any permutation of the $x$, $y$, $z$ and $w$ fugacities. 
We will verify the symmetry enhancement below in some simplified examples. 

Our basic conjecture is that the full index $Z_N(x,y,z,w)$ would admits a giant graviton expansion in terms of an improved collection of dual indices $\tilde Z_k(\tilde x,y,z,w)$,
which coincide with the superconformal index for the 6d theory of $k$ coincident M5 branes, with $y$, $z$ and $w$ being roughly fugacities for rotational symmetries 
and $\tilde x$, $q$ for R-symmetry generators.

If we could ignore the contribution from monopole sectors, the index would reduce to an $U(N)$ fugacity integral which could be analyzed as 
we did in 4d. The resulting dual index for an expansion based on the $x$ fugacity would be built from  
\begin{align}
1-\tilde f &=\frac{(1-\tilde x)(1-q)}{(1-y)(1-z w)} \cr
\tilde v &= -\sqrt{\tilde x y} \frac{1}{1-y} \cr
\tilde{ \bar v} &= -\sqrt{\tilde x y} \frac{1}{1-y}
\end{align}
with $y z w = \tilde x q$. The adjoint contributions have a denominator with two factors, which is naturally associated to the presence of two planes of rotation. 
The (anti)fundamentals, though, only have a single denominator factor and look like the contributions of 2d chiral fermions. 

Two-dimensional chiral fermions can appear when M5 branes are compactified on a Taub-NUT geometry. A reduction along the circle fiber of Taub-NUT reduces the 6d SCFT to 5d $U(k)$ SYM away from the origin but leaves a co-dimension three defect where 2d chiral fermions are coupled to the 5d gauge fields. BPS operators in the 6d SCFT 
will descend to BPS operators in the 5d theory, but operators which are charged under rotations of the fiber will descend to operators which carry instanton charge. 
Concretely, that means that operators which are perturbative in 5d couple only to the combination $zw$ of the $z$ and $w$ fugacities associated to rotations within Taub-NUT.

We thus conjecture that the above $\tilde f$ and $\tilde v$, $\tilde{ \bar v}$ arise from a formulation of the 6d index in terms of 5d SYM with a co-dimension 3 defect
associated to a Taub-NUT compactification. The corresponding $U(k)$ index should be reproduced by a perturbative 5d calculation while the monopole corrections in 3d should give rise to instanton corrections in 5d. We leave a detailed match to future work.

\subsection{Very simplified indices}

There are several simplified versions of the index where one sends to zero certain fugacities. An extreme choice is to send $y$, $z$ and $w$ to zero. 
This eliminates monopoles and anti-fundamentals. Gauge invariance also eliminates fundamentals. We are left with the familiar count of a single matrix 
$X$ with fugacity $x$. 

It is interesting to observe the mirror limit sending $x$, $y$, $w$ to zero. Then all matter contributions drop out and 
we are left with pure monopole operators. 
Monopole sectors are labelled by a collections of $N$ integers $m_i$. We only need to consider non-negative $m_i$ here, weighed by a power $z^{\sum_i m_i}$:
negative $m_i$ would contribute powers of $w=0$. 

Non-zero monopole charges modify the Vandermonde determinant. 
In this particular limit they simply eliminate $(1-\mu_i/\mu_j)$ factors when $m_i \neq m_j$. If the magnetic charges consist of $n_m$ copies of $m$, 
we effectively get an $\prod_m U(n_m)$ Vandermonde integration measure. 

Including an overall ${N \choose n_1, \cdots}$ combinatorial factor we find that the $\prod_m U(n_m)$ integral gives simply $1$ for every choice of a partition $N=\sum_m n_m$.
We are left with the overall power of $z$. We can readily compute the generating function for all $Z_N$ indices:
\begin{equation}
\sum_{N=0}^\infty \zeta^N Z_N(z) = \sum_{n_m\geq 0} \zeta^{\sum_m n_m} z^{\sum_m m n_m} = \frac{1}{\prod_{m\geq 0} (1- \zeta z^m)}
\end{equation}
which indeed is a generating function for the index
\begin{equation}
Z_N(z) = \frac{1}{\prod_{m=1}^N (1- z^m)}
\end{equation}
expected from mirror symmetry.

We are left with this important insight: the combinatorics of a single matrix $X$ can be reproduced by the 
combinatorics of pure monopole operators of positive charge. It is actually possible to match individual mirror-symmetric operators.
In particular, the operator $\det X$ is mirror to a purely Abelian monopole, with all $m_i=1$. 

The monopole version of the counting of determinant modifications is rather intuitive. In the large $N$ limit, operators of finite charge must have 
a large value of $n_0$. The sum over the remaining, unconstrained, $n_m$ gives the usual $Z_\infty$. If we were to study fluctuations of 
a purely Abelian monopole, with all $m_i=k$, we would consider a situation where $n_k$ is very large and all other $n_m$ are finite. 
Summing over these unconstrained $n_m$ gives formally the expected $Z_\infty \hat Z_k$. 

\subsection{Higgs branch index}
The limit sending $z$ and $w$ to 0 fully suppresses monopole contributions. The reduced index counts ``Higgs'' branch operators. 
We have a variant of a familiar problem
\begin{align}
1-f &=(1-x)(1-y) \cr
v &= \sqrt{x y}  \cr
\bar v &= \sqrt{x y} 
\end{align}
and thus 
\begin{equation}
Z_{\infty}(x;y) =  \frac{1}{\prod_{n=1}^\infty \left(1-x^n \right)\left(1-y^n \right)} \PE\left[ \frac{x y}{ \left(1-x \right)\left(1-y \right)}\right]
\end{equation}
i.e. 
\begin{equation}
Z_{\infty}(x;y) =  \frac{1}{\prod_{n,m \geq 0 | (n,m) \neq (0,0)}\left(1-x^n y^m \right)}
\end{equation}

The giant graviton expansion proceeds as discussed in previous sections, with 
\begin{align}
1-\tilde f &=\frac{1-\tilde x}{1-y} \cr
v &= -\frac{\sqrt{y}}{1-y}  \cr
\bar v &= -\frac{\sqrt{y}}{1-y} 
\end{align}
and large $k$ limit
\begin{equation}
\tilde Z_{\infty}(\tilde x;y) = \prod_{n=1}^\infty \frac{1-y^n}{1-\tilde x^n} \PE\left[ \frac{y}{ \left(1-\tilde x \right)\left(1-y \right)}\right]
\end{equation}
i.e. 
\begin{equation}
\tilde Z_{\infty}(\tilde x;y) =  \frac{1}{\prod_{n>0,m \geq 0}\left(1-\tilde x^n y^m \right)}
\end{equation}
This is indeed the large $N$ limit of an index counting certain BPS operators of the form $\partial_z^m W_n$ in the M5 SCFT,
which give rise to a 2d W-algebra upon $\Omega$ deformation \cite{Beem:2014kka}. 

At this point we could readily conjecture explicit expressions for $Z_N(x;y)$ and $\tilde Z_k(\tilde x;y)$. Instead, we will derive these expressions from 
the mirror description of the problem.

\subsection{Coulomb branch index}
In the limit $x$, $y$ to zero the matter contributions to the index are still mostly suppressed. We have an adjoint scalar $\Phi$ of fugacity 
$z w$ and monopole contributions with $m_i$ which can be any integer. As $q=0$, the effect of non-zero $m_i$ on the integrand is still to 
reduce it to the integrand to that of $\prod_m U(n_m)$, with the same $f$. We can thus write explicitly a generating function for all indices:
\begin{equation}
\sum_{N=0}^\infty \zeta^N Z_N(z;w) = \sum_{n_m\geq 0} \zeta^{\sum_m n_m} z^{\sum_{m>0} m n_m} w^{\sum_{m<0} (-m) n_m} \frac{1}{\prod_m \prod_{k_m=1}^{n_m} (1-(z w)^{k_m})}
\end{equation}
which factorizes into three parts
\begin{align}
&\sum_{n_0\geq 0} \zeta^{n_0} \frac{1}{\prod_{k_0=1}^{n_0} (1-(z w)^{k_0})}\cr
&\prod_{m>0} \left[\sum_{n_m\geq 0} (z^m \zeta)^{n_m}  \frac{1}{\prod_{k_m=1}^{n_m} (1-(z w)^{k_m})}\right]\cr
&\prod_{m<0} \left[\sum_{n_m\geq 0} (w^{-m} \zeta)^{n_m}  \frac{1}{\prod_{k_m=1}^{n_m} (1-(z w)^{k_m})}\right]
\end{align}
giving 
\begin{equation}
\sum_{N=0}^\infty \zeta^N Z_N(z;w) = \frac{1}{\prod_{k_0\geq0}(1- \zeta (z w)^{k_0})}\prod_{m>0} \frac{1}{\prod_{k_m\geq0} (1-z^m \zeta (z w)^{k_m})}\prod_{m<0} \frac{1}{\prod_{k_m\geq 0} (1-w^{-m} \zeta (z w)^{k_m})}
\end{equation}
which can be rearranged to a very simple form
\begin{equation}
\sum_{N=0}^\infty \zeta^N Z_N(z;w) = \frac{1}{\prod_{a\geq 0,b\geq 0}(1- \zeta z^a w^b)}
\end{equation}
These $Z_N(z;w)$ indices actually coincide with the Higgs branch indices from the previous section. 

We will now take a small detour which gives some insight on different possible expansions for the index. Suppose for a moment that $z$ and $w$ have finite values in the unit disk. 
We can now compute $Z_N$ by a contour integral on the unit circle in the $\zeta$ plane, which can be closed at infinity leading to a sum over poles 
at $z^{-k} w^{-k'}$. The contribution from the $(k,k')$ pole is 
\begin{equation}
Z_N^{k,k'}(z;w) = z^{k N} w^{k' N} \frac{1}{\prod_{a\geq 0,b\geq 0|(a,b)\neq (k,k')}(1- z^{a-k} w^{b-k'})}
\end{equation}
The natural holographic interpretation of this sum would involve contributions from giant gravitons wrapping two different 
sub-manifolds in S$^7$. In a mirror picture, that would represent finite modifications of a product $(\det X)^k (\det Y)^{k'}$ of determinants of different fields. 
Notice that $Z_N^{k,k'}$ contains denominator factors of the form $(z-w)$ which are somewhat ambiguous when we attempt a power series expansion in $z$ and $w$. 
Presumably, the ambiguity disappears when multiple terms are added together. 

The terms $(k,0)$ have the form we expect from a giant graviton expansion:
\begin{equation}
Z_N^{k,0}(z;w) = z^{k N}\frac{1}{\prod_{a= 1}^{k}\prod_{b\geq 0}(1- z^{-a} w^{b})}
\frac{1}{\prod_{a\geq 0,b\geq 0|(a,b)\neq (0,0)}(1- z^{a} w^{b})}
\end{equation}
with the expected $Z_\infty(z;w)$ and 
\begin{equation}
\tilde Z_k(\tilde z;w) = \frac{1}{\prod_{a= 1}^{k}\prod_{b\geq 0}(1- \tilde z^a w^{b})}
\end{equation}
Experimentally, we find that the giant graviton expansion works at finite $N$ with the above $\tilde Z_k(\tilde z;w)$:
the sum over $Z_N^{k,0}(z;w)$ only, when expanded first in powers of $w$ and then of $z$, reproduces $Z_N$. 

These $\tilde Z_k(\tilde z;w)$ match the expectation from the M5 brane SCFT: we see the contribution of chiral algebra operators
$\partial_z^b W_a$ for $a=1, \cdots k$. 

We can also run the giant graviton expansion in reverse. We can write 
\begin{equation}
\frac{\tilde Z_N(\tilde z;w)}{\tilde Z_\infty(\tilde z;w)} =\prod_{a= 1}^{\infty}\prod_{b\geq 0}(1- \tilde z^{N+a} w^{b})
\end{equation}
and expand out
\begin{equation}
\frac{\tilde Z_N(\tilde z;w)}{\tilde Z_\infty(\tilde z;w)} =\sum_{k_b=0}^\infty \tilde z^{N \sum_b k_b} w^{\sum_b b k_b} \prod_{b\geq 0} \frac{1}{\prod_{n_b=1}^{k_b}(1-\tilde z^{-1})^{n_b}}
\end{equation}
so that
\begin{equation}
\hat{\tilde Z}_k(\tilde z;w)=\sum_{k_b=0 | \sum_b k_b=k}^\infty w^{\sum_b b k_b} \prod_{b\geq 0} \frac{1}{\prod_{n_b=1}^{k_b}(1-\tilde z^{-1})^{n_b}}
\end{equation}
In the next section, we will recognize this as being the same as $Z_k(z;w)$, as expected. 

\subsection{A mixed index}
Finally, we can send $y$ and $w$ to zero for a peculiar mixed limit which keeps both $X$ and positive monopole charges. 
Proceeding in the same manner as before, we get
\begin{equation}
\sum_{N=0}^\infty \zeta^N Z_N(x,z) = \sum_{n_m\geq 0} \zeta^{\sum_m n_m} z^{\sum_{m} m n_m} \frac{1}{\prod_m \prod_{k_m=1}^{n_m} (1-x^{k_m})}
\end{equation}
and thus 
\begin{equation}
\sum_{N=0}^\infty \zeta^N Z_N(x,z) = \prod_{m\geq 0} \frac{1}{\prod_{k_m\geq0} (1-z^m \zeta x^{k_m})}
\end{equation}
which is just another version of the usual 
\begin{equation}
\sum_{N=0}^\infty \zeta^N Z_N(x,z) = \frac{1}{\prod_{a\geq 0,b\geq 0}(1- \zeta x^a z^b)}
\end{equation}
Indeed, this limit is related to the previous ones by the symmetries of the IR theory. 

\subsection{A minimally reduced index}
We could also consider setting $w$ to $0$ only. We now have a rich matter contribution
\begin{align}
1-f &=(1-x)(1-y) \cr
v &= \sqrt{x y}  \cr
 \bar v &= \sqrt{x y} 
\end{align}
but also include the effect of positive monopole operators. 
Proceeding in the same manner as before, we get
\begin{equation}
\sum_{N=0}^\infty \zeta^N Z_N(x,y,z) = \sum_{n_m\geq 0} \zeta^{\sum_m n_m} z^{\sum_{m} m n_m} Z^{\mathrm{Higgs}}_{n_m}(x,y)
\end{equation}
i.e. 
\begin{equation}
\sum_{N=0}^\infty \zeta^N Z_N(x,y,z) = \prod_{m\geq 0} \frac{1}{\prod_{k_m\geq0, k'_m\geq0} (1-z^m \zeta x^{k_m}y^{k'_m})}
\end{equation}
which gives a nicely symmetric expression
\begin{equation}
\sum_{N=0}^\infty \zeta^N Z_N(x,y,z) =  \frac{1}{\prod_{a\geq0, b\geq0, c\geq0} (1-\zeta x^a y^b z^c)}
\end{equation}

The same contour integral analysis as before gives a tentative
\begin{equation}
\tilde Z_k(\tilde x;y,z) = \frac{1}{\prod_{a= 1}^{k}\prod_{b\geq 0,c\geq 0}(1- \tilde x^a y^{b} z^c)}
\end{equation}
It would be nice to verify that an appropriate reduced sector of the M5 theory has corresponding generators $\partial_{z_1}^b \partial_{z_2}^c W_a$. 

\subsection{The full index}
In the full index, the monopole charges modify the integrand by a rescaling $\mu_a/\mu_b \to \mu_a/\mu_b q^{\frac{|m_a-m_b|}{2}}$
of the Vandermonde and adjoint contributions, $\mu_a \to \mu_a q^{\frac{|m_a|}{2}}$ of the fundamentals, 
$\mu^{-1}_a \to \mu^{-1}_a q^{\frac{|m_a|}{2}}$  of the anti-fundamentals. The powers of $z$ and $w$ are as in the simplified models,
the sum of positive $m_i$ and $-m_i$ respectively.  

The index in each monopole sector thus takes the form of a complicated $\prod_m U(n_m)$ quiver gauge theory, 
with 
\begin{align}
f_{m,m'} &= q^{\frac{|m-m'|}{2}} (f-1) + \delta_{m,m'}\cr
v_m &= q^{\frac{|m|}{2}} v \cr
\bar v_m &= q^{\frac{|m|}{2}} \bar v
\end{align}
The following analysis will be blind to the precise form of $f$ and $v$.

If we consider a large $N$ limit with operators of finite charge, only a finite number of $m_i$ can be non-zero. Thus only $n_0$ will be large. 
We can deal with this as we learned in previous sections. The $U(n_0)$ theory has (anti)fundamental indices 
\begin{equation}
v + \sum_m q^{\frac{|m|}{2}} (f-1) \bar \chi_m \qquad \qquad \bar v + \sum_m q^{\frac{|m|}{2}} (f-1) \chi_m
\end{equation}
where $\chi_m$ and $\bar \chi_m$ are the appropriate characters of gauge fugacities for $U(n_m)$. The large $n_0$ limit leaves behind new effective adjoint and 
(anti)fundamental indices for the remaining $U(n_m)$ nodes at $m\neq 0$:
\begin{align}
f_{m,m'} &= (q^{\frac{|m-m'|}{2}} - q^{\frac{|m|+|m'|}{2}})(f-1)+ \delta_{m,m'}\cr
v_m &= 0 \cr
\bar v_m &= 0
\end{align}
In other words, nodes with different signs for $m$ and $m'$ are completely disconnected. The (anti)fundamental degrees of freedom are effectively gone.
The adjoint contributions and the bifundamentals between nodes with the same sign have a simplification which eliminates the $1-q$ denominator:
a finite collection of letters appear in the effective quivers. 

Unfortunately, the simplification is not sufficient to make $Z_\infty$ fully explicit. This makes a further analysis trickier. We leave it to future work and conclude with a few observations. 

\subsection{Determinants and giant gravitons}
The natural way to insert a determinant in the large $N$ limit is to add (anti)fundamental fermions and antifields 
coupled to the $U(n_0)$ node of the effective quiver. Indeed, in the presence of the monopole, only the $n_0 \times n_0$ components 
of a scalar field $X$ are available to build an operator. 

The heuristic analysis of determinant fluctuations in the large $n_0$ limit then proceeds as before, leading to an effective $U(k)$ gauge theory 
which is coupled to the remaining $U(n_m)$ nodes at $m\neq 0$
 The $U(n_0)$ theory has (anti)fundamental indices 
\begin{equation}
v +(x^{\frac12}-x^{-\frac12}) \bar \chi_\lambda + \sum_m q^{\frac{|m|}{2}} (f-1) \bar \chi_m \qquad \qquad \bar v + (x^{\frac12}-x^{-\frac12}) \chi_\lambda + \sum_m q^{\frac{|m|}{2}} (f-1) \chi_m
\end{equation}
so the new effective bifundamental fields have $f_{m,*} = -(x^{\frac12}-x^{-\frac12})q^{\frac{|m|}{2}}$.

\subsection{Abelian monopoles and giant gravitons}
The Coulomb branch perspective on giant gravitons is also interesting. We can study finite modification of an Abelian monopole operator 
of charge $k$ simply by taking $n_k$ to be large and all other $n_m$ finite. The $U(n_k)$ theory has (anti)fundamental indices 
\begin{equation}
q^{\frac{|k|}{2}} v + \sum_m q^{\frac{|m-k|}{2}} (f-1) \bar \chi_m \qquad \qquad q^{\frac{|k|}{2}} \bar v + \sum_m q^{\frac{|m-k|}{2}} (f-1) \chi_m
\end{equation}
where $\chi_m$ and $\bar \chi_m$ are the appropriate characters of gauge fugacities for $U(n_m)$. The large $n_k$ limit leaves behind new effective adjoint and 
(anti)fundamental indices for the remaining $U(n_m)$ nodes at $m\neq k$:
\begin{align}
f_{m,m'} &= (q^{\frac{|m-m'|}{2}} - q^{\frac{|m-k|+|m'-k|}{2}})(f-1)+ \delta_{m,m'}\cr
v_m &= (q^{\frac{|m|}{2}}-q^{\frac{|k|}{2}} q^{\frac{|m-k|}{2}}) v \cr
\bar v_m &= (q^{\frac{|m|}{2}}-q^{\frac{|k|}{2}} q^{\frac{|m-k|}{2}}) \bar v
\end{align}

\subsection*{Acknowledgements}
This research was supported in part by a grant from the Krembil Foundation. J.H.L. and D.G. are supported by the NSERC Discovery Grant program and by the Perimeter Institute for Theoretical Physics. Research at Perimeter Institute is supported in part by the Government of Canada through the Department of Innovation, Science and Economic Development Canada and by the Province of Ontario through the Ministry of Colleges and Universities.

\appendix
\section{Reflections}\label{app:reflex}
Consider the $U(N)$ integrand for $f(x)=x$:
\begin{equation}
Z_N(x) = \frac{(-1)^{\frac{N(N-1)}{2}}}{N!} \int \left[\prod_a \frac{d\mu_a}{2 \pi i \mu_a (1-x)}\right] \prod_{a \neq b}\frac{\mu_b -\mu_a}{\mu_b - x \mu_a}
\end{equation}
A basic determinant identity for $\det_{a,b} \left[\frac{1}{\mu_b - \nu_a}\right]$ allows one to rewrite that as 
\begin{equation}
Z_N(x) = \frac{x^{-\frac{N(N-1)}{2}}}{N!} \int \left[\prod_a \frac{d\mu_a}{2 \pi i}\right] \det_{a,b} \frac{1}{\mu_b - x \mu_a}
\end{equation}
i.e.
\begin{equation}
Z_N(x) = \frac{x^{-\frac{N(N-1)}{2}}}{N!}\sum_{\sigma \in S_N} (-1)^\sigma \int \left[\prod_a \frac{d\mu_a}{2 \pi i}\right] \frac{1}{\mu_{\sigma(a)} - x \mu_a}
\end{equation}
For every cycle in the permutation $\sigma$ we have a sequence of convolutions. Notice that
\begin{equation}
\int \frac{d\mu_2}{2 \pi i}\frac{1}{\mu_{2} - x^a \mu_1}\frac{1}{\mu_{3} - x^b \mu_2} = \frac{1}{\mu_{3} - x^{a+b} \mu_1}
\end{equation}
so that a cycle of length $\ell$ in $\sigma$ ultimately gives a factor of $(-1)^{\ell-1} \frac{1}{1-x^\ell}$ to the final answer: 
\begin{equation}
Z_N(x) =x^{-\frac{N(N-1)}{2}}(-1)^N \sum_{n_k | \sum_k k n_k=N} \prod_k (-1)^{n_k} \frac{1}{n_k! k^{n_k}} \frac{1}{(1-x^k)^{n_k}}
\end{equation}
We can compute a generating function 
\begin{equation}\label{eq:genapp}
\sum_N \zeta^N x^{\frac{N(N+1)}{2}} Z_N(x) = \sum_{n_k } \prod_k (-1)^{n_k} (-x \zeta)^{k n_k} \frac{1}{n_k! k^{n_k}} \frac{1}{(1-x^k)^{n_k}}
\end{equation}
i.e.
\begin{equation}
\log[\sum_k \zeta^k x^{\frac{k(k+1)}{2}} Z_k(x)] = - \sum_k (x \zeta)^{k} \frac{1}{k} \frac{1}{(1-x^k)} = \sum_{n=1}^\infty \log (1+\zeta x^n)
\end{equation}
which we used with $\zeta = - x^N$ to derive the giant graviton expansion in the simplest toy model. 

Consider now how the answer would change if we inserted in the integrand the character of some representation of $U(N)$. For example, 
consider an adjoint $\sum_{u,v} \mu_u \mu_v^{-1}$. These extra factors will have interesting effects when inserted in the cycles of $\sigma$. 
For example, the insertion of a single power of $\mu_a$ along a cycle gives a vanishing result. For positive $n$ we have 
\begin{equation}
\int \frac{d\mu_2}{2 \pi i}\frac{\mu_2^n }{\mu_{2} - x^a \mu_1}\frac{1}{\mu_{3} - x^b \mu_2} = \frac{x^{a n} \mu_1^n}{\mu_{3} - x^{a+b} \mu_1}
\end{equation}
and for negative $n$ we have 
\begin{equation}
\int \frac{d\mu_2}{2 \pi i}\frac{\mu_2^n }{\mu_{2} - x^a \mu_1}\frac{1}{\mu_{3} - x^b \mu_2} = \frac{x^{-b n} \mu_3^n}{\mu_{3} - x^{a+b} \mu_1}
\end{equation}
If both $\mu_u$ and $\mu_v^{-1}$ are inserted along the same cycle we get a power of $x$ equal to the difference between the location of the insertion of $\mu_u$ and $\mu^{-1}_v$. 

The final answer will thus be a sum of terms where the two eigenvalues fall within a cycle of length $k$. We get an overall factor of $k n_k \frac{1-x^k}{1-x}$. 
Thus in the analogue of \eqref{eq:genapp} we can shift $n_k \to n_k+1$ to reabsorb most of this. We are left with an overall factor of $-\frac{(- x \zeta)^k}{1-x}$.
Summing over $k$, 
this gives an overall factor of 
\begin{equation}
\frac{x \zeta}{(1-x)(1+x \zeta)}
\end{equation}
so that 
\begin{equation}
\sum_k \zeta^k x^{\frac{k(k+1)}{2}} Z^{\mathrm{Adj}}_k(x) =\frac{x \zeta}{1-x} \prod_{n=2}^\infty \log (1+\zeta x^n)
\end{equation}
i.e. 
\begin{equation}
\sum_k \zeta^k x^{\frac{k(k+1)}{2}} Z^{\mathrm{Adj}}_k(x) =\frac{x \zeta}{1-x} \sum_k x^k \zeta^k x^{\frac{k(k+1)}{2}} Z_k(x) 
\end{equation}
or 
\begin{equation}
\sum_k \zeta^k x^{\frac{k(k+1)}{2}} Z^{\mathrm{Adj}}_k(x) =\frac{1}{1-x} \sum_k \zeta^{k} x^{\frac{k(k+1)}{2}} Z_{k-1}(x) 
\end{equation}
so that 
\begin{equation}
Z^{\mathrm{Adj}}_N(x) = \frac{1}{1-x} Z_{N-1}(x) 
\end{equation}

In particular, we have 
\begin{equation}
\frac{Z^{\mathrm{Adj}}_N(x)}{Z_\infty(x)} =  \sum_{k=0}^\infty x^{k N} \frac{x^{-k}}{1-x} Z_k(x^{-1}) 
\end{equation}

We could dub the functions on the right hand side of this as
\begin{equation}
\tilde Z^{\mathrm{Adj}}_k(\tilde x)=  -\frac{\tilde x^{k+1}}{1-\tilde x} Z_k(\tilde x) 
\end{equation}
We have a reciprocal relation:
\begin{equation}
\frac{\tilde Z^{\mathrm{Adj}}_N(x)}{Z_\infty(x)}=  -\frac{x^{N+1}}{1-x} \sum_{k=0}^\infty x^{k N} Z_k(x^{-1}) =  \sum_{k=1}^\infty x^{k N} Z^{\mathrm{Adj}}_{k}(x^{-1})
\end{equation}

We expect a similar pattern to persist when we replace the adjoint representation by other representations. This would imply that for any index-like quantity $Z_N(x;y)$ 
we can write a giant-graviton-like expansion for the ratio $Z_N(x;y)/Z_\infty(x)$ and thus for $Z_N(x;y)/Z_\infty(x;y)$. 

\section{Schur index}\label{section: schur}

The Schur index $Z_N(q,x)$ is a specialization of the $\mathcal{N}=4$ index. It is also the torus partition function of the $U(N)$-gauged adjoint $\beta\gamma$ system, which arises as the 2d chiral algebra associated to $\mathcal{N}=4$ super Yang-Mills. We consider Schur index in the Ramond sector for simplicity, to avoid powers of $q^{1/2}$ in our formulae. The index in Neveu-Schwarz sector can be obtained from the Ramond index by a shift of the $SU(2)_R$ fugacity $x \to \sqrt{q} x$.

The full expression for the Schur index in Ramond sector is
\begin{equation}
Z_N = \frac{1}{N!} \oint \prod_i \frac{d \sigma_i}{2 \pi i \sigma_i}  \frac{\prod_{i<j} (1-\sigma_i \sigma_j^{-1})(1-\sigma_j \sigma_i^{-1})}{\prod_{i,j} (1 - x \sigma_i \sigma_j^{-1})} \frac{\prod_{i, j}  \prod_{n=1}^\infty (1-\sigma_i/\sigma_j q^n)(1-\sigma_j/\sigma_i q^n)}{\prod_{i,j}  \prod_{n=1}^\infty (1-x \sigma_i/\sigma_j q^n)(1-x^{-1} \sigma_j/\sigma_i q^n) }
\end{equation}
This can be understood as the BRST reduction of a module built from a highest weight annihilated by the zero mode of one of the two symplectic bosons, say $Y_0$. The zero-energy states are built from the action of $\mathrm{Tr} X_0^n$ on that vector. 

Here, we cannot solve the index exactly in $N$. We can, however, evaluate the indices perturbatively for a given $N$ and see if the statement we made above for the toy index generalizes.

An explicit formula for a family of generating functions of the Schur index is
\begin{equation} \label{schur generating fn}
\sum_N \frac{\theta(u x^N)}{\theta(u)} Z_N(q;x) x^{\frac{N^2}{2}} z^N = \prod_{n=-\infty}^\infty \left( 1 + \frac{z x^n}{1-u q^n} \right)
\end{equation}
to be understood as expanded in the regime $|q|<|x|<1$. This equality holds for all values of $u$ and can be derived by an application of a determinant identity for theta functions followed by a Fermi Gas analysis, which rewrites the generating function as the determinant of an integral operator.  

%If we set, say, $u=q^{-\frac12}$ and send $q \to 0$, we get
%\begin{equation}
%\sum_N c_N(x) x^{\frac{N(N+1)}{2}} z^N = \prod_{n=1}^\infty \left( 1 +z x^n \right).
%\end{equation}
%This is a generating function for the toy index that is different from the one presented in Section \ref{section: toy index}.\footnote{If we keep $u$ fixed and send $q \to 0$, we get 
%\begin{equation}
%\sum_N \frac{1-u x^N}{1-u} c_N(x) x^{\frac{N(N-1)}{2}} z^N = \left( 1 + \frac{z}{1-u} \right) \prod_{n=1}^\infty \left( 1 +z x^n \right)
%\end{equation}
%which reduces to the above.}

\section{Tests for the Schur index} \label{appendix: check of schur proposal}

In this section, we provide evidence in support of the giant graviton expansion of the Schur index. For convenience, we rewrite the proposal here:
\begin{equation}
    Z_N(x;q) = Z_\infty(x;q) \left[ 1 + x^N \hat{Z}_1(x;q) + x^{2 N} \hat{Z}_2(x;q) + \cdots \right].
\end{equation}
In practice, let us consider a truncation of the formula at the $k$-th correction:
\begin{equation} \label{schurtruncated-appendix}
    Z_N(x;q) = Z_\infty(x;q) \left[ 1 + x^N \hat{Z}_1(x;q) + \cdots + x^{k N} \hat{Z}_k(x;q) \right],
\end{equation}
where we also define
\[
Z_0 = 1, \quad Z_{-1} = Z_{-2} = \cdots = 0.
\]

The following tables enumerate the series coefficients of $x$ of the left and right hand sides of \eqref{schurtruncated-appendix} for a given $N$, order of $q$, and truncation level $k$. Integers $n$ denote the power of $x$ in the series. The top row lists the powers $n$ of $x$. The middle rows tabulate the coefficients of $x^n$ of the right hand side of \eqref{schurtruncated-appendix} for different values of the truncation $k$. The final row tabulates the coefficients of $x^n$ of $Z_N(q,x)$.

From the results below, we observe that the order of $x$ to which the formula holds, at a given order $s$ of $q$, is
\begin{equation}
N + \frac{k}{2} (k+3+2N) - (2+k)s.
\end{equation}
Checks beyond those presented below can be readily conducted by using the generating function \eqref{schur generating fn} for the Schur index.

\begin{landscape}

\newpage

\noindent At $N=1$ and $O(q^{0})$,

\noindent
\begin{tabular}{|c|ccccccccccccccccccccc|}
\hline
$n$ & 0 & 1 & 2 & 3 & 4 & 5 & 6 & 7 & 8 & 9 & 10 & 11 & 12 & 13 & 14 & 15 & 16 & 17 & 18 & 19 & 20 \\
\hline \hline
$k=0$ & 1 & 1 & 2 & 3 & 5 & 7 & 11 & 15 & 22 & 30 & 42 & 56 & 77 & 101 & 135 & 176 & 231 & 297 & 385 & 490 & 627 \\
$k=1$ & 1 & 1 & 1 & 1 & 1 & 0 & -1 & -4 & -8 & -15 & -25 & -41 & -62 & -94 & -137 & -197 & -277 & -387 & -530 & -722 & -970 \\
$k=2$ & 1 & 1 & 1 & 1 & 1 & 1 & 1 & 1 & 1 & 2 & 3 & 6 & 11 & 20 & 33 & 56 & 88 & 138 & 208 & 311 & 452 \\
$k=3$ & 1 & 1 & 1 & 1 & 1 & 1 & 1 & 1 & 1 & 1 & 1 & 1 & 1 & 1 & 0 & -1 & -4 & -9 & -19 & -34 & -60 \\
$k=4$ & 1 & 1 & 1 & 1 & 1 & 1 & 1 & 1 & 1 & 1 & 1 & 1 & 1 & 1 & 1 & 1 & 1 & 1 & 1 & 1 & 2 \\
$k=5$ & 1 & 1 & 1 & 1 & 1 & 1 & 1 & 1 & 1 & 1 & 1 & 1 & 1 & 1 & 1 & 1 & 1 & 1 & 1 & 1 & 1 \\
\hline
LHS & 1 & 1 & 1 & 1 & 1 & 1 & 1 & 1 & 1 & 1 & 1 & 1 & 1 & 1 & 1 & 1 & 1 & 1 & 1 & 1 & 1 \\
\hline
\end{tabular}

\noindent At $N=1$ and $O(q^{1})$,

\noindent \begin{tabular}{|c|cccccccccccccccccccccc|}
\hline
$n$ & -1 & 0 & 1 & 2 & 3 & 4 & 5 & 6 & 7 & 8 & 9 & 10 & 11 & 12 & 13 & 14 & 15 & 16 & 17 & 18 & 19 & 20 \\
\hline \hline
$k=0$ & 1 & 0 & 1 & 1 & 2 & 2 & 4 & 4 & 7 & 8 & 12 & 14 & 21 & 24 & 34 & 41 & 55 & 66 & 88 & 105 & 137 & 165 \\
$k=1$ & 1 & -1 & 0 & -1 & -1 & -3 & -3 & -7 & -8 & -14 & -18 & -28 & -35 & -53 & -67 & -94 & -121 & -165 & -209 & -280 & -353 & -462 \\
$k=2$ & 1 & -1 & 0 & 0 & 0 & 0 & 1 & 1 & 3 & 5 & 8 & 13 & 21 & 30 & 45 & 66 & 92 & 130 & 180 & 246 & 333 & 449 \\
$k=3$ & 1 & -1 & 0 & 0 & 0 & 0 & 0 & 0 & 0 & 0 & -1 & -1 & -3 & -5 & -10 & -14 & -26 & -37 & -60 & -85 & -129 & -180 \\
$k=4$ & 1 & -1 & 0 & 0 & 0 & 0 & 0 & 0 & 0 & 0 & 0 & 0 & 0 & 0 & 0 & 1 & 1 & 3 & 5 & 10 & 16 & 27 \\
$k=5$ & 1 & -1 & 0 & 0 & 0 & 0 & 0 & 0 & 0 & 0 & 0 & 0 & 0 & 0 & 0 & 0 & 0 & 0 & 0 & 0 & 0 & -1 \\
\hline
LHS & 1 & -1 & 0 & 0 & 0 & 0 & 0 & 0 & 0 & 0 & 0 & 0 & 0 & 0 & 0 & 0 & 0 & 0 & 0 & 0 & 0 & 0 \\
\hline
\end{tabular}

\noindent At $N=1$ and $O(q^{2})$,

\noindent \begin{tabular}{|c|ccccccccccccccccccccccc|}
\hline
$n$ & -2 & -1 & 0 & 1 & 2 & 3 & 4 & 5 & 6 & 7 & 8 & 9 & 10 & 11 & 12 & 13 & 14 & 15 & 16 & 17 & 18 & 19 & 20 \\
\hline \hline
$k=0$ & 2 & 1 & 2 & 3 & 5 & 6 & 10 & 12 & 18 & 23 & 32 & 40 & 56 & 69 & 92 & 116 & 151 & 187 & 242 & 298 & 379 & 467 & 585 \\
$k=1$ & 1 & -1 & -1 & -3 & -4 & -8 & -11 & -19 & -26 & -40 & -55 & -80 & -107 & -151 & -200 & -272 & -357 & -476 & -616 & -808 & -1035 & -1337 & -1699 \\
$k=2$ & 1 & 0 & 0 & 0 & 1 & 2 & 4 & 7 & 12 & 20 & 31 & 48 & 72 & 106 & 152 & 218 & 304 & 423 & 578 & 787 & 1056 & 1413 & 1866 \\
$k=3$ & 1 & 0 & 0 & -1 & 0 & 0 & -1 & -1 & -2 & -4 & -7 & -11 & -20 & -30 & -49 & -74 & -113 & -164 & -245 & -346 & -496 & -692 & -967 \\
$k=4$ & 1 & 0 & 0 & -1 & 0 & 0 & 0 & 0 & 0 & 0 & 1 & 1 & 2 & 4 & 7 & 11 & 19 & 30 & 47 & 73 & 111 & 165 & 245 \\
$k=5$ & 1 & 0 & 0 & -1 & 0 & 0 & 0 & 0 & 0 & 0 & 0 & 0 & 0 & 0 & 0 & -1 & -1 & -2 & -4 & -7 & -11 & -19 & -29 \\
\hline
LHS & 1 & 0 & 0 & -1 & 0 & 0 & 0 & 0 & 0 & 0 & 0 & 0 & 0 & 0 & 0 & 0 & 0 & 0 & 0 & 0 & 0 & 0 & 0 \\
\hline
\end{tabular}

\newpage

\noindent At $N=1$ and $O(q^{3})$,

\noindent \begin{tabular}{|c|cccccccccccccccccccccc|}
\hline
$n$ & -4 & -3 & -2 & -1 & 0 & 1 & 2 & 3 & 4 & 5 & 6 & 7 & 8 & 9 & 10 & 11 & 12 & 13 & 14 & 15 & 16 & 17 \\
\hline \hline
$k=0$ & 0 & 3 & 1 & 3 & 4 & 7 & 8 & 14 & 16 & 25 & 31 & 44 & 54 & 77 & 93 & 126 & 157 & 206 & 253 & 330 & 403 & 516 \\
$k=1$ & -1 & 1 & -2 & -3 & -5 & -8 & -14 & -19 & -31 & -43 & -63 & -87 & -124 & -165 & -229 & -304 & -407 & -534 & -706 & -911 & -1187 & -1519 \\
$k=2$ & 0 & 2 & 1 & 0 & 4 & 4 & 8 & 12 & 21 & 29 & 49 & 67 & 103 & 144 & 210 & 285 & 408 & 547 & 755 & 1006 & 1360 & 1786 \\
$k=3$ & 0 & 1 & 0 & -2 & 0 & -2 & -3 & -6 & -7 & -14 & -19 & -32 & -45 & -71 & -99 & -151 & -209 & -304 & -421 & -596 & -811 & -1125 \\
$k=4$ & 0 & 1 & 0 & -1 & 1 & 0 & 0 & 1 & 2 & 3 & 5 & 9 & 13 & 21 & 31 & 48 & 69 & 104 & 146 & 215 & 300 & 429 \\
$k=5$ & 0 & 1 & 0 & -1 & 1 & 0 & -1 & 0 & 0 & 0 & -1 & -1 & -2 & -3 & -6 & -8 & -15 & -20 & -34 & -47 & -73 & -102 \\
\hline
LHS & 0 & 1 & 0 & -1 & 1 & 0 & -1 & 0 & 0 & 0 & 0 & 0 & 0 & 0 & 0 & 0 & 0 & 0 & 0 & 0 & 0 & 0 \\
\hline
\end{tabular}

\noindent At $N=1$ and $O(q^{4})$,

\noindent \begin{tabular}{|c|ccccccccccccccccccccc|}
\hline
$n$ & -7 & -6 & -5 & -4 & -3 & -2 & -1 & 0 & 1 & 2 & 3 & 4 & 5 & 6 & 7 & 8 & 9 & 10 & 11 & 12 & 13 \\
\hline \hline
$k=0$ & 0 & 0 & 0 & 5 & 2 & 5 & 7 & 12 & 14 & 24 & 28 & 43 & 54 & 76 & 94 & 133 & 162 & 218 & 273 & 357 & 440 \\
$k=1$ & 0 & -1 & -2 & 1 & -5 & -7 & -11 & -17 & -29 & -39 & -62 & -85 & -123 & -169 & -238 & -315 & -435 & -574 & -765 & -1000 & -1317 \\
$k=2$ & 1 & 0 & 1 & 4 & 3 & 5 & 11 & 14 & 24 & 35 & 55 & 78 & 120 & 165 & 242 & 335 & 472 & 640 & 889 & 1186 & 1609 \\
$k=3$ & 0 & -1 & -1 & 0 & -2 & -4 & -6 & -10 & -14 & -24 & -33 & -51 & -73 & -109 & -152 & -223 & -308 & -439 & -603 & -840 & -1139 \\
$k=4$ & 0 & 0 & 0 & 2 & 1 & 2 & 2 & 4 & 7 & 10 & 15 & 24 & 33 & 52 & 73 & 105 & 149 & 215 & 294 & 417 & 572 \\
$k=5$ & 0 & 0 & 0 & 1 & 0 & 0 & -1 & -1 & -2 & -3 & -5 & -7 & -12 & -16 & -25 & -37 & -53 & -74 & -111 & -151 & -215 \\
\hline
LHS & 0 & 0 & 0 & 1 & 0 & 0 & 0 & 0 & 0 & 0 & -1 & 0 & 0 & 0 & 0 & 0 & 0 & 0 & 0 & 0 & 0 \\
\hline
\end{tabular}

\noindent At $N=2$ and $O(q^{0})$,

\noindent \begin{tabular}{|c|ccccccccccccccccccccc|}
\hline
$n$ & 0 & 1 & 2 & 3 & 4 & 5 & 6 & 7 & 8 & 9 & 10 & 11 & 12 & 13 & 14 & 15 & 16 & 17 & 18 & 19 & 20 \\
\hline \hline
$k=0$ & 1 & 1 & 2 & 3 & 5 & 7 & 11 & 15 & 22 & 30 & 42 & 56 & 77 & 101 & 135 & 176 & 231 & 297 & 385 & 490 & 627 \\
$k=1$ & 1 & 1 & 2 & 2 & 3 & 3 & 4 & 3 & 3 & 0 & -3 & -11 & -20 & -38 & -60 & -96 & -142 & -211 & -299 & -425 & -585 \\
$k=2$ & 1 & 1 & 2 & 2 & 3 & 3 & 4 & 4 & 5 & 5 & 6 & 6 & 8 & 9 & 13 & 18 & 28 & 42 & 66 & 100 & 153 \\
$k=3$ & 1 & 1 & 2 & 2 & 3 & 3 & 4 & 4 & 5 & 5 & 6 & 6 & 7 & 7 & 8 & 8 & 9 & 9 & 9 & 8 & 6 \\
$k=4$ & 1 & 1 & 2 & 2 & 3 & 3 & 4 & 4 & 5 & 5 & 6 & 6 & 7 & 7 & 8 & 8 & 9 & 9 & 10 & 10 & 11 \\
$k=5$ & 1 & 1 & 2 & 2 & 3 & 3 & 4 & 4 & 5 & 5 & 6 & 6 & 7 & 7 & 8 & 8 & 9 & 9 & 10 & 10 & 11 \\
\hline
LHS & 1 & 1 & 2 & 2 & 3 & 3 & 4 & 4 & 5 & 5 & 6 & 6 & 7 & 7 & 8 & 8 & 9 & 9 & 10 & 10 & 11 \\
\hline
\end{tabular}

\newpage

\noindent At $N=2$ and $O(q^{1})$,

\noindent \begin{tabular}{|c|ccccccccccccccccccccc|}
\hline
$n$ & -1 & 0 & 1 & 2 & 3 & 4 & 5 & 6 & 7 & 8 & 9 & 10 & 11 & 12 & 13 & 14 & 15 & 16 & 17 & 18 & 19 \\
\hline \hline
$k=0$ & 1 & 0 & 1 & 1 & 2 & 2 & 4 & 4 & 7 & 8 & 12 & 14 & 21 & 24 & 34 & 41 & 55 & 66 & 88 & 105 & 137 \\
$k=1$ & 1 & 0 & 0 & 0 & 0 & -1 & -1 & -3 & -4 & -7 & -10 & -16 & -21 & -32 & -43 & -60 & -80 & -110 & -143 & -192 & -248 \\
$k=2$ & 1 & 0 & 0 & 0 & 0 & 0 & 0 & 0 & 0 & 1 & 1 & 3 & 5 & 9 & 13 & 23 & 32 & 50 & 70 & 103 & 141 \\
$k=3$ & 1 & 0 & 0 & 0 & 0 & 0 & 0 & 0 & 0 & 0 & 0 & 0 & 0 & 0 & -1 & -1 & -3 & -5 & -10 & -15 & -26 \\
$k=4$ & 1 & 0 & 0 & 0 & 0 & 0 & 0 & 0 & 0 & 0 & 0 & 0 & 0 & 0 & 0 & 0 & 0 & 0 & 0 & 0 & 1 \\
$k=5$ & 1 & 0 & 0 & 0 & 0 & 0 & 0 & 0 & 0 & 0 & 0 & 0 & 0 & 0 & 0 & 0 & 0 & 0 & 0 & 0 & 0 \\
\hline
LHS & 1 & 0 & 0 & 0 & 0 & 0 & 0 & 0 & 0 & 0 & 0 & 0 & 0 & 0 & 0 & 0 & 0 & 0 & 0 & 0 & 0 \\
\hline
\end{tabular}

\noindent At $N=2$ and $O(q^{2})$,

\noindent \begin{tabular}{|c|ccccccccccccccccccccc|}
\hline
$n$ & -2 & -1 & 0 & 1 & 2 & 3 & 4 & 5 & 6 & 7 & 8 & 9 & 10 & 11 & 12 & 13 & 14 & 15 & 16 & 17 & 18 \\
\hline \hline
$k=0$ & 2 & 1 & 2 & 3 & 5 & 6 & 10 & 12 & 18 & 23 & 32 & 40 & 56 & 69 & 92 & 116 & 151 & 187 & 242 & 298 & 379 \\
$k=1$ & 2 & 0 & 0 & 0 & -1 & -3 & -4 & -9 & -13 & -21 & -31 & -47 & -64 & -94 & -128 & -176 & -237 & -321 & -421 & -560 & -727 \\
$k=2$ & 2 & 0 & 0 & 1 & 0 & 0 & 1 & 1 & 2 & 5 & 7 & 13 & 22 & 34 & 51 & 81 & 115 & 169 & 240 & 339 & 467 \\
$k=3$ & 2 & 0 & 0 & 1 & 0 & 0 & 0 & 0 & 0 & 0 & -1 & -1 & -2 & -4 & -8 & -11 & -21 & -32 & -52 & -78 & -120 \\
$k=4$ & 2 & 0 & 0 & 1 & 0 & 0 & 0 & 0 & 0 & 0 & 0 & 0 & 0 & 0 & 0 & 1 & 1 & 2 & 4 & 7 & 12 \\
$k=5$ & 2 & 0 & 0 & 1 & 0 & 0 & 0 & 0 & 0 & 0 & 0 & 0 & 0 & 0 & 0 & 0 & 0 & 0 & 0 & 0 & 0 \\
\hline
LHS & 2 & 0 & 0 & 1 & 0 & 0 & 0 & 0 & 0 & 0 & 0 & 0 & 0 & 0 & 0 & 0 & 0 & 0 & 0 & 0 & 0 \\
\hline
\end{tabular}

\noindent At $N=2$ and $O(q^{3})$,

\noindent \begin{tabular}{|c|ccccccccccccccccccccc|}
\hline
$n$ & -3 & -2 & -1 & 0 & 1 & 2 & 3 & 4 & 5 & 6 & 7 & 8 & 9 & 10 & 11 & 12 & 13 & 14 & 15 & 16 & 17 \\
\hline \hline
$k=0$ & 3 & 1 & 3 & 4 & 7 & 8 & 14 & 16 & 25 & 31 & 44 & 54 & 77 & 93 & 126 & 157 & 206 & 253 & 330 & 403 & 516 \\
$k=1$ & 2 & -1 & 0 & -2 & -2 & -7 & -8 & -17 & -22 & -37 & -50 & -77 & -101 & -149 & -196 & -273 & -358 & -487 & -629 & -838 & -1074 \\
$k=2$ & 2 & 0 & 1 & 1 & 1 & 2 & 4 & 5 & 9 & 15 & 22 & 35 & 53 & 78 & 113 & 166 & 231 & 328 & 452 & 623 & 843 \\
$k=3$ & 2 & 0 & 1 & 0 & 0 & 0 & 0 & -1 & -2 & -3 & -6 & -8 & -15 & -21 & -35 & -49 & -78 & -108 & -165 & -228 & -333 \\
$k=4$ & 2 & 0 & 1 & 0 & 0 & 0 & 1 & 0 & 0 & 0 & 1 & 1 & 2 & 3 & 6 & 9 & 14 & 22 & 34 & 50 & 75 \\
$k=5$ & 2 & 0 & 1 & 0 & 0 & 0 & 1 & 0 & 0 & 0 & 0 & 0 & 0 & 0 & 0 & -1 & -1 & -2 & -3 & -6 & -9 \\
\hline
LHS & 2 & 0 & 1 & 0 & 0 & 0 & 1 & 0 & 0 & 0 & 0 & 0 & 0 & 0 & 0 & 0 & 0 & 0 & 0 & 0 & 0 \\
\hline
\end{tabular}

\newpage

\noindent At $N=3$ and $O(q^{0})$,

\noindent \begin{tabular}{|c|ccccccccccccccccccccc|}
\hline
$n$ & 0 & 1 & 2 & 3 & 4 & 5 & 6 & 7 & 8 & 9 & 10 & 11 & 12 & 13 & 14 & 15 & 16 & 17 & 18 & 19 & 20 \\
\hline \hline
$k=0$ & 1 & 1 & 2 & 3 & 5 & 7 & 11 & 15 & 22 & 30 & 42 & 56 & 77 & 101 & 135 & 176 & 231 & 297 & 385 & 490 & 627 \\
$k=1$ & 1 & 1 & 2 & 3 & 4 & 5 & 7 & 8 & 10 & 11 & 12 & 11 & 10 & 4 & -4 & -19 & -41 & -76 & -123 & -194 & -288 \\
$k=2$ & 1 & 1 & 2 & 3 & 4 & 5 & 7 & 8 & 10 & 12 & 14 & 16 & 19 & 21 & 24 & 28 & 32 & 38 & 47 & 59 & 77 \\
$k=3$ & 1 & 1 & 2 & 3 & 4 & 5 & 7 & 8 & 10 & 12 & 14 & 16 & 19 & 21 & 24 & 27 & 30 & 33 & 37 & 40 & 44 \\
$k=4$ & 1 & 1 & 2 & 3 & 4 & 5 & 7 & 8 & 10 & 12 & 14 & 16 & 19 & 21 & 24 & 27 & 30 & 33 & 37 & 40 & 44 \\
$k=5$ & 1 & 1 & 2 & 3 & 4 & 5 & 7 & 8 & 10 & 12 & 14 & 16 & 19 & 21 & 24 & 27 & 30 & 33 & 37 & 40 & 44 \\
\hline
LHS & 1 & 1 & 2 & 3 & 4 & 5 & 7 & 8 & 10 & 12 & 14 & 16 & 19 & 21 & 24 & 27 & 30 & 33 & 37 & 40 & 44 \\
\hline
\end{tabular}

\noindent At $N=3$ and $O(q^{1})$,

\noindent \begin{tabular}{|c|ccccccccccccccccccccc|}
\hline
$n$ & -1 & 0 & 1 & 2 & 3 & 4 & 5 & 6 & 7 & 8 & 9 & 10 & 11 & 12 & 13 & 14 & 15 & 16 & 17 & 18 & 19 \\
\hline \hline
$k=0$ & 1 & 0 & 1 & 1 & 2 & 2 & 4 & 4 & 7 & 8 & 12 & 14 & 21 & 24 & 34 & 41 & 55 & 66 & 88 & 105 & 137 \\
$k=1$ & 1 & 0 & 1 & 0 & 1 & 0 & 1 & -1 & 0 & -3 & -3 & -8 & -9 & -18 & -22 & -36 & -46 & -69 & -88 & -126 & -160 \\
$k=2$ & 1 & 0 & 1 & 0 & 1 & 0 & 1 & 0 & 1 & 0 & 1 & 0 & 2 & 1 & 4 & 5 & 10 & 14 & 24 & 34 & 53 \\
$k=3$ & 1 & 0 & 1 & 0 & 1 & 0 & 1 & 0 & 1 & 0 & 1 & 0 & 1 & 0 & 1 & 0 & 1 & 0 & 0 & -1 & -2 \\
$k=4$ & 1 & 0 & 1 & 0 & 1 & 0 & 1 & 0 & 1 & 0 & 1 & 0 & 1 & 0 & 1 & 0 & 1 & 0 & 1 & 0 & 1 \\
$k=5$ & 1 & 0 & 1 & 0 & 1 & 0 & 1 & 0 & 1 & 0 & 1 & 0 & 1 & 0 & 1 & 0 & 1 & 0 & 1 & 0 & 1 \\
\hline
LHS & 1 & 0 & 1 & 0 & 1 & 0 & 1 & 0 & 1 & 0 & 1 & 0 & 1 & 0 & 1 & 0 & 1 & 0 & 1 & 0 & 1 \\
\hline
\end{tabular}

\noindent At $N=3$ and $O(q^{2})$,

\noindent \begin{tabular}{|c|ccccccccccccccccccccc|}
\hline
$n$ & -2 & -1 & 0 & 1 & 2 & 3 & 4 & 5 & 6 & 7 & 8 & 9 & 10 & 11 & 12 & 13 & 14 & 15 & 16 & 17 & 18 \\
\hline \hline
$k=0$ & 2 & 1 & 2 & 3 & 5 & 6 & 10 & 12 & 18 & 23 & 32 & 40 & 56 & 69 & 92 & 116 & 151 & 187 & 242 & 298 & 379 \\
$k=1$ & 2 & 1 & 1 & 1 & 2 & 0 & 1 & -2 & -3 & -8 & -12 & -23 & -31 & -51 & -71 & -104 & -141 & -201 & -266 & -365 & -479 \\
$k=2$ & 2 & 1 & 1 & 1 & 2 & 1 & 2 & 1 & 2 & 2 & 3 & 3 & 7 & 9 & 15 & 24 & 38 & 56 & 86 & 125 & 182 \\
$k=3$ & 2 & 1 & 1 & 1 & 2 & 1 & 2 & 1 & 2 & 1 & 2 & 1 & 2 & 1 & 1 & 0 & 0 & -3 & -6 & -11 & -19 \\
$k=4$ & 2 & 1 & 1 & 1 & 2 & 1 & 2 & 1 & 2 & 1 & 2 & 1 & 2 & 1 & 2 & 1 & 2 & 1 & 2 & 1 & 3 \\
$k=5$ & 2 & 1 & 1 & 1 & 2 & 1 & 2 & 1 & 2 & 1 & 2 & 1 & 2 & 1 & 2 & 1 & 2 & 1 & 2 & 1 & 2 \\
\hline
LHS & 2 & 1 & 1 & 1 & 2 & 1 & 2 & 1 & 2 & 1 & 2 & 1 & 2 & 1 & 2 & 1 & 2 & 1 & 2 & 1 & 2 \\
\hline
\end{tabular}

\newpage

\noindent At $N=3$ and $O(q^{3})$,

\noindent \begin{tabular}{|c|ccccccccccccccccccccc|}
\hline
$n$ & -3 & -2 & -1 & 0 & 1 & 2 & 3 & 4 & 5 & 6 & 7 & 8 & 9 & 10 & 11 & 12 & 13 & 14 & 15 & 16 & 17 \\
\hline \hline
$k=0$ & 3 & 1 & 3 & 4 & 7 & 8 & 14 & 16 & 25 & 31 & 44 & 54 & 77 & 93 & 126 & 157 & 206 & 253 & 330 & 403 & 516 \\
$k=1$ & 3 & 0 & 1 & 1 & 1 & -1 & -1 & -6 & -8 & -16 & -24 & -40 & -54 & -85 & -116 & -165 & -224 & -311 & -410 & -556 & -725 \\
$k=2$ & 3 & 0 & 1 & 2 & 2 & 2 & 2 & 3 & 4 & 6 & 7 & 12 & 18 & 27 & 38 & 62 & 85 & 128 & 179 & 259 & 356 \\
$k=3$ & 3 & 0 & 1 & 2 & 2 & 2 & 1 & 2 & 2 & 2 & 1 & 1 & 0 & -1 & -5 & -6 & -14 & -20 & -36 & -50 & -80 \\
$k=4$ & 3 & 0 & 1 & 2 & 2 & 2 & 1 & 2 & 2 & 2 & 2 & 2 & 2 & 2 & 2 & 3 & 3 & 4 & 5 & 8 & 12 \\
$k=5$ & 3 & 0 & 1 & 2 & 2 & 2 & 1 & 2 & 2 & 2 & 2 & 2 & 2 & 2 & 2 & 2 & 2 & 2 & 2 & 2 & 2 \\
\hline
LHS & 3 & 0 & 1 & 2 & 2 & 2 & 1 & 2 & 2 & 2 & 2 & 2 & 2 & 2 & 2 & 2 & 2 & 2 & 2 & 2 & 2 \\
\hline
\end{tabular}

The pattern of convergence continues, where the expansion in terms of giant graviton branes on the right hand side converges faster in $k$ to $Z_N$ for larger values of $N$. Some representative examples with $N=10$ are shown:

\noindent At $N=10$ and $O(q^{0})$,

\noindent \begin{tabular}{|c|ccccccccccccccccccccc|}
\hline
$n$ & 0 & 1 & 2 & 3 & 4 & 5 & 6 & 7 & 8 & 9 & 10 & 11 & 12 & 13 & 14 & 15 & 16 & 17 & 18 & 19 & 20 \\
\hline \hline
$k=0$ & 1 & 1 & 2 & 3 & 5 & 7 & 11 & 15 & 22 & 30 & 42 & 56 & 77 & 101 & 135 & 176 & 231 & 297 & 385 & 490 & 627 \\
$k=1$ & 1 & 1 & 2 & 3 & 5 & 7 & 11 & 15 & 22 & 30 & 42 & 55 & 75 & 97 & 128 & 164 & 212 & 267 & 340 & 423 & 530 \\
$k=2$ & 1 & 1 & 2 & 3 & 5 & 7 & 11 & 15 & 22 & 30 & 42 & 55 & 75 & 97 & 128 & 164 & 212 & 267 & 340 & 423 & 530 \\
$k=3$ & 1 & 1 & 2 & 3 & 5 & 7 & 11 & 15 & 22 & 30 & 42 & 55 & 75 & 97 & 128 & 164 & 212 & 267 & 340 & 423 & 530 \\
\hline
LHS & 1 & 1 & 2 & 3 & 5 & 7 & 11 & 15 & 22 & 30 & 42 & 55 & 75 & 97 & 128 & 164 & 212 & 267 & 340 & 423 & 530 \\
\hline
\end{tabular}

\noindent At $N=10$ and $O(q^{1})$,

\noindent \begin{tabular}{|c|ccccccccccccccccccccc|}
\hline
$n$ & -1 & 0 & 1 & 2 & 3 & 4 & 5 & 6 & 7 & 8 & 9 & 10 & 11 & 12 & 13 & 14 & 15 & 16 & 17 & 18 & 19 \\
\hline \hline
$k=0$ & 1 & 0 & 1 & 1 & 2 & 2 & 4 & 4 & 7 & 8 & 12 & 14 & 21 & 24 & 34 & 41 & 55 & 66 & 88 & 105 & 137 \\
$k=1$ & 1 & 0 & 1 & 1 & 2 & 2 & 4 & 4 & 7 & 8 & 11 & 13 & 19 & 21 & 29 & 34 & 44 & 51 & 66 & 75 & 95 \\
$k=2$ & 1 & 0 & 1 & 1 & 2 & 2 & 4 & 4 & 7 & 8 & 11 & 13 & 19 & 21 & 29 & 34 & 44 & 51 & 66 & 75 & 95 \\
$k=3$ & 1 & 0 & 1 & 1 & 2 & 2 & 4 & 4 & 7 & 8 & 11 & 13 & 19 & 21 & 29 & 34 & 44 & 51 & 66 & 75 & 95 \\
\hline
LHS & 1 & 0 & 1 & 1 & 2 & 2 & 4 & 4 & 7 & 8 & 11 & 13 & 19 & 21 & 29 & 34 & 44 & 51 & 66 & 75 & 95 \\
\hline
\end{tabular}

\newpage

\noindent At $N=10$ and $O(q^{2})$,

\noindent \begin{tabular}{|c|ccccccccccccccccccccc|}
\hline
$n$ & -2 & -1 & 0 & 1 & 2 & 3 & 4 & 5 & 6 & 7 & 8 & 9 & 10 & 11 & 12 & 13 & 14 & 15 & 16 & 17 & 18 \\
\hline \hline
$k=0$ & 2 & 1 & 2 & 3 & 5 & 6 & 10 & 12 & 18 & 23 & 32 & 40 & 56 & 69 & 92 & 116 & 151 & 187 & 242 & 298 & 379 \\
$k=1$ & 2 & 1 & 2 & 3 & 5 & 6 & 10 & 12 & 18 & 22 & 30 & 37 & 50 & 60 & 78 & 95 & 120 & 143 & 179 & 211 & 259 \\
$k=2$ & 2 & 1 & 2 & 3 & 5 & 6 & 10 & 12 & 18 & 22 & 30 & 37 & 50 & 60 & 78 & 95 & 120 & 143 & 179 & 212 & 260 \\
$k=3$ & 2 & 1 & 2 & 3 & 5 & 6 & 10 & 12 & 18 & 22 & 30 & 37 & 50 & 60 & 78 & 95 & 120 & 143 & 179 & 212 & 260 \\
\hline
LHS & 2 & 1 & 2 & 3 & 5 & 6 & 10 & 12 & 18 & 22 & 30 & 37 & 50 & 60 & 78 & 95 & 120 & 143 & 179 & 212 & 260 \\
\hline
\end{tabular}

\noindent At $N=10$ and $O(q^{3})$,

\noindent \begin{tabular}{|c|ccccccccccccccccccccc|}
\hline
$n$ & -3 & -2 & -1 & 0 & 1 & 2 & 3 & 4 & 5 & 6 & 7 & 8 & 9 & 10 & 11 & 12 & 13 & 14 & 15 & 16 & 17 \\
\hline \hline
$k=0$ & 3 & 1 & 3 & 4 & 7 & 8 & 14 & 16 & 25 & 31 & 44 & 54 & 77 & 93 & 126 & 157 & 206 & 253 & 330 & 403 & 516 \\
$k=1$ & 3 & 1 & 3 & 4 & 7 & 8 & 14 & 16 & 24 & 29 & 41 & 48 & 68 & 78 & 104 & 124 & 159 & 185 & 236 & 272 & 338 \\
$k=2$ & 3 & 1 & 3 & 4 & 7 & 8 & 14 & 16 & 24 & 29 & 41 & 48 & 68 & 78 & 104 & 124 & 159 & 186 & 237 & 275 & 341 \\
$k=3$ & 3 & 1 & 3 & 4 & 7 & 8 & 14 & 16 & 24 & 29 & 41 & 48 & 68 & 78 & 104 & 124 & 159 & 186 & 237 & 275 & 341 \\
\hline
LHS & 3 & 1 & 3 & 4 & 7 & 8 & 14 & 16 & 24 & 29 & 41 & 48 & 68 & 78 & 104 & 124 & 159 & 186 & 237 & 275 & 341 \\
\hline
\end{tabular}

The expansion also holds at $N \leq 0$:

\noindent At $N=0$ and $O(q^{0})$,

\noindent \begin{tabular}{|c|ccccccccccccccccccccc|}
\hline
$n$ & 0 & 1 & 2 & 3 & 4 & 5 & 6 & 7 & 8 & 9 & 10 & 11 & 12 & 13 & 14 & 15 & 16 & 17 & 18 & 19 & 20 \\
\hline \hline
$k=0$ & 1 & 1 & 2 & 3 & 5 & 7 & 11 & 15 & 22 & 30 & 42 & 56 & 77 & 101 & 135 & 176 & 231 & 297 & 385 & 490 & 627 \\
$k=1$ & 1 & 0 & 0 & -1 & -2 & -5 & -8 & -15 & -23 & -37 & -55 & -83 & -118 & -171 & -238 & -332 & -453 & -618 & -827 & -1107 & -1460 \\
$k=2$ & 1 & 0 & 0 & 0 & 0 & 0 & 1 & 2 & 5 & 10 & 18 & 31 & 52 & 82 & 127 & 193 & 285 & 415 & 595 & 841 & 1174 \\
$k=3$ & 1 & 0 & 0 & 0 & 0 & 0 & 0 & 0 & 0 & 0 & -1 & -2 & -5 & -10 & -20 & -34 & -60 & -97 & -157 & -242 & -371 \\
$k=4$ & 1 & 0 & 0 & 0 & 0 & 0 & 0 & 0 & 0 & 0 & 0 & 0 & 0 & 0 & 0 & 1 & 2 & 5 & 10 & 20 & 36 \\
$k=5$ & 1 & 0 & 0 & 0 & 0 & 0 & 0 & 0 & 0 & 0 & 0 & 0 & 0 & 0 & 0 & 0 & 0 & 0 & 0 & 0 & 0 \\
\hline
LHS & 1 & 0 & 0 & 0 & 0 & 0 & 0 & 0 & 0 & 0 & 0 & 0 & 0 & 0 & 0 & 0 & 0 & 0 & 0 & 0 & 0 \\
\hline
\end{tabular}

\newpage

\noindent At $N=0$ and $O(q^{1})$,

\noindent \begin{tabular}{|c|ccccccccccccccccccccc|}
\hline
$n$ & -1 & 0 & 1 & 2 & 3 & 4 & 5 & 6 & 7 & 8 & 9 & 10 & 11 & 12 & 13 & 14 & 15 & 16 & 17 & 18 & 19 \\
\hline \hline
$k=0$ & 1 & 0 & 1 & 1 & 2 & 2 & 4 & 4 & 7 & 8 & 12 & 14 & 21 & 24 & 34 & 41 & 55 & 66 & 88 & 105 & 137 \\
$k=1$ & 0 & -1 & -1 & -2 & -3 & -5 & -7 & -11 & -15 & -22 & -30 & -42 & -56 & -77 & -101 & -135 & -176 & -231 & -297 & -385 & -490 \\
$k=2$ & 0 & 0 & 0 & 1 & 1 & 3 & 4 & 8 & 11 & 19 & 26 & 41 & 56 & 83 & 112 & 160 & 213 & 295 & 389 & 526 & 686 \\
$k=3$ & 0 & 0 & 0 & 0 & 0 & 0 & -1 & -1 & -3 & -5 & -9 & -14 & -24 & -35 & -55 & -80 & -118 & -167 & -240 & -331 & -462 \\
$k=4$ & 0 & 0 & 0 & 0 & 0 & 0 & 0 & 0 & 0 & 0 & 1 & 1 & 3 & 5 & 10 & 15 & 27 & 40 & 65 & 95 & 145 \\
$k=5$ & 0 & 0 & 0 & 0 & 0 & 0 & 0 & 0 & 0 & 0 & 0 & 0 & 0 & 0 & 0 & -1 & -1 & -3 & -5 & -10 & -16 \\
$k=6$ & 0 & 0 & 0 & 0 & 0 & 0 & 0 & 0 & 0 & 0 & 0 & 0 & 0 & 0 & 0 & 0 & 0 & 0 & 0 & 0 & 0 \\
\hline
LHS & 0 & 0 & 0 & 0 & 0 & 0 & 0 & 0 & 0 & 0 & 0 & 0 & 0 & 0 & 0 & 0 & 0 & 0 & 0 & 0 & 0 \\
\hline
\end{tabular}

\noindent At $N=0$ and $O(q^{2})$,

\noindent \begin{tabular}{|c|ccccccccccccccccccccc|}
\hline
$n$ & -3 & -2 & -1 & 0 & 1 & 2 & 3 & 4 & 5 & 6 & 7 & 8 & 9 & 10 & 11 & 12 & 13 & 14 & 15 & 16 & 17 \\
\hline \hline
$k=0$ & 0 & 2 & 1 & 2 & 3 & 5 & 6 & 10 & 12 & 18 & 23 & 32 & 40 & 56 & 69 & 92 & 116 & 151 & 187 & 242 & 298 \\
$k=1$ & -1 & 0 & -2 & -4 & -6 & -9 & -15 & -21 & -32 & -45 & -64 & -88 & -123 & -164 & -223 & -296 & -392 & -512 & -671 & -864 & -1116 \\
$k=2$ & 0 & 1 & 1 & 1 & 4 & 6 & 11 & 17 & 28 & 41 & 64 & 91 & 134 & 188 & 267 & 365 & 507 & 682 & 924 & 1227 & 1634 \\
$k=3$ & 0 & 0 & 0 & -1 & -1 & -2 & -3 & -7 & -10 & -18 & -28 & -45 & -67 & -104 & -150 & -222 & -316 & -451 & -628 & -878 & -1199 \\
$k=4$ & 0 & 0 & 0 & 0 & 0 & 0 & 1 & 1 & 2 & 4 & 6 & 11 & 18 & 28 & 44 & 70 & 103 & 156 & 229 & 334 & 477 \\
$k=5$ & 0 & 0 & 0 & 0 & 0 & 0 & 0 & 0 & 0 & 0 & -1 & -1 & -2 & -4 & -7 & -10 & -19 & -28 & -45 & -68 & -105 \\
$k=6$ & 0 & 0 & 0 & 0 & 0 & 0 & 0 & 0 & 0 & 0 & 0 & 0 & 0 & 0 & 0 & 1 & 1 & 2 & 4 & 7 & 11 \\
$k=7$ & 0 & 0 & 0 & 0 & 0 & 0 & 0 & 0 & 0 & 0 & 0 & 0 & 0 & 0 & 0 & 0 & 0 & 0 & 0 & 0 & 0 \\
\hline
LHS & 0 & 0 & 0 & 0 & 0 & 0 & 0 & 0 & 0 & 0 & 0 & 0 & 0 & 0 & 0 & 0 & 0 & 0 & 0 & 0 & 0 \\
\hline
\end{tabular}

\newpage

\noindent At $N=0$ and $O(q^{3})$,

\noindent \begin{tabular}{|c|ccccccccccccccccccccccc|}
\hline
$n$ & -6 & -5 & -4 & -3 & -2 & -1 & 0 & 1 & 2 & 3 & 4 & 5 & 6 & 7 & 8 & 9 & 10 & 11 & 12 & 13 & 14 & 15 & 16  \\
\hline \hline
$k=0$ & 0 & 0 & 0 & 3 & 1 & 3 & 4 & 7 & 8 & 14 & 16 & 25 & 31 & 44 & 54 & 77 & 93 & 126 & 157 & 206 & 253 & 330 & 403 \\
$k=1$ & 0 & -1 & -2 & 0 & -5 & -6 & -11 & -15 & -25 & -33 & -52 & -69 & -100 & -134 & -188 & -245 & -337 & -438 & -583 & -753 & -988 & -1260 & -1632 \\
$k=2$ & 1 & 0 & 1 & 3 & 4 & 6 & 11 & 16 & 27 & 39 & 60 & 85 & 127 & 175 & 251 & 344 & 478 & 643 & 878 & 1164 & 1559 & 2045 & 2695 \\
$k=3$ & 0 & -1 & -1 & -1 & -2 & -5 & -7 & -12 & -16 & -29 & -39 & -63 & -88 & -134 & -185 & -273 & -373 & -533 & -724 & -1007 & -1352 & -1849 & -2454 \\
$k=4$ & 0 & 0 & 0 & 1 & 1 & 2 & 2 & 5 & 8 & 12 & 19 & 29 & 42 & 65 & 93 & 135 & 194 & 278 & 387 & 547 & 755 & 1043 & 1424 \\
$k=5$ & 0 & 0 & 0 & 0 & 0 & 0 & -1 & -1 & -2 & -3 & -5 & -8 & -14 & -19 & -31 & -45 & -68 & -95 & -144 & -198 & -288 & -397 & -563 \\
$k=6$ & 0 & 0 & 0 & 0 & 0 & 0 & 0 & 0 & 0 & 0 & 1 & 1 & 2 & 3 & 6 & 8 & 14 & 21 & 32 & 47 & 70 & 101 & 147 \\
$k=7$ & 0 & 0 & 0 & 0 & 0 & 0 & 0 & 0 & 0 & 0 & 0 & 0 & 0 & 0 & 0 & -1 & -1 & -2 & -3 & -6 & -9 & -14 & -21 \\
$k=8$ & 0 & 0 & 0 & 0 & 0 & 0 & 0 & 0 & 0 & 0 & 0 & 0 & 0 & 0 & 0 & 0 & 0 & 0 & 0 & 0 & 0 & 1 & 1 \\
\hline
LHS & 0 & 0 & 0 & 0 & 0 & 0 & 0 & 0 & 0 & 0 & 0 & 0 & 0 & 0 & 0 & 0 & 0 & 0 & 0 & 0 & 0 & 0 & 0 \\
\hline
\end{tabular}

\noindent At $N=-1$ and $O(q^{0})$,

\noindent \begin{tabular}{|c|ccccccccccccccccccccc|}
\hline
$n$ & 0 & 1 & 2 & 3 & 4 & 5 & 6 & 7 & 8 & 9 & 10 & 11 & 12 & 13 & 14 & 15 & 16 & 17 & 18 & 19 & 20 \\
\hline \hline
$k=0$ & 1 & 1 & 2 & 3 & 5 & 7 & 11 & 15 & 22 & 30 & 42 & 56 & 77 & 101 & 135 & 176 & 231 & 297 & 385 & 490 & 627 \\
$k=1$ & 0 & -1 & -2 & -4 & -7 & -12 & -19 & -30 & -45 & -67 & -97 & -139 & -195 & -272 & -373 & -508 & -684 & -915 & -1212 & -1597 & -2086 \\
$k=2$ & 0 & 0 & 0 & 1 & 2 & 5 & 9 & 17 & 28 & 47 & 73 & 114 & 170 & 253 & 365 & 525 & 738 & 1033 & 1422 & 1939 & 2616 \\
$k=3$ & 0 & 0 & 0 & 0 & 0 & 0 & -1 & -2 & -5 & -10 & -19 & -33 & -57 & -92 & -147 & -227 & -345 & -512 & -733 & -1043 & -1462 \\
$k=4$ & 0 & 0 & 0 & 0 & 0 & 0 & 0 & 0 & 0 & 0 & 1 & 2 & 5 & 10 & 20 & 35 & 62 & 91 & 141 & 207 & 303 \\
$k=5$ & 0 & 0 & 0 & 0 & 0 & 0 & 0 & 0 & 0 & 0 & 0 & 0 & 0 & 0 & 0 & -1 & -1 & -2 & -3 & -5 & -7 \\
$k=6$ & 0 & 0 & 0 & 0 & 0 & 0 & 0 & 0 & 0 & 0 & 0 & 0 & 0 & 0 & 0 & 0 & 0 & 0 & 0 & 0 & 0 \\
\hline
LHS & 0 & 0 & 0 & 0 & 0 & 0 & 0 & 0 & 0 & 0 & 0 & 0 & 0 & 0 & 0 & 0 & 0 & 0 & 0 & 0 & 0 \\
\hline
\end{tabular}

\newpage

\noindent At $N=-1$ and $O(q^{1})$,

\noindent \begin{tabular}{|c|ccccccccccccccccccccc|}
\hline
$n$ & -2 & -1 & 0 & 1 & 2 & 3 & 4 & 5 & 6 & 7 & 8 & 9 & 10 & 11 & 12 & 13 & 14 & 15 & 16 & 17 & 18 \\
\hline \hline
$k=0$ & 0 & 1 & 0 & 1 & 1 & 2 & 2 & 4 & 4 & 7 & 8 & 12 & 14 & 21 & 24 & 34 & 41 & 55 & 66 & 88 & 105 \\
$k=1$ & -1 & 0 & -2 & -2 & -4 & -5 & -9 & -11 & -18 & -23 & -34 & -44 & -63 & -80 & -111 & -142 & -190 & -242 & -319 & -402 & -522 \\
$k=2$ & 0 & 1 & 1 & 2 & 4 & 6 & 10 & 15 & 23 & 33 & 49 & 68 & 97 & 133 & 184 & 247 & 336 & 444 & 592 & 774 & 1007 \\
$k=3$ & 0 & 0 & 0 & -1 & -1 & -3 & -4 & -9 & -12 & -22 & -31 & -50 & -70 & -107 & -147 & -215 & -293 & -413 & -556 & -747 & -996 \\
$k=4$ & 0 & 0 & 0 & 0 & 0 & 0 & 1 & 1 & 3 & 5 & 9 & 15 & 25 & 38 & 60 & 90 & 133 & 194 & 269 & 379 & 523 \\
$k=5$ & 0 & 0 & 0 & 0 & 0 & 0 & 0 & 0 & 0 & 0 & -1 & -1 & -3 & -5 & -10 & -15 & -28 & -40 & -62 & -89 & -131 \\
$k=6$ & 0 & 0 & 0 & 0 & 0 & 0 & 0 & 0 & 0 & 0 & 0 & 0 & 0 & 0 & 0 & 1 & 1 & 2 & 3 & 5 & 7 \\
$k=7$ & 0 & 0 & 0 & 0 & 0 & 0 & 0 & 0 & 0 & 0 & 0 & 0 & 0 & 0 & 0 & 0 & 0 & 0 & 0 & 0 & 0 \\
\hline
LHS & 0 & 0 & 0 & 0 & 0 & 0 & 0 & 0 & 0 & 0 & 0 & 0 & 0 & 0 & 0 & 0 & 0 & 0 & 0 & 0 & 0 \\
\hline
\end{tabular}

\noindent At $N=-1$ and $O(q^{2})$,

\noindent \begin{tabular}{|c|ccccccccccccccccccccc|}
\hline
$n$ & -5 & -4 & -3 & -2 & -1 & 0 & 1 & 2 & 3 & 4 & 5 & 6 & 7 & 8 & 9 & 10 & 11 & 12 & 13 & 14 & 15 \\
\hline \hline
$k=0$ & 0 & 0 & 0 & 2 & 1 & 2 & 3 & 5 & 6 & 10 & 12 & 18 & 23 & 32 & 40 & 56 & 69 & 92 & 116 & 151 & 187 \\
$k=1$ & 0 & -1 & -2 & -1 & -5 & -7 & -11 & -16 & -25 & -34 & -51 & -69 & -97 & -131 & -180 & -236 & -319 & -416 & -547 & -707 & -919 \\
$k=2$ & 1 & 0 & 1 & 4 & 5 & 8 & 15 & 22 & 35 & 52 & 77 & 110 & 160 & 221 & 310 & 425 & 580 & 778 & 1048 & 1384 & 1831 \\
$k=3$ & 0 & -1 & -1 & -1 & -3 & -6 & -9 & -16 & -24 & -40 & -59 & -91 & -132 & -196 & -277 & -398 & -553 & -774 & -1057 & -1449 & -1949 \\
$k=4$ & 0 & 0 & 0 & 1 & 1 & 2 & 3 & 6 & 10 & 16 & 26 & 41 & 62 & 96 & 142 & 209 & 304 & 438 & 619 & 874 & 1193 \\
$k=5$ & 0 & 0 & 0 & 0 & 0 & 0 & -1 & -1 & -2 & -4 & -6 & -10 & -18 & -26 & -42 & -65 & -98 & -144 & -218 & -310 & -445 \\
$k=6$ & 0 & 0 & 0 & 0 & 0 & 0 & 0 & 0 & 0 & 0 & 1 & 1 & 2 & 4 & 7 & 10 & 18 & 28 & 43 & 65 & 94 \\
$k=7$ & 0 & 0 & 0 & 0 & 0 & 0 & 0 & 0 & 0 & 0 & 0 & 0 & 0 & 0 & 0 & -1 & -1 & -2 & -4 & -6 & -9 \\
$k=8$ & 0 & 0 & 0 & 0 & 0 & 0 & 0 & 0 & 0 & 0 & 0 & 0 & 0 & 0 & 0 & 0 & 0 & 0 & 0 & 0 & 0 \\
\hline
LHS & 0 & 0 & 0 & 0 & 0 & 0 & 0 & 0 & 0 & 0 & 0 & 0 & 0 & 0 & 0 & 0 & 0 & 0 & 0 & 0 & 0 \\
\hline
\end{tabular}

\newpage

\noindent At $N=-2$ and $O(q^{0})$,

\noindent \begin{tabular}{|c|ccccccccccccccccccccc|}
\hline
$n$ & -1 & 0 & 1 & 2 & 3 & 4 & 5 & 6 & 7 & 8 & 9 & 10 & 11 & 12 & 13 & 14 & 15 & 16 & 17 & 18 & 19 \\
\hline \hline
$k=0$ & 0 & 1 & 1 & 2 & 3 & 5 & 7 & 11 & 15 & 22 & 30 & 42 & 56 & 77 & 101 & 135 & 176 & 231 & 297 & 385 & 490 \\
$k=1$ & -1 & -1 & -3 & -5 & -9 & -14 & -23 & -34 & -52 & -75 & -109 & -153 & -216 & -296 & -407 & -549 & -739 & -981 & -1300 & -1702 & -2223 \\
$k=2$ & 0 & 1 & 2 & 4 & 8 & 14 & 24 & 39 & 62 & 95 & 144 & 212 & 309 & 442 & 626 & 873 & 1209 & 1653 & 2236 & 3001 & 3998 \\
$k=3$ & 0 & 0 & 0 & -1 & -2 & -5 & -9 & -18 & -30 & -52 & -83 & -133 & -203 & -310 & -457 & -672 & -946 & -1337 & -1850 & -2547 & -3450 \\
$k=4$ & 0 & 0 & 0 & 0 & 0 & 0 & 1 & 2 & 5 & 10 & 19 & 34 & 59 & 97 & 144 & 217 & 317 & 456 & 643 & 898 & 1232 \\
$k=5$ & 0 & 0 & 0 & 0 & 0 & 0 & 0 & 0 & 0 & 0 & -1 & -2 & -3 & -5 & -8 & -12 & -18 & -26 & -37 & -52 & -72 \\
$k=6$ & 0 & 0 & 0 & 0 & 0 & 0 & 0 & 0 & 0 & 0 & 0 & 0 & 0 & 0 & 0 & 0 & 0 & 0 & 0 & 0 & 0 \\
\hline
LHS & 0 & 0 & 0 & 0 & 0 & 0 & 0 & 0 & 0 & 0 & 0 & 0 & 0 & 0 & 0 & 0 & 0 & 0 & 0 & 0 & 0 \\
\hline
\end{tabular}

\noindent At $N=-2$ and $O(q^{1})$,

\noindent \begin{tabular}{|c|ccccccccccccccccccccc|}
\hline
$n$ & -4 & -3 & -2 & -1 & 0 & 1 & 2 & 3 & 4 & 5 & 6 & 7 & 8 & 9 & 10 & 11 & 12 & 13 & 14 & 15 & 16 \\
\hline \hline
$k=0$ & 0 & 0 & 0 & 1 & 0 & 1 & 1 & 2 & 2 & 4 & 4 & 7 & 8 & 12 & 14 & 21 & 24 & 34 & 41 & 55 & 66 \\
$k=1$ & 0 & -1 & -1 & -1 & -3 & -4 & -6 & -9 & -13 & -18 & -26 & -35 & -48 & -65 & -87 & -114 & -152 & -197 & -256 & -330 & -424 \\
$k=2$ & 1 & 0 & 2 & 3 & 5 & 7 & 13 & 17 & 28 & 38 & 57 & 77 & 112 & 148 & 208 & 275 & 374 & 489 & 655 & 846 & 1105 \\
$k=3$ & 0 & -1 & -1 & -2 & -4 & -7 & -11 & -18 & -27 & -42 & -61 & -90 & -128 & -183 & -254 & -354 & -483 & -659 & -866 & -1165 & -1528 \\
$k=4$ & 0 & 0 & 0 & 1 & 1 & 3 & 4 & 9 & 13 & 23 & 34 & 55 & 79 & 122 & 172 & 253 & 340 & 483 & 651 & 893 & 1185 \\
$k=5$ & 0 & 0 & 0 & 0 & 0 & 0 & -1 & -1 & -3 & -5 & -9 & -15 & -26 & -39 & -61 & -88 & -129 & -182 & -258 & -354 & -488 \\
$k=6$ & 0 & 0 & 0 & 0 & 0 & 0 & 0 & 0 & 0 & 0 & 1 & 1 & 3 & 4 & 7 & 10 & 16 & 22 & 33 & 45 & 64 \\
$k=7$ & 0 & 0 & 0 & 0 & 0 & 0 & 0 & 0 & 0 & 0 & 0 & 0 & 0 & 0 & 0 & 0 & 0 & 0 & 0 & 0 & 0 \\
\hline
LHS & 0 & 0 & 0 & 0 & 0 & 0 & 0 & 0 & 0 & 0 & 0 & 0 & 0 & 0 & 0 & 0 & 0 & 0 & 0 & 0 & 0 \\
\hline
\end{tabular}

\newpage

\noindent At $N=-2$ and $O(q^{2})$,

\noindent \begin{tabular}{|c|cccccccccccccccccccccc|}
\hline
$n$ & -8 & -7 & -6 & -5 & -4 & -3 & -2 & -1 & 0 & 1 & 2 & 3 & 4 & 5 & 6 & 7 & 8 & 9 & 10 & 11 & 12 & 13 \\
\hline \hline
$k=0$ & 0 & 0 & 0 & 0 & 0 & 0 & 2 & 1 & 2 & 3 & 5 & 6 & 10 & 12 & 18 & 23 & 32 & 40 & 56 & 69 & 92 & 116 \\
$k=1$ & 0 & 0 & 0 & -1 & -2 & -3 & -4 & -8 & -12 & -18 & -26 & -38 & -53 & -75 & -102 & -140 & -188 & -252 & -332 & -439 & -571 & -742 \\
$k=2$ & 0 & 1 & 1 & 2 & 3 & 7 & 11 & 18 & 26 & 42 & 60 & 90 & 126 & 182 & 250 & 350 & 473 & 647 & 862 & 1156 & 1520 & 2008 \\
$k=3$ & -1 & 0 & -1 & -3 & -5 & -7 & -13 & -20 & -33 & -50 & -76 & -111 & -166 & -235 & -337 & -473 & -660 & -905 & -1243 & -1677 & -2260 & -3012  \\
$k=4$ & 0 & 1 & 1 & 1 & 3 & 5 & 9 & 14 & 23 & 35 & 56 & 83 & 126 & 184 & 270 & 384 & 552 & 771 & 1080 & 1487 & 2046 & 2773  \\
$k=5$ & 0 & 0 & 0 & -1 & -1 & -2 & -3 & -6 & -9 & -16 & -24 & -39 & -58 & -90 & -132 & -198 & -285 & -415 & -589 & -838 & -1169 & -1634  \\
$k=6$ & 0 & 0 & 0 & 0 & 0 & 0 & 1 & 1 & 2 & 4 & 6 & 10 & 17 & 26 & 40 & 63 & 93 & 139 & 205 & 299 & 427 & 609  \\
$k=7$ & 0 & 0 & 0 & 0 & 0 & 0 & 0 & 0 & 0 & 0 & -1 & -1 & -2 & -4 & -7 & -10 & -18 & -27 & -43 & -64 & -93 & -132  \\
$k=8$ & 0 & 0 & 0 & 0 & 0 & 0 & 0 & 0 & 0 & 0 & 0 & 0 & 0 & 0 & 0 & 1 & 1 & 2 & 3 & 5 & 7 & 11 \\
$k=9$ & 0 & 0 & 0 & 0 & 0 & 0 & 0 & 0 & 0 & 0 & 0 & 0 & 0 & 0 & 0 & 0 & 0 & 0 & 0 & 0 & 0 & 0  \\
\hline
LHS & 0 & 0 & 0 & 0 & 0 & 0 & 0 & 0 & 0 & 0 & 0 & 0 & 0 & 0 & 0 & 0 & 0 & 0 & 0 & 0 & 0 & 0  \\
\hline
\end{tabular}

\end{landscape}

\section{Character formula for $U(N)$ indices} \label{app: character formula for U(N)}

We review the derivation of \cite{Dolan:2007rq,Dutta:2007ws}, following closely \cite{Murthy:2020rbd}, showing that $U(N)$ gauge theory indices can be written in terms of characters of symmetric groups $S_n$. This will dramatically simplify the computations of $\mathcal{N}=4$ indices.

We will consider a $U(N)$ gauge theory index of the form
\[
Z_N = \int_{U(N)} DU \exp \left( \sum_{j=1}^{\infty} \frac{1}{j} f(x^j) \Tr U^{j} \Tr U^{-j} \right)
\]
where $x$ denotes a collection of fugacities, each of which is exponentiated by $j$.

Let us rewrite the plethystic exponential as
\[
\exp \left( \sum_{j=1}^{\infty} \frac{1}{j} a(x^j) \Tr U^{j} \Tr U^{-j} \right) = \left( \sum_{k_1 = 0}^\infty \sum_{k_2 = 0}^\infty \cdots \right) \prod_{j=1}^\infty \frac{1}{j^{k_j} k_j!} \left( a(x^j) \Tr U^{j} \Tr U^{-j} \right)^{k_j}.
\]
If we interpret $j$ as an element of a partition $P$ and $k_j$ as the multiplicity of $j$ in $P$, the collection of sums on the right can be recognized as the sum over \textit{all} partitions $P$
\[
\sum_P = \sum_{k_1 = 0}^\infty \sum_{k_2 = 0}^\infty \cdots,
\]
so that the index becomes
\[
Z_N = \sum_P \int_{U(N)} DU \prod_{j=1}^\infty \frac{1}{j^{k_j} k_j!} \left( f(x^j) \Tr U^{j} \Tr U^{-j} \right)^{k_j}.
\]
Note that the product over $j$ must be finite for any partition $P$ of a finite number.

Representations of $U(n)$ and of symmetric group $S_n$ share the property that they are both labelled by a partition $\lambda$ of $n$. While the characters $\tilde{\chi}_\lambda (U)$ of $U(n)$ are functions of the conjugacy class of $U$ in representation $\lambda$, the characters $\chi^\lambda(P)$ of $S_n$ are functions of another partition $P$ of $n$ labelling the conjugacy class (i.e. a cycle shape) of a symmetric group element in $S_n$. For given $U$ and $P$, the Frobenius character formula for $U(N)$ relates the characters of $U(n)$ and of $S_n$ as
\[
\prod_{j=1}^{m} (\Tr U^{j})^{k_j} = \sum_{\substack{\lambda \ \vdash \ |P| \\ l(\lambda) \leq N}} \tilde{\chi}_\lambda (U) \chi^\lambda (P).
\]
Here, the weight $|P|$ of $P$ is the integer being partitioned by $P$, and $l(\lambda)$ is the number of elements (or parts) in a partition $\lambda$. The partition $P$ is encoded on the left hand side through the set of positive integers $j$ in $P$ and their multiplicities $k_j$. Since the product over $j$ must be finite for any partition of a finite number, we denoted the largest integer in $P$ as $m$.

The Frobenius formula allows us to evaluate the integral over $U(N)$ in the gauge theory index using the orthogonality relation for characters
\[
\int DU \tilde{\chi}_\lambda(U) \tilde{\chi}_{\lambda'}(U^\dagger) = \delta_{\lambda \lambda'},
\]
so that
\[
\int DU \prod_{j=1}^m \left( \Tr U^{j} \Tr U^{-j} \right)^{k_j} = \sum_{\substack{\lambda \ \vdash \ |P| \\ l(\lambda) \leq N}} \chi^\lambda(P)^2.
\]
The expression on the right does not involve a complex conjugation because symmetric group characters are integers.

Combining, we have
\begin{equation} \label{index as characters}
    Z_N(x) = \sum_P f(x)_P \frac{1}{z_P} \sum_{\substack{\lambda \ \vdash \ |P| \\ l(\lambda) \leq N}} \chi^\lambda(P)^2
\end{equation}
where we defined
\[
z_P := \prod_{j=1}^m j^{k_j} k_j !
\]
and
\[
f(x)_P := \prod_{j=1}^m f(x^j)^{k_j}.
\]
The result of the derivation states that we need only generate character tables of symmetric groups to compute $U(N)$ gauge theory indices perturbatively. As in the Schur case, an appropriate resummation is necessary to obtain $\hat{Z}_N$ from a perturbative expression of $Z_N$.

\section{Tests for the specialized $\mathcal{N}=4$ index} \label{appendix: check of reduced proposal}

In this section, we provide evidence in support of the giant graviton expansion for the specialized $\mathcal{N}=4$ index discussed in Section \ref{subsection: specialized N=4 index}. We take $x=w^2$ to avoid square roots in the expansion:
\begin{equation}
    Z_N(w;r) = Z_\infty(w;r) \left[ 1 + w^{2N} \hat{Z}_1(w;r) + w^{4 N} \hat{Z}_2(w;r) + \cdots \right],
\end{equation}
where our prescription for the giant graviton corrections $\hat{Z}_k$ amounts to
\begin{equation}
    \widehat \ : \quad w \to \frac{1}{w} , \quad r \to \frac{r}{w}.
\end{equation}

To check our proposal, we truncate the expansion at the $k$-th correction as
\begin{equation} \label{reduced truncated-appendix}
    Z_N(w;r) = Z_\infty(w;r) \left[ 1 + w^{2N} \hat{Z}_1(w;r) + \cdots +  w^{2 k N} \hat{Z}_k(w;r) \right],
\end{equation}
where we also defined
\[
Z_0 = 1, \quad Z_{-1} = Z_{-2} = \cdots = 0.
\]

The following tables enumerate the series coefficients of $w$ of the left and right hand sides of \eqref{reduced truncated-appendix} for given $N$, order of $r$, and truncation level $k$. Integers $n$ denote the power of $w$ in the series. The top row lists the powers $n$ of $w$. The middle rows tabulate the coefficients of $w^n$ of the right hand side of \eqref{reduced truncated-appendix} for different values of the truncation $k$. The final row tabulates the coefficients of $w^n$ of $Z_N(w;r)$.

From the results below, we observe that the order of $w$ to which the formula holds, at a given order $s$ of $r$, is
\begin{equation}
1 + k^2 + 2 N (1 + k) + k (3 - 2 s) - 3 s.
\end{equation}
Checks beyond those presented below can be conducted by using the relation of $U(N)$ indices to symmetric group characters as reviewed in Appendix \ref{app: character formula for U(N)}.

\begin{landscape}

\noindent At $N=1$ and $O(r^{0})$,

\noindent
\begin{tabular}{|c|cccccccccccccccccccccccccc|}
\hline
$n$ & 0 & 1 & 2 & 3 & 4 & 5 & 6 & 7 & 8 & 9 & 10 & 11 & 12 & 13 & 14 & 15 & 16 & 17 & 18 & 19 & 20 & 21 & 22 & 23 & 24 & 25 \\
\hline \hline
$k=0$ & 1 & 0 & 1 & 0 & 2 & 0 & 3 & 0 & 5 & 0 & 7 & 0 & 11 & 0 & 15 & 0 & 22 & 0 & 30 & 0 & 42 & 0 & 56 & 0 & 77 & 0 \\
$k=1$ & 1 & 0 & 1 & 0 & 1 & 0 & 1 & 0 & 1 & 0 & 0 & 0 & -1 & 0 & -4 & 0 & -8 & 0 & -15 & 0 & -25 & 0 & -41 & 0 & -62 & 0 \\
$k=2$ & 1 & 0 & 1 & 0 & 1 & 0 & 1 & 0 & 1 & 0 & 1 & 0 & 1 & 0 & 1 & 0 & 1 & 0 & 2 & 0 & 3 & 0 & 6 & 0 & 11 & 0 \\
$k=3$ & 1 & 0 & 1 & 0 & 1 & 0 & 1 & 0 & 1 & 0 & 1 & 0 & 1 & 0 & 1 & 0 & 1 & 0 & 1 & 0 & 1 & 0 & 1 & 0 & 1 & 0 \\
\hline
LHS & 1 & 0 & 1 & 0 & 1 & 0 & 1 & 0 & 1 & 0 & 1 & 0 & 1 & 0 & 1 & 0 & 1 & 0 & 1 & 0 & 1 & 0 & 1 & 0 & 1 & 0 \\
\hline
\end{tabular}

\noindent At $N=1$ and $O(r^{1})$,

\noindent \begin{tabular}{|c|ccccccccccccccccccccccccc|}
\hline
$n$ & -1 & 0 & 1 & 2 & 3 & 4 & 5 & 6 & 7 & 8 & 9 & 10 & 11 & 12 & 13 & 14 & 15 & 16 & 17 & 18 & 19 & 20 & 21 & 22 & 23 \\
\hline \hline
$k=0$ & 2 & -2 & 2 & -2 & 4 & -4 & 6 & -6 & 10 & -10 & 14 & -14 & 22 & -22 & 30 & -30 & 44 & -44 & 60 & -60 & 84 & -84 & 112 & -112 & 154 \\
$k=1$ & 2 & -2 & 0 & 0 & 0 & 0 & -2 & 2 & -4 & 4 & -10 & 10 & -16 & 16 & -30 & 30 & -46 & 46 & -74 & 74 & -110 & 110 & -166 & 166 & -236 \\
$k=2$ & 2 & -2 & 0 & 0 & 0 & 0 & 0 & 0 & 0 & 0 & 0 & 0 & 2 & -2 & 4 & -4 & 10 & -10 & 20 & -20 & 36 & -36 & 62 & -62 & 104 \\
$k=3$ & 2 & -2 & 0 & 0 & 0 & 0 & 0 & 0 & 0 & 0 & 0 & 0 & 0 & 0 & 0 & 0 & 0 & 0 & 0 & 0 & -2 & 2 & -4 & 4 & -10 \\
$k=4$ & 2 & -2 & 0 & 0 & 0 & 0 & 0 & 0 & 0 & 0 & 0 & 0 & 0 & 0 & 0 & 0 & 0 & 0 & 0 & 0 & 0 & 0 & 0 & 0 & 0 \\
\hline
LHS & 2 & -2 & 0 & 0 & 0 & 0 & 0 & 0 & 0 & 0 & 0 & 0 & 0 & 0 & 0 & 0 & 0 & 0 & 0 & 0 & 0 & 0 & 0 & 0 & 0 \\
\hline
\end{tabular}

\noindent At $N=1$ and $O(r^{2})$,

\noindent \begin{tabular}{|c|ccccccccccccccccccccc|}
\hline
$n$ & -2 & -1 & 0 & 1 & 2 & 3 & 4 & 5 & 6 & 7 & 8 & 9 & 10 & 11 & 12 & 13 & 14 & 15 & 16 & 17 & 18 \\
\hline \hline
$k=0$ & 5 & -4 & 4 & -4 & 9 & -8 & 13 & -12 & 22 & -20 & 30 & -28 & 48 & -44 & 64 & -60 & 95 & -88 & 128 & -120 & 180 \\
$k=1$ & 2 & 0 & -6 & 8 & -11 & 12 & -21 & 24 & -36 & 40 & -63 & 68 & -97 & 104 & -155 & 164 & -229 & 240 & -341 & 356 & -490 \\
$k=2$ & 2 & 0 & -3 & 4 & -3 & 0 & 3 & -4 & 8 & -12 & 22 & -28 & 45 & -56 & 85 & -100 & 146 & -172 & 247 & -280 & 388 \\
$k=3$ & 2 & 0 & -3 & 4 & -3 & 0 & 0 & 0 & 0 & 0 & 0 & 0 & -3 & 4 & -8 & 12 & -22 & 28 & -46 & 60 & -94 \\
$k=4$ & 2 & 0 & -3 & 4 & -3 & 0 & 0 & 0 & 0 & 0 & 0 & 0 & 0 & 0 & 0 & 0 & 0 & 0 & 0 & 0 & 3 \\
\hline
LHS & 2 & 0 & -3 & 4 & -3 & 0 & 0 & 0 & 0 & 0 & 0 & 0 & 0 & 0 & 0 & 0 & 0 & 0 & 0 & 0 & 0 \\
\hline
\end{tabular}

\newpage

\noindent At $N=2$ and $O(r^{0})$,

\noindent \begin{tabular}{|c|ccccccccccccccccccccccccc|}
\hline
$n$ & 0 & 1 & 2 & 3 & 4 & 5 & 6 & 7 & 8 & 9 & 10 & 11 & 12 & 13 & 14 & 15 & 16 & 17 & 18 & 19 & 20 & 21 & 22 & 23 & 24 \\
\hline \hline
$k=0$ & 1 & 0 & 1 & 0 & 2 & 0 & 3 & 0 & 5 & 0 & 7 & 0 & 11 & 0 & 15 & 0 & 22 & 0 & 30 & 0 & 42 & 0 & 56 & 0 & 77 \\
$k=1$ & 1 & 0 & 1 & 0 & 2 & 0 & 2 & 0 & 3 & 0 & 3 & 0 & 4 & 0 & 3 & 0 & 3 & 0 & 0 & 0 & -3 & 0 & -11 & 0 & -20 \\
$k=2$ & 1 & 0 & 1 & 0 & 2 & 0 & 2 & 0 & 3 & 0 & 3 & 0 & 4 & 0 & 4 & 0 & 5 & 0 & 5 & 0 & 6 & 0 & 6 & 0 & 8 \\
$k=3$ & 1 & 0 & 1 & 0 & 2 & 0 & 2 & 0 & 3 & 0 & 3 & 0 & 4 & 0 & 4 & 0 & 5 & 0 & 5 & 0 & 6 & 0 & 6 & 0 & 7 \\
\hline
LHS & 1 & 0 & 1 & 0 & 2 & 0 & 2 & 0 & 3 & 0 & 3 & 0 & 4 & 0 & 4 & 0 & 5 & 0 & 5 & 0 & 6 & 0 & 6 & 0 & 7 \\
\hline
\end{tabular}

\noindent At $N=2$ and $O(r^{1})$,

\noindent \begin{tabular}{|c|ccccccccccccccccccccccccc|}
\hline
$n$ & -1 & 0 & 1 & 2 & 3 & 4 & 5 & 6 & 7 & 8 & 9 & 10 & 11 & 12 & 13 & 14 & 15 & 16 & 17 & 18 & 19 & 20 & 21 & 22 & 23 \\
\hline \hline
$k=0$ & 2 & -2 & 2 & -2 & 4 & -4 & 6 & -6 & 10 & -10 & 14 & -14 & 22 & -22 & 30 & -30 & 44 & -44 & 60 & -60 & 84 & -84 & 112 & -112 & 154 \\
$k=1$ & 2 & -2 & 2 & -2 & 2 & -2 & 2 & -2 & 2 & -2 & 0 & 0 & -2 & 2 & -8 & 8 & -16 & 16 & -30 & 30 & -50 & 50 & -82 & 82 & -124  \\
$k=2$ & 2 & -2 & 2 & -2 & 2 & -2 & 2 & -2 & 2 & -2 & 2 & -2 & 2 & -2 & 2 & -2 & 2 & -2 & 4 & -4 & 6 & -6 & 12 & -12 & 22  \\
$k=3$ & 2 & -2 & 2 & -2 & 2 & -2 & 2 & -2 & 2 & -2 & 2 & -2 & 2 & -2 & 2 & -2 & 2 & -2 & 2 & -2 & 2 & -2 & 2 & -2 & 2 \\
\hline
LHS & 2 & -2 & 2 & -2 & 2 & -2 & 2 & -2 & 2 & -2 & 2 & -2 & 2 & -2 & 2 & -2 & 2 & -2 & 2 & -2 & 2 & -2 & 2 & -2 & 2 \\
\hline
\end{tabular}

\noindent At $N=2$ and $O(r^{2})$,

\noindent \begin{tabular}{|c|ccccccccccccccccccccccc|}
\hline
$n$ & -2 & -1 & 0 & 1 & 2 & 3 & 4 & 5 & 6 & 7 & 8 & 9 & 10 & 11 & 12 & 13 & 14 & 15 & 16 & 17 & 18 & 19 & 20 \\
\hline \hline
$k=0$ & 5 & -4 & 4 & -4 & 9 & -8 & 13 & -12 & 22 & -20 & 30 & -28 & 48 & -44 & 64 & -60 & 95 & -88 & 128 & -120 & 180 & -168 & 238 \\
$k=1$ & 5 & -4 & 1 & 0 & -1 & 4 & -7 & 8 & -12 & 16 & -28 & 32 & -45 & 52 & -81 & 88 & -124 & 136 & -196 & 208 & -289 & 308 & -432 \\
$k=2$ & 5 & -4 & 1 & 0 & -1 & 4 & -4 & 4 & -4 & 4 & -4 & 4 & -1 & 0 & 4 & -8 & 18 & -24 & 44 & -56 & 86 & -104 & 156 \\
$k=3$ & 5 & -4 & 1 & 0 & -1 & 4 & -4 & 4 & -4 & 4 & -4 & 4 & -4 & 4 & -4 & 4 & -4 & 4 & -4 & 4 & -7 & 8 & -12 \\
$k=4$ & 5 & -4 & 1 & 0 & -1 & 4 & -4 & 4 & -4 & 4 & -4 & 4 & -4 & 4 & -4 & 4 & -4 & 4 & -4 & 4 & -4 & 4 & -4 \\
\hline
LHS & 5 & -4 & 1 & 0 & -1 & 4 & -4 & 4 & -4 & 4 & -4 & 4 & -4 & 4 & -4 & 4 & -4 & 4 & -4 & 4 & -4 & 4 & -4 \\
\hline
\end{tabular}

\newpage

\noindent At $N=2$ and $O(r^{3})$,

\noindent \begin{tabular}{|c|cccccccccccccccccccccc|}
\hline
$n$ & -3 & -2 & -1 & 0 & 1 & 2 & 3 & 4 & 5 & 6 & 7 & 8 & 9 & 10 & 11 & 12 & 13 & 14 & 15 & 16 & 17 & 18 \\
\hline \hline
$k=0$ & 10 & -10 & 8 & -8 & 18 & -18 & 26 & -26 & 44 & -44 & 60 & -60 & 96 & -96 & 128 & -128 & 190 & -190 & 256 & -256 & 360 & -360 \\
$k=1$ & 6 & -4 & -8 & 18 & -26 & 32 & -44 & 56 & -78 & 96 & -136 & 164 & -214 & 250 & -338 & 394 & -508 & 578 & -750 & 852 & -1086 & 1218 \\
$k=2$ & 6 & -4 & -4 & 12 & -14 & 12 & -2 & -6 & 14 & -24 & 46 & -64 & 96 & -132 & 192 & -246 & 334 & -424 & 576 & -708 & 914 & -1110 \\
$k=3$ & 6 & -4 & -4 & 12 & -14 & 12 & -6 & 0 & 2 & -4 & 8 & -8 & 4 & -2 & -4 & 12 & -30 & 48 & -80 & 116 & -184 & 254 \\
$k=4$ & 6 & -4 & -4 & 12 & -14 & 12 & -6 & 0 & 2 & -4 & 8 & -8 & 8 & -8 & 8 & -8 & 8 & -8 & 8 & -8 & 12 & -14 \\
\hline
LHS & 6 & -4 & -4 & 12 & -14 & 12 & -6 & 0 & 2 & -4 & 8 & -8 & 8 & -8 & 8 & -8 & 8 & -8 & 8 & -8 & 8 & -8 \\
\hline
\end{tabular}

\noindent At $N=2$ and $O(r^{4})$,

\noindent \begin{tabular}{|c|ccccccccccccccccccccc|}
\hline
$n$ & -6 & -5 & -4 & -3 & -2 & -1 & 0 & 1 & 2 & 3 & 4 & 5 & 6 & 7 & 8 & 9 & 10 & 11 & 12 & 13 & 14 \\
\hline \hline
$k=0$ & 0 & 0 & 20 & -20 & 15 & -16 & 36 & -36 & 51 & -52 & 87 & -88 & 118 & -120 & 190 & -192 & 252 & -256 & 376 & -380 & 505 \\
$k=1$ & -5 & 8 & -2 & 20 & -62 & 88 & -111 & 132 & -177 & 224 & -298 & 364 & -489 & 584 & -736 & 868 & -1128 & 1316 & -1641 & 1888 & -2376 \\
$k=2$ & 0 & 0 & 14 & -8 & -2 & -8 & 41 & -92 & 171 & -236 & 319 & -420 & 574 & -740 & 979 & -1240 & 1606 & -1988 & 2519 & -3088 & 3889 \\
$k=3$ & 0 & 0 & 9 & 0 & -18 & 20 & -13 & -8 & 29 & -20 & -14 & 52 & -108 & 184 & -304 & 436 & -623 & 860 & -1198 & 1576 & -2092 \\
$k=4$ & 0 & 0 & 9 & 0 & -18 & 20 & -8 & -16 & 45 & -48 & 40 & -32 & 28 & -20 & 19 & -28 & 51 & -76 & 128 & -196 & 304 \\
\hline
LHS & 0 & 0 & 9 & 0 & -18 & 20 & -8 & -16 & 45 & -48 & 40 & -32 & 23 & -12 & 3 & 0 & -3 & 8 & -8 & 8 & -8 \\
\hline
\end{tabular}

\noindent At $N=3$ and $O(r^{0})$,

\noindent \begin{tabular}{|c|cccccccccccccccccccccccccc|}
\hline
$n$ & 0 & 1 & 2 & 3 & 4 & 5 & 6 & 7 & 8 & 9 & 10 & 11 & 12 & 13 & 14 & 15 & 16 & 17 & 18 & 19 & 20 & 21 & 22 & 23 & 24 & 25 \\
\hline \hline
$k=0$ & 1 & 0 & 1 & 0 & 2 & 0 & 3 & 0 & 5 & 0 & 7 & 0 & 11 & 0 & 15 & 0 & 22 & 0 & 30 & 0 & 42 & 0 & 56 & 0 & 77 & 0 \\
$k=1$ & 1 & 0 & 1 & 0 & 2 & 0 & 3 & 0 & 4 & 0 & 5 & 0 & 7 & 0 & 8 & 0 & 10 & 0 & 11 & 0 & 12 & 0 & 11 & 0 & 10 & 0 \\
$k=2$ & 1 & 0 & 1 & 0 & 2 & 0 & 3 & 0 & 4 & 0 & 5 & 0 & 7 & 0 & 8 & 0 & 10 & 0 & 12 & 0 & 14 & 0 & 16 & 0 & 19 & 0 \\
$k=3$ & 1 & 0 & 1 & 0 & 2 & 0 & 3 & 0 & 4 & 0 & 5 & 0 & 7 & 0 & 8 & 0 & 10 & 0 & 12 & 0 & 14 & 0 & 16 & 0 & 19 & 0 \\
$k=4$ & 1 & 0 & 1 & 0 & 2 & 0 & 3 & 0 & 4 & 0 & 5 & 0 & 7 & 0 & 8 & 0 & 10 & 0 & 12 & 0 & 14 & 0 & 16 & 0 & 19 & 0 \\
\hline
LHS & 1 & 0 & 1 & 0 & 2 & 0 & 3 & 0 & 4 & 0 & 5 & 0 & 7 & 0 & 8 & 0 & 10 & 0 & 12 & 0 & 14 & 0 & 16 & 0 & 19 & 0 \\
\hline
\end{tabular}

\newpage

\noindent At $N=3$ and $O(r^{1})$,

\noindent \begin{tabular}{|c|cccccccccccccccccccccccccc|}
\hline
$n$ & -1 & 0 & 1 & 2 & 3 & 4 & 5 & 6 & 7 & 8 & 9 & 10 & 11 & 12 & 13 & 14 & 15 & 16 & 17 & 18 & 19 & 20 & 21 & 22 & 23 & 24 \\
\hline \hline
$k=0$ & 2 & -2 & 2 & -2 & 4 & -4 & 6 & -6 & 10 & -10 & 14 & -14 & 22 & -22 & 30 & -30 & 44 & -44 & 60 & -60 & 84 & -84 & 112 & -112 & 154 & -154 \\
$k=1$ & 2 & -2 & 2 & -2 & 4 & -4 & 4 & -4 & 6 & -6 & 6 & -6 & 8 & -8 & 6 & -6 & 6 & -6 & 0 & 0 & -6 & 6 & -22 & 22 & -40 & 40 \\
$k=2$ & 2 & -2 & 2 & -2 & 4 & -4 & 4 & -4 & 6 & -6 & 6 & -6 & 8 & -8 & 8 & -8 & 10 & -10 & 10 & -10 & 12 & -12 & 12 & -12 & 16 & -16 \\
$k=3$ & 2 & -2 & 2 & -2 & 4 & -4 & 4 & -4 & 6 & -6 & 6 & -6 & 8 & -8 & 8 & -8 & 10 & -10 & 10 & -10 & 12 & -12 & 12 & -12 & 14 & -14 \\
$k=4$ & 2 & -2 & 2 & -2 & 4 & -4 & 4 & -4 & 6 & -6 & 6 & -6 & 8 & -8 & 8 & -8 & 10 & -10 & 10 & -10 & 12 & -12 & 12 & -12 & 14 & -14 \\
\hline
LHS & 2 & -2 & 2 & -2 & 4 & -4 & 4 & -4 & 6 & -6 & 6 & -6 & 8 & -8 & 8 & -8 & 10 & -10 & 10 & -10 & 12 & -12 & 12 & -12 & 14 & -14 \\
\hline
\end{tabular}

\noindent At $N=3$ and $O(r^{2})$,

\noindent \begin{tabular}{|c|cccccccccccccccccccccccc|}
\hline
$n$ & -2 & -1 & 0 & 1 & 2 & 3 & 4 & 5 & 6 & 7 & 8 & 9 & 10 & 11 & 12 & 13 & 14 & 15 & 16 & 17 & 18 & 19 & 20 & 21 \\
\hline \hline
$k=0$ & 5 & -4 & 4 & -4 & 9 & -8 & 13 & -12 & 22 & -20 & 30 & -28 & 48 & -44 & 64 & -60 & 95 & -88 & 128 & -120 & 180 & -168 & 238 & -224 \\
$k=1$ & 5 & -4 & 4 & -4 & 6 & -4 & 3 & 0 & 2 & 0 & -4 & 8 & -10 & 16 & -29 & 36 & -50 & 60 & -91 & 104 & -144 & 160 & -231 & 252 \\
$k=2$ & 5 & -4 & 4 & -4 & 6 & -4 & 3 & 0 & 2 & 0 & -1 & 4 & -2 & 4 & -5 & 8 & -6 & 8 & -6 & 8 & -2 & 0 & 9 & -12 \\
$k=3$ & 5 & -4 & 4 & -4 & 6 & -4 & 3 & 0 & 2 & 0 & -1 & 4 & -2 & 4 & -5 & 8 & -6 & 8 & -9 & 12 & -10 & 12 & -13 & 16 \\
$k=4$ & 5 & -4 & 4 & -4 & 6 & -4 & 3 & 0 & 2 & 0 & -1 & 4 & -2 & 4 & -5 & 8 & -6 & 8 & -9 & 12 & -10 & 12 & -13 & 16 \\
\hline
LHS & 5 & -4 & 4 & -4 & 6 & -4 & 3 & 0 & 2 & 0 & -1 & 4 & -2 & 4 & -5 & 8 & -6 & 8 & -9 & 12 & -10 & 12 & -13 & 16 \\
\hline
\end{tabular}

\noindent At $N=3$ and $O(r^{3})$,

\noindent \begin{tabular}{|c|ccccccccccccccccccccccc|}
\hline
$n$ & -3 & -2 & -1 & 0 & 1 & 2 & 3 & 4 & 5 & 6 & 7 & 8 & 9 & 10 & 11 & 12 & 13 & 14 & 15 & 16 & 17 & 18 & 19 \\
\hline \hline
$k=0$ & 10 & -10 & 8 & -8 & 18 & -18 & 26 & -26 & 44 & -44 & 60 & -60 & 96 & -96 & 128 & -128 & 190 & -190 & 256 & -256 & 360 & -360 & 476 \\
$k=1$ & 10 & -10 & 4 & -2 & 2 & 8 & -18 & 24 & -26 & 38 & -62 & 80 & -100 & 128 & -182 & 218 & -276 & 332 & -442 & 512 & -646 & 748 & -970  \\
$k=2$ & 10 & -10 & 4 & -2 & 2 & 8 & -14 & 18 & -14 & 18 & -20 & 18 & -8 & 8 & 0 & -10 & 34 & -50 & 88 & -128 & 196 & -254 & 356  \\
$k=3$ & 10 & -10 & 4 & -2 & 2 & 8 & -14 & 18 & -14 & 18 & -20 & 18 & -12 & 14 & -12 & 10 & -4 & 6 & -4 & 2 & 0 & 4 & -8  \\
$k=4$ & 10 & -10 & 4 & -2 & 2 & 8 & -14 & 18 & -14 & 18 & -20 & 18 & -12 & 14 & -12 & 10 & -4 & 6 & -4 & 2 & 4 & -2 & 4  \\
\hline
LHS & 10 & -10 & 4 & -2 & 2 & 8 & -14 & 18 & -14 & 18 & -20 & 18 & -12 & 14 & -12 & 10 & -4 & 6 & -4 & 2 & 4 & -2 & 4  \\
\hline
\end{tabular}

\newpage

\noindent At $N=3$ and $O(r^{4})$,

\noindent \begin{tabular}{|c|ccccccccccccccccccccc|}
\hline
$n$ & -4 & -3 & -2 & -1 & 0 & 1 & 2 & 3 & 4 & 5 & 6 & 7 & 8 & 9 & 10 & 11 & 12 & 13 & 14 & 15 & 16 \\
\hline \hline
$k=0$ & 20 & -20 & 15 & -16 & 36 & -36 & 51 & -52 & 87 & -88 & 118 & -120 & 190 & -192 & 252 & -256 & 376 & -380 & 505 & -512 & 712 \\
$k=1$ & 15 & -12 & -7 & 24 & -41 & 68 & -96 & 116 & -141 & 188 & -267 & 332 & -417 & 512 & -674 & 804 & -1004 & 1192 & -1512 & 1756 & -2169 \\
$k=2$ & 15 & -12 & -2 & 16 & -25 & 40 & -36 & 20 & 11 & -36 & 81 & -128 & 200 & -272 & 389 & -520 & 711 & -916 & 1222 & -1548 & 1991 \\
$k=3$ & 15 & -12 & -2 & 16 & -25 & 40 & -41 & 28 & -5 & -8 & 27 & -44 & 58 & -56 & 56 & -48 & 29 & 8 & -61 & 128 & -238 \\
$k=4$ & 15 & -12 & -2 & 16 & -25 & 40 & -41 & 28 & -5 & -8 & 27 & -44 & 63 & -64 & 72 & -76 & 83 & -76 & 75 & -76 & 85 \\
\hline
LHS & 15 & -12 & -2 & 16 & -25 & 40 & -41 & 28 & -5 & -8 & 27 & -44 & 63 & -64 & 72 & -76 & 83 & -76 & 75 & -76 & 80 \\
\hline
\end{tabular}
The pattern of convergence persists, where the expansion in terms of giant gravitons on the right hand side converges faster in $k$ to $Z_N$ for larger values of $N$.

The expansion also holds at $N \leq 0$:

\noindent At $N=0$ and $O(r^{0})$,

\noindent \begin{tabular}{|c|cccccccccccccccccccccccccc|}
\hline
$n$ & 0 & 1 & 2 & 3 & 4 & 5 & 6 & 7 & 8 & 9 & 10 & 11 & 12 & 13 & 14 & 15 & 16 & 17 & 18 & 19 & 20 & 21 & 22 & 23 & 24 & 25 \\
\hline \hline
$k=0$ & 1 & 0 & 1 & 0 & 2 & 0 & 3 & 0 & 5 & 0 & 7 & 0 & 11 & 0 & 15 & 0 & 22 & 0 & 30 & 0 & 42 & 0 & 56 & 0 & 77 & 0 \\
$k=1$ & 1 & 0 & 0 & 0 & 0 & 0 & -1 & 0 & -2 & 0 & -5 & 0 & -8 & 0 & -15 & 0 & -23 & 0 & -37 & 0 & -55 & 0 & -83 & 0 & -118 & 0 \\
$k=2$ & 1 & 0 & 0 & 0 & 0 & 0 & 0 & 0 & 0 & 0 & 0 & 0 & 1 & 0 & 2 & 0 & 5 & 0 & 10 & 0 & 18 & 0 & 31 & 0 & 52 & 0 \\
$k=3$ & 1 & 0 & 0 & 0 & 0 & 0 & 0 & 0 & 0 & 0 & 0 & 0 & 0 & 0 & 0 & 0 & 0 & 0 & 0 & 0 & -1 & 0 & -2 & 0 & -5 & 0 \\
$k=4$ & 1 & 0 & 0 & 0 & 0 & 0 & 0 & 0 & 0 & 0 & 0 & 0 & 0 & 0 & 0 & 0 & 0 & 0 & 0 & 0 & 0 & 0 & 0 & 0 & 0 & 0 \\
\hline
LHS & 1 & 0 & 0 & 0 & 0 & 0 & 0 & 0 & 0 & 0 & 0 & 0 & 0 & 0 & 0 & 0 & 0 & 0 & 0 & 0 & 0 & 0 & 0 & 0 & 0 & 0 \\
\hline
\end{tabular}

\noindent At $N=0$ and $O(r^{1})$,

\noindent \begin{tabular}{|c|ccccccccccccccccccccc|}
\hline
$n$ & -1 & 0 & 1 & 2 & 3 & 4 & 5 & 6 & 7 & 8 & 9 & 10 & 11 & 12 & 13 & 14 & 15 & 16 & 17 & 18 & 19 \\
\hline \hline
$k=0$ & 2 & -2 & 2 & -2 & 4 & -4 & 6 & -6 & 10 & -10 & 14 & -14 & 22 & -22 & 30 & -30 & 44 & -44 & 60 & -60 & 84 \\
$k=1$ & 0 & 0 & -2 & 2 & -4 & 4 & -8 & 8 & -14 & 14 & -24 & 24 & -38 & 38 & -60 & 60 & -90 & 90 & -134 & 134 & -194 \\
$k=2$ & 0 & 0 & 0 & 0 & 0 & 0 & 2 & -2 & 4 & -4 & 10 & -10 & 18 & -18 & 34 & -34 & 56 & -56 & 94 & -94 & 146 \\
$k=3$ & 0 & 0 & 0 & 0 & 0 & 0 & 0 & 0 & 0 & 0 & 0 & 0 & -2 & 2 & -4 & 4 & -10 & 10 & -20 & 20 & -38 \\
$k=4$ & 0 & 0 & 0 & 0 & 0 & 0 & 0 & 0 & 0 & 0 & 0 & 0 & 0 & 0 & 0 & 0 & 0 & 0 & 0 & 0 & 2 \\
\hline
LHS & 0 & 0 & 0 & 0 & 0 & 0 & 0 & 0 & 0 & 0 & 0 & 0 & 0 & 0 & 0 & 0 & 0 & 0 & 0 & 0 & 0 \\
\hline
\end{tabular}

\newpage

\noindent At $N=0$ and $O(r^{2})$,

\noindent \begin{tabular}{|c|ccccccccccccccccccccc|}
\hline
$n$ & -4 & -3 & -2 & -1 & 0 & 1 & 2 & 3 & 4 & 5 & 6 & 7 & 8 & 9 & 10 & 11 & 12 & 13 & 14 & 15 & 16 \\
\hline \hline
$k=0$ & 0 & 0 & 5 & -4 & 4 & -4 & 9 & -8 & 13 & -12 & 22 & -20 & 30 & -28 & 48 & -44 & 64 & -60 & 95 & -88 & 128 \\
$k=1$ & -3 & 4 & -5 & 8 & -16 & 16 & -25 & 28 & -45 & 48 & -71 & 76 & -115 & 120 & -171 & 180 & -260 & 268 & -374 & 388 & -542 \\
$k=2$ & 0 & 0 & 3 & -4 & 8 & -12 & 19 & -24 & 40 & -48 & 71 & -84 & 125 & -144 & 204 & -232 & 328 & -368 & 504 & -560 & 767 \\
$k=3$ & 0 & 0 & 0 & 0 & 0 & 0 & -3 & 4 & -8 & 12 & -22 & 28 & -43 & 56 & -89 & 108 & -154 & 188 & -271 & 320 & -442 \\
$k=4$ & 0 & 0 & 0 & 0 & 0 & 0 & 0 & 0 & 0 & 0 & 0 & 0 & 3 & -4 & 8 & -12 & 22 & -28 & 46 & -60 & 92 \\
$k=5$ & 0 & 0 & 0 & 0 & 0 & 0 & 0 & 0 & 0 & 0 & 0 & 0 & 0 & 0 & 0 & 0 & 0 & 0 & 0 & 0 & -3 \\
\hline
LHS & 0 & 0 & 0 & 0 & 0 & 0 & 0 & 0 & 0 & 0 & 0 & 0 & 0 & 0 & 0 & 0 & 0 & 0 & 0 & 0 & 0 \\
\hline
\end{tabular}

\noindent At $N=0$ and $O(r^{3})$,

\noindent \begin{tabular}{|c|ccccccccccccccccccccc|}
\hline
$n$ & -9 & -8 & -7 & -6 & -5 & -4 & -3 & -2 & -1 & 0 & 1 & 2 & 3 & 4 & 5 & 6 & 7 & 8 & 9 & 10 & 11 \\
\hline \hline
$k=0$ & 0 & 0 & 0 & 0 & 0 & 0 & 10 & -10 & 8 & -8 & 18 & -18 & 26 & -26 & 44 & -44 & 60 & -60 & 96 & -96 & 128 \\
$k=1$ & 0 & 0 & -4 & 6 & -16 & 26 & -34 & 40 & -62 & 74 & -104 & 122 & -170 & 198 & -266 & 302 & -406 & 462 & -602 & 672 & -878 \\
$k=2$ & 4 & -6 & 8 & -14 & 26 & -36 & 58 & -80 & 120 & -154 & 206 & -260 & 360 & -442 & 576 & -700 & 920 & -1098 & 1398 & -1656 & 2112 \\
$k=3$ & 0 & 0 & -4 & 6 & -12 & 20 & -34 & 50 & -76 & 104 & -158 & 212 & -296 & 382 & -522 & 664 & -878 & 1094 & -1434 & 1766 & -2258 \\
$k=4$ & 0 & 0 & 0 & 0 & 0 & 0 & 4 & -6 & 12 & -20 & 38 & -56 & 84 & -118 & 180 & -242 & 342 & -454 & 630 & -810 & 1080 \\
$k=5$ & 0 & 0 & 0 & 0 & 0 & 0 & 0 & 0 & 0 & 0 & 0 & 0 & -4 & 6 & -12 & 20 & -38 & 56 & -88 & 124 & -188 \\
$k=6$ & 0 & 0 & 0 & 0 & 0 & 0 & 0 & 0 & 0 & 0 & 0 & 0 & 0 & 0 & 0 & 0 & 0 & 0 & 0 & 0 & 4 \\
\hline
LHS & 0 & 0 & 0 & 0 & 0 & 0 & 0 & 0 & 0 & 0 & 0 & 0 & 0 & 0 & 0 & 0 & 0 & 0 & 0 & 0 & 0 \\
\hline
\end{tabular}

\noindent At $N=-1$ and $O(r^{0})$,

\noindent \begin{tabular}{|c|ccccccccccccccccccccccccc|}
\hline
$n$ & 0 & 1 & 2 & 3 & 4 & 5 & 6 & 7 & 8 & 9 & 10 & 11 & 12 & 13 & 14 & 15 & 16 & 17 & 18 & 19 & 20 & 21 & 22 & 23 & 24 \\
\hline \hline
$k=0$ & 1 & 0 & 1 & 0 & 2 & 0 & 3 & 0 & 5 & 0 & 7 & 0 & 11 & 0 & 15 & 0 & 22 & 0 & 30 & 0 & 42 & 0 & 56 & 0 & 77 \\
$k=1$ & 0 & 0 & -1 & 0 & -2 & 0 & -4 & 0 & -7 & 0 & -12 & 0 & -19 & 0 & -30 & 0 & -45 & 0 & -67 & 0 & -97 & 0 & -139 & 0 & -195 \\
$k=2$ & 0 & 0 & 0 & 0 & 0 & 0 & 1 & 0 & 2 & 0 & 5 & 0 & 9 & 0 & 17 & 0 & 28 & 0 & 47 & 0 & 73 & 0 & 114 & 0 & 170 \\
$k=3$ & 0 & 0 & 0 & 0 & 0 & 0 & 0 & 0 & 0 & 0 & 0 & 0 & -1 & 0 & -2 & 0 & -5 & 0 & -10 & 0 & -19 & 0 & -33 & 0 & -57 \\
$k=4$ & 0 & 0 & 0 & 0 & 0 & 0 & 0 & 0 & 0 & 0 & 0 & 0 & 0 & 0 & 0 & 0 & 0 & 0 & 0 & 0 & 1 & 0 & 2 & 0 & 5 \\
\hline
LHS & 0 & 0 & 0 & 0 & 0 & 0 & 0 & 0 & 0 & 0 & 0 & 0 & 0 & 0 & 0 & 0 & 0 & 0 & 0 & 0 & 0 & 0 & 0 & 0 & 0 \\
\hline
\end{tabular}

\newpage

\noindent At $N=-1$ and $O(r^{1})$,

\noindent \begin{tabular}{|c|ccccccccccccccccccccc|}
\hline
$n$ & -3 & -2 & -1 & 0 & 1 & 2 & 3 & 4 & 5 & 6 & 7 & 8 & 9 & 10 & 11 & 12 & 13 & 14 & 15 & 16 & 17 \\
\hline \hline
$k=0$ & 0 & 0 & 2 & -2 & 2 & -2 & 4 & -4 & 6 & -6 & 10 & -10 & 14 & -14 & 22 & -22 & 30 & -30 & 44 & -44 & 60 \\
$k=1$ & -2 & 2 & -2 & 2 & -6 & 6 & -10 & 10 & -18 & 18 & -28 & 28 & -46 & 46 & -68 & 68 & -104 & 104 & -150 & 150 & -218 \\
$k=2$ & 0 & 0 & 2 & -2 & 4 & -4 & 8 & -8 & 16 & -16 & 28 & -28 & 48 & -48 & 78 & -78 & 124 & -124 & 190 & -190 & 288 \\
$k=3$ & 0 & 0 & 0 & 0 & 0 & 0 & -2 & 2 & -4 & 4 & -10 & 10 & -18 & 18 & -36 & 36 & -60 & 60 & -104 & 104 & -166 \\
$k=4$ & 0 & 0 & 0 & 0 & 0 & 0 & 0 & 0 & 0 & 0 & 0 & 0 & 2 & -2 & 4 & -4 & 10 & -10 & 20 & -20 & 38 \\
$k=5$ & 0 & 0 & 0 & 0 & 0 & 0 & 0 & 0 & 0 & 0 & 0 & 0 & 0 & 0 & 0 & 0 & 0 & 0 & 0 & 0 & -2 \\
\hline
LHS & 0 & 0 & 0 & 0 & 0 & 0 & 0 & 0 & 0 & 0 & 0 & 0 & 0 & 0 & 0 & 0 & 0 & 0 & 0 & 0 & 0 \\
\hline
\end{tabular}

\noindent At $N=-1$ and $O(r^{2})$,

\noindent \begin{tabular}{|c|ccccccccccccccccccccc|}
\hline
$n$ & -8 & -7 & -6 & -5 & -4 & -3 & -2 & -1 & 0 & 1 & 2 & 3 & 4 & 5 & 6 & 7 & 8 & 9 & 10 & 11 & 12 \\
\hline \hline
$k=0$ & 0 & 0 & 0 & 0 & 0 & 0 & 5 & -4 & 4 & -4 & 9 & -8 & 13 & -12 & 22 & -20 & 30 & -28 & 48 & -44 & 64 \\
$k=1$ & 0 & 0 & -3 & 4 & -10 & 12 & -15 & 16 & -30 & 32 & -49 & 52 & -80 & 84 & -123 & 128 & -189 & 196 & -276 & 284 & -405 \\
$k=2$ & 3 & -4 & 5 & -8 & 14 & -16 & 29 & -36 & 55 & -64 & 93 & -108 & 160 & -180 & 252 & -284 & 399 & -440 & 602 & -664 & 904 \\
$k=3$ & 0 & 0 & -3 & 4 & -8 & 12 & -19 & 24 & -38 & 48 & -75 & 92 & -133 & 160 & -230 & 272 & -376 & 440 & -607 & 700 & -940 \\
$k=4$ & 0 & 0 & 0 & 0 & 0 & 0 & 3 & -4 & 8 & -12 & 22 & -28 & 43 & -56 & 87 & -108 & 158 & -196 & 279 & -336 & 466 \\
$k=5$ & 0 & 0 & 0 & 0 & 0 & 0 & 0 & 0 & 0 & 0 & 0 & 0 & -3 & 4 & -8 & 12 & -22 & 28 & -46 & 60 & -92 \\
$k=6$ & 0 & 0 & 0 & 0 & 0 & 0 & 0 & 0 & 0 & 0 & 0 & 0 & 0 & 0 & 0 & 0 & 0 & 0 & 0 & 0 & 3 \\
\hline
LHS & 0 & 0 & 0 & 0 & 0 & 0 & 0 & 0 & 0 & 0 & 0 & 0 & 0 & 0 & 0 & 0 & 0 & 0 & 0 & 0 & 0 \\
\hline
\end{tabular}

\noindent At $N=-2$ and $O(r^{0})$,

\noindent \begin{tabular}{|c|ccccccccccccccccccccc|}
\hline
$n$ & -2 & -1 & 0 & 1 & 2 & 3 & 4 & 5 & 6 & 7 & 8 & 9 & 10 & 11 & 12 & 13 & 14 & 15 & 16 & 17 & 18 \\
\hline \hline
$k=0$ & 0 & 0 & 1 & 0 & 1 & 0 & 2 & 0 & 3 & 0 & 5 & 0 & 7 & 0 & 11 & 0 & 15 & 0 & 22 & 0 & 30 \\
$k=1$ & -1 & 0 & -1 & 0 & -3 & 0 & -5 & 0 & -9 & 0 & -14 & 0 & -23 & 0 & -34 & 0 & -52 & 0 & -75 & 0 & -109 \\
$k=2$ & 0 & 0 & 1 & 0 & 2 & 0 & 4 & 0 & 8 & 0 & 14 & 0 & 24 & 0 & 39 & 0 & 62 & 0 & 95 & 0 & 144 \\
$k=3$ & 0 & 0 & 0 & 0 & 0 & 0 & -1 & 0 & -2 & 0 & -5 & 0 & -9 & 0 & -18 & 0 & -30 & 0 & -52 & 0 & -83 \\
$k=4$ & 0 & 0 & 0 & 0 & 0 & 0 & 0 & 0 & 0 & 0 & 0 & 0 & 1 & 0 & 2 & 0 & 5 & 0 & 10 & 0 & 19 \\
$k=5$ & 0 & 0 & 0 & 0 & 0 & 0 & 0 & 0 & 0 & 0 & 0 & 0 & 0 & 0 & 0 & 0 & 0 & 0 & 0 & 0 & -1 \\
\hline
LHS & 0 & 0 & 0 & 0 & 0 & 0 & 0 & 0 & 0 & 0 & 0 & 0 & 0 & 0 & 0 & 0 & 0 & 0 & 0 & 0 & 0 \\
\hline
\end{tabular}%

\end{landscape}

\section{Tests for a general class of $U(N)$ indices} \label{appendix: check of general proposal}

Here, we provide evidence in support of the giant graviton expansion for a more general class of $U(N)$ indices
\begin{equation}
    Z_N = \int_{U(N)} DU \exp \left( \sum_{j=1}^{\infty} \frac{1}{j} f(x^j) \Tr U^{j} \Tr U^{-j} \right)
\end{equation}
with
\begin{equation}
    f(x^j) = x^j + (1 - x^j) h_j.
\end{equation}
Under tilde, the quantity $h_j$ transforms as
\[
\sim \ : \ h_j \to \tilde{h}_j = \frac{h_j}{h_j - 1}.
\]
This class of indices namely includes the $U(N)$ $\mathcal{N}=4$ SYM index.

We consider the truncated expansion
\begin{equation} \label{general conjecture-2 truncated - appendix}
    Z_N / Z_\infty =  1 + x^N \hat{Z}_1 + x^{2 N} \hat{Z}_2 + \cdots + x^{k N} \hat{Z}_k.
\end{equation}
with
\[
Z_0 = 1, \quad Z_{-1} = Z_{-2} = \cdots = 0.
\]
We compare the perturbative series in $x$ that arise as coefficients of monomials of $h_j$, on each side of \eqref{general conjecture-2 truncated - appendix}.

We compare the series expansion in $x$ on left and right sides of \eqref{general conjecture-2 truncated - appendix} for given $N$, truncation level $k$, and monomials of $h_j$ for which the $x$-series is a coefficient. For convenience, we write the left and right sides as $L_{N}[\{ h_j \}]$ and $R_{N}^{(k)}[\{ h_j \}]$, where $\{ h_j \}$ denotes the monomial in $h_j$.

Checks beyond those presented below can be conducted by using the relation of $U(N)$ indices to symmetric group characters reviewed in Appendix \ref{app: character formula for U(N)}.

For $N=1$ and coefficient of $1$, we have
\begin{align*}
    L_{1}[1] &= 1-x^2-x^3-x^4+x^7+x^8+x^9+x^{10}+x^{11}-x^{15}+O\left(x^{16}\right) \\
    R_{1}^{(0)}[1] &= 1 + O\left(x^{16}\right) \\
    R_{1}^{(1)}[1] &= 1-x^2-x^3-x^4-x^5-x^6-x^7-x^8-x^9-x^{10}-x^{11}-x^{12}-x^{13}-x^{14}-x^{15}+O\left(x^{16}\right) \\
    R_{1}^{(2)}[1] &= 1-x^2-x^3-x^4+x^7+x^8+2 x^9+2 x^{10}+3 x^{11}+3 x^{12}+4 x^{13}+4 x^{14}+5 x^{15}+O\left(x^{16}\right) \\
    R_{1}^{(3)}[1] &= 1-x^2-x^3-x^4+x^7+x^8+x^9+x^{10}+x^{11}-x^{14}-2 x^{15}+O\left(x^{16}\right) \\
    R_{1}^{(4)}[1] &= 1-x^2-x^3-x^4+x^7+x^8+x^9+x^{10}+x^{11}-x^{15}+O\left(x^{16}\right) \\
    R_{1}^{(5)}[1] &= 1-x^2-x^3-x^4+x^7+x^8+x^9+x^{10}+x^{11}-x^{15}+O\left(x^{16}\right) 
\end{align*}
For $N=1$ and coefficient of $h_1$, we have
\begin{align*}
    L_{1}[h_1] &= -x+x^3+x^4+x^5-x^8-x^9-x^{10}-x^{11}-x^{12}+O\left(x^{15}\right) \\
    R_{1}^{(0)}[h_1] &= O\left(x^{15}\right) \\
    R_{1}^{(1)}[h_1] &= -x+O\left(x^{15}\right) \\
    R_{1}^{(2)}[h_1] &= -x+x^3+x^4+x^5+x^6+x^7+x^8+x^9+x^{10}+x^{11}+x^{12}+x^{13}+x^{14}+O\left(x^{15}\right) \\
    R_{1}^{(3)}[h_1] &= -x+x^3+x^4+x^5-x^8-x^9-2 x^{10}-2 x^{11}-3 x^{12}-3 x^{13}-4 x^{14}+O\left(x^{15}\right) \\
    R_{1}^{(4)}[h_1] &= -x+x^3+x^4+x^5-x^8-x^9-x^{10}-x^{11}-x^{12}+O\left(x^{15}\right) \\
    R_{1}^{(5)}[h_1] &= -x+x^3+x^4+x^5-x^8-x^9-x^{10}-x^{11}-x^{12}+O\left(x^{15}\right) 
\end{align*}
For $N=1$ and coefficient of $h_1^2$, we have
\begin{align*}
    L_{1}[h_1^2] &= -\frac{1}{2}+x^2+\frac{x^3}{2}-\frac{x^5}{2}-\frac{x^6}{2}-\frac{x^7}{2}-\frac{x^8}{2}+\frac{x^{12}}{2}+\frac{x^{13}}{2}+O\left(x^{14}\right) \\
    R_{1}^{(0)}[h_1^2] &= O\left(x^{14}\right) \\
    R_{1}^{(1)}[h_1^2] &= -\frac{1}{2}-\frac{x}{2}+O\left(x^{14}\right) \\
    R_{1}^{(2)}[h_1^2] &= -\frac{1}{2}+x^2+x^3+x^4+x^5+x^6+x^7+x^8+x^9+x^{10}+x^{11}+x^{12}+x^{13}+O\left(x^{14}\right) \\
    R_{1}^{(3)}[h_1^2] &= -\frac{1}{2}+x^2+\frac{x^3}{2}-\frac{x^5}{2}-x^6-\frac{3 x^7}{2}-2 x^8-\frac{5 x^9}{2}-3 x^{10}-\frac{7 x^{11}}{2}-4 x^{12}-\frac{9 x^{13}}{2}+O\left(x^{14}\right) \\
    R_{1}^{(4)}[h_1^2] &= -\frac{1}{2}+x^2+\frac{x^3}{2}-\frac{x^5}{2}-\frac{x^6}{2}-\frac{x^7}{2}-\frac{x^8}{2}+\frac{x^{10}}{2}+x^{11}+2 x^{12}+3 x^{13}+O\left(x^{14}\right) \\
    R_{1}^{(5)}[h_1^2] &= -\frac{1}{2}+x^2+\frac{x^3}{2}-\frac{x^5}{2}-\frac{x^6}{2}-\frac{x^7}{2}-\frac{x^8}{2}+\frac{x^{12}}{2}+\frac{x^{13}}{2}+O\left(x^{14}\right) 
\end{align*}
For $N=1$ and coefficient of $h_2$, we have
\begin{align*}
    L_{1}[h_2] &= -\frac{1}{2}+\frac{x^3}{2}+x^4+\frac{x^5}{2}+\frac{x^6}{2}-\frac{x^7}{2}-\frac{x^8}{2}-x^9-x^{10}-x^{11}-\frac{x^{12}}{2}-\frac{x^{13}}{2}+O\left(x^{14}\right) \\
    R_{1}^{(0)}[h_2] &= O\left(x^{14}\right) \\
    R_{1}^{(1)}[h_2] &= -\frac{1}{2}-\frac{x}{2}+O\left(x^{14}\right) \\
    R_{1}^{(2)}[h_2] &= -\frac{1}{2}+x^3+x^4+x^5+x^6+x^7+x^8+x^9+x^{10}+x^{11}+x^{12}+x^{13}+O\left(x^{14}\right) \\
    R_{1}^{(3)}[h_2] &= -\frac{1}{2}+\frac{x^3}{2}+x^4+\frac{x^5}{2}-\frac{x^7}{2}-x^8-\frac{3 x^9}{2}-2 x^{10}-\frac{5 x^{11}}{2}-3 x^{12}-\frac{7 x^{13}}{2}+O\left(x^{14}\right) \\
    R_{1}^{(4)}[h_2] &= -\frac{1}{2}+\frac{x^3}{2}+x^4+\frac{x^5}{2}+\frac{x^6}{2}-\frac{x^7}{2}-\frac{x^8}{2}-x^9-\frac{x^{10}}{2}-x^{11}+O\left(x^{14}\right) \\
    R_{1}^{(5)}[h_2] &= -\frac{1}{2}+\frac{x^3}{2}+x^4+\frac{x^5}{2}+\frac{x^6}{2}-\frac{x^7}{2}-\frac{x^8}{2}-x^9-x^{10}-x^{11}-\frac{x^{12}}{2}-\frac{x^{13}}{2}+O\left(x^{14}\right) 
\end{align*}
For $N=1$ and coefficient of $h_1 h_2$, we have
\begin{align*}
    L_{1}[h_1 h_2] &= \frac{x}{2}-\frac{x^4}{2}-x^5-\frac{x^6}{2}-\frac{x^7}{2}+\frac{x^8}{2}+\frac{x^9}{2}+x^{10}+x^{11}+x^{12}+O\left(x^{13}\right) \\
    R_{1}^{(0)}[h_1 h_2] &= O\left(x^{13}\right) \\
    R_{1}^{(1)}[h_1 h_2] &= -\frac{1}{2 x}+\frac{x}{2}+O\left(x^{13}\right) \\
    R_{1}^{(2)}[h_1 h_2] &= \frac{1}{2}+x+\frac{x^2}{2}+O\left(x^{13}\right) \\
    R_{1}^{(3)}[h_1 h_2] &= \frac{x}{2}-\frac{x^2}{2}-\frac{x^3}{2}-\frac{3 x^4}{2}-\frac{3 x^5}{2}-\frac{3 x^6}{2}-\frac{3 x^7}{2}-\frac{3 x^8}{2}-\frac{3 x^9}{2}-\frac{3 x^{10}}{2}-\frac{3 x^{11}}{2}-\frac{3 x^{12}}{2}+O\left(x^{13}\right) \\
    R_{1}^{(4)}[h_1 h_2] &= \frac{x}{2}-\frac{x^4}{2}-\frac{x^5}{2}+\frac{x^7}{2}+\frac{3 x^8}{2}+2 x^9+\frac{5 x^{10}}{2}+\frac{7 x^{11}}{2}+4 x^{12}+O\left(x^{13}\right) \\
    R_{1}^{(5)}[h_1 h_2] &= \frac{x}{2}-\frac{x^4}{2}-x^5-\frac{x^6}{2}-\frac{x^7}{2}+\frac{x^8}{2}+\frac{x^{10}}{2}+O\left(x^{13}\right)
\end{align*}
For $N=1$ and coefficient of $h_3$, we have
\begin{align*}
    L_{1}[h_3] &= -\frac{2}{3}+\frac{2 x^2}{3}+\frac{x^3}{3}+\frac{2 x^4}{3}+\frac{x^5}{3}+\frac{x^6}{3}-\frac{x^7}{3}-\frac{2 x^8}{3}-\frac{2 x^9}{3}-x^{10}-x^{11}-\frac{x^{12}}{3}+O\left(x^{13}\right) \\
    R_{1}^{(0)}[h_3] &= O\left(x^{13}\right) \\
    R_{1}^{(1)}[h_3] &= -\frac{1}{3 x}-\frac{1}{3}-\frac{x}{3}+O\left(x^{13}\right) \\
    R_{1}^{(2)}[h_3] &= -\frac{1}{3}+\frac{x^2}{3}+x^3+x^4+x^5+x^6+x^7+x^8+x^9+x^{10}+x^{11}+x^{12}+O\left(x^{13}\right) \\
    R_{1}^{(3)}[h_3] &= -\frac{2}{3}+\frac{x^2}{3}+\frac{x^3}{3}+\frac{2 x^4}{3}+\frac{x^5}{3}-\frac{x^6}{3}-\frac{2 x^7}{3}-\frac{4 x^8}{3}-\frac{5 x^9}{3}-\frac{7 x^{10}}{3}-\frac{8 x^{11}}{3}-\frac{10 x^{12}}{3}+O\left(x^{13}\right) \\
    R_{1}^{(4)}[h_3] &= -\frac{2}{3}+\frac{2 x^2}{3}+\frac{x^3}{3}+\frac{2 x^4}{3}+\frac{2 x^5}{3}+\frac{x^6}{3}-\frac{x^7}{3}-\frac{x^8}{3}-\frac{2 x^9}{3}-\frac{x^{10}}{3}-\frac{x^{11}}{3}+\frac{x^{12}}{3}+O\left(x^{13}\right) \\
    R_{1}^{(5)}[h_3] &= -\frac{2}{3}+\frac{2 x^2}{3}+\frac{x^3}{3}+\frac{2 x^4}{3}+\frac{x^5}{3}+\frac{x^6}{3}-\frac{x^7}{3}-\frac{2 x^8}{3}-x^9-x^{10}-x^{11}-\frac{2 x^{12}}{3}+O\left(x^{13}\right) 
\end{align*}
For $N=2$ and coefficient of $1$, we have
\begin{align*}
    L_{2}[1] &= 1-x^3-x^4-x^5-x^6+x^9+x^{10}+2 x^{11}+x^{12}+2 x^{13}+x^{14}+x^{15}+O\left(x^{16}\right) \\
    R_{2}^{(0)}[1] &= 1+O\left(x^{16}\right) \\
    R_{2}^{(1)}[1] &= 1-x^3-x^4-x^5-x^6-x^7-x^8-x^9-x^{10}-x^{11}-x^{12}-x^{13}-x^{14}-x^{15}+O\left(x^{16}\right) \\
    R_{2}^{(2)}[1] &= 1-x^3-x^4-x^5-x^6+x^9+x^{10}+2 x^{11}+2 x^{12}+3 x^{13}+3 x^{14}+4 x^{15}+O\left(x^{16}\right) \\
    R_{2}^{(3)}[1] &= 1-x^3-x^4-x^5-x^6+x^9+x^{10}+2 x^{11}+x^{12}+2 x^{13}+x^{14}+x^{15}+O\left(x^{16}\right) \\
    R_{2}^{(4)}[1] &= 1-x^3-x^4-x^5-x^6+x^9+x^{10}+2 x^{11}+x^{12}+2 x^{13}+x^{14}+x^{15}+O\left(x^{16}\right)
\end{align*}
For $N=2$ and coefficient of $h_1$, we have
\begin{align*}
    L_{2}[h_1] &= -x^2+x^5+x^6+x^7+x^8-x^{11}-x^{12}-2 x^{13}-x^{14}+O\left(x^{15}\right) \\
    R_{2}^{(0)}[h_1] &= O\left(x^{15}\right) \\
    R_{2}^{(1)}[h_1] &= -x^2+O\left(x^{15}\right) \\
    R_{2}^{(2)}[h_1] &= -x^2+x^5+x^6+x^7+x^8+x^9+x^{10}+x^{11}+x^{12}+x^{13}+x^{14}+O\left(x^{15}\right) \\
    R_{2}^{(3)}[h_1] &= -x^2+x^5+x^6+x^7+x^8-x^{11}-x^{12}-2 x^{13}-2 x^{14}+O\left(x^{15}\right) \\
    R_{2}^{(4)}[h_1] &= -x^2+x^5+x^6+x^7+x^8-x^{11}-x^{12}-2 x^{13}-x^{14}+O\left(x^{15}\right)
\end{align*}
For $N=2$ and coefficient of $h_1^2$, we have
\begin{align*}
    L_{2}[h_1^2] &= -\frac{x}{2}-\frac{x^2}{2}+\frac{x^3}{2}+x^4+x^5+\frac{x^6}{2}-\frac{x^8}{2}-x^9-x^{10}-x^{11}-x^{12}-\frac{x^{13}}{2}+O\left(x^{14}\right) \\
    R_{2}^{(0)}[h_1^2] &= O\left(x^{14}\right) \\
    R_{2}^{(1)}[h_1^2] &= -\frac{x}{2}-\frac{x^2}{2}+O\left(x^{14}\right) \\
    R_{2}^{(2)}[h_1^2] &= -\frac{x}{2}-\frac{x^2}{2}+\frac{x^3}{2}+x^4+x^5+x^6+x^7+x^8+x^9+x^{10}+x^{11}+x^{12}+x^{13}+O\left(x^{14}\right) \\
    R_{2}^{(3)}[h_1^2] &= -\frac{x}{2}-\frac{x^2}{2}+\frac{x^3}{2}+x^4+x^5+\frac{x^6}{2}-\frac{x^8}{2}-x^9-\frac{3 x^{10}}{2}-2 x^{11}-\frac{5 x^{12}}{2}-3 x^{13}+O\left(x^{14}\right) \\
    R_{2}^{(4)}[h_1^2] &= -\frac{x}{2}-\frac{x^2}{2}+\frac{x^3}{2}+x^4+x^5+\frac{x^6}{2}-\frac{x^8}{2}-x^9-x^{10}-x^{11}-x^{12}-\frac{x^{13}}{2}+O\left(x^{14}\right)
\end{align*}
For $N=2$ and coefficient of $h_2$, we have
\begin{align*}
    L_{2}[h_2] &= -\frac{x}{2}-\frac{x^2}{2}+\frac{x^3}{2}+x^5+\frac{x^6}{2}+x^7+\frac{x^8}{2}-x^{11}-x^{12}-\frac{3 x^{13}}{2}+O\left(x^{14}\right) \\
    R_{2}^{(0)}[h_2] &= O\left(x^{14}\right) \\
    R_{2}^{(1)}[h_2] &= -\frac{x}{2}-\frac{x^2}{2}+O\left(x^{14}\right) \\
    R_{2}^{(2)}[h_2] &= -\frac{x}{2}-\frac{x^2}{2}+\frac{x^3}{2}+x^5+x^6+x^7+x^8+x^9+x^{10}+x^{11}+x^{12}+x^{13}+O\left(x^{14}\right) \\
    R_{2}^{(3)}[h_2] &= -\frac{x}{2}-\frac{x^2}{2}+\frac{x^3}{2}+x^5+\frac{x^6}{2}+x^7+\frac{x^8}{2}-\frac{x^{10}}{2}-x^{11}-\frac{3 x^{12}}{2}-2 x^{13}+O\left(x^{14}\right) \\
    R_{2}^{(4)}[h_2] &= -\frac{x}{2}-\frac{x^2}{2}+\frac{x^3}{2}+x^5+\frac{x^6}{2}+x^7+\frac{x^8}{2}-x^{11}-x^{12}-\frac{3 x^{13}}{2}+O\left(x^{14}\right)
\end{align*}
For $N=2$ and coefficient of $h_1 h_2$, we have
\begin{align*}
    L_{2}[h_1 h_2] &= -\frac{1}{2}+\frac{x}{2}+x^2-x^5-x^7-\frac{x^8}{2}-\frac{x^9}{2}-\frac{x^{10}}{2}+\frac{x^{12}}{2}+O\left(x^{13}\right) \\
    R_{2}^{(0)}[h_1 h_2] &= O\left(x^{13}\right) \\
    R_{2}^{(1)}[h_1 h_2] &= -\frac{1}{2}+\frac{x^2}{2}+O\left(x^{13}\right) \\
    R_{2}^{(2)}[h_1 h_2] &= -\frac{1}{2}+\frac{x}{2}+x^2+\frac{x^3}{2}+\frac{x^4}{2}+O\left(x^{13}\right) \\
    R_{2}^{(3)}[h_1 h_2] &= -\frac{1}{2}+\frac{x}{2}+x^2-x^5-\frac{x^6}{2}-\frac{3 x^7}{2}-\frac{3 x^8}{2}-\frac{3 x^9}{2}-\frac{3 x^{10}}{2}-\frac{3 x^{11}}{2}-\frac{3 x^{12}}{2}+O\left(x^{13}\right) \\
    R_{2}^{(4)}[h_1 h_2] &= -\frac{1}{2}+\frac{x}{2}+x^2-x^5-x^7-\frac{x^8}{2}-\frac{x^9}{2}+\frac{x^{11}}{2}+\frac{3 x^{12}}{2}+O\left(x^{13}\right) \\
    R_{2}^{(5)}[h_1 h_2] &= -\frac{1}{2}+\frac{x}{2}+x^2-x^5-x^7-\frac{x^8}{2}-\frac{x^9}{2}-\frac{x^{10}}{2}+\frac{x^{12}}{2}+O\left(x^{13}\right)
\end{align*}
For $N=2$ and coefficient of $h_3$, we have
\begin{align*}
    L_{2}[h_3] &= -\frac{1}{3}-\frac{x^2}{3}+\frac{x^4}{3}+x^5+\frac{2 x^6}{3}+\frac{2 x^7}{3}+\frac{x^8}{3}-\frac{x^{10}}{3}-x^{11}-\frac{2 x^{12}}{3}+O\left(x^{13}\right) \\
    R_{2}^{(0)}[h_3] &= O\left(x^{13}\right) \\
    R_{2}^{(1)}[h_3] &= -\frac{1}{3}-\frac{x}{3}-\frac{x^2}{3}+O\left(x^{13}\right) \\
    R_{2}^{(2)}[h_3] &= -\frac{1}{3}-\frac{x^2}{3}+\frac{x^3}{3}+\frac{x^4}{3}+x^5+x^6+x^7+x^8+x^9+x^{10}+x^{11}+x^{12}+O\left(x^{13}\right) \\
    R_{2}^{(3)}[h_3] &= -\frac{1}{3}-\frac{x^2}{3}+\frac{x^4}{3}+x^5+\frac{x^6}{3}+\frac{2 x^7}{3}+\frac{x^8}{3}-\frac{x^9}{3}-\frac{2 x^{10}}{3}-\frac{4 x^{11}}{3}-\frac{5 x^{12}}{3}+O\left(x^{13}\right) \\
    R_{2}^{(4)}[h_3] &= -\frac{1}{3}-\frac{x^2}{3}+\frac{x^4}{3}+x^5+\frac{2 x^6}{3}+\frac{2 x^7}{3}+\frac{x^8}{3}-x^{11}-\frac{2 x^{12}}{3}+O\left(x^{13}\right) \\
    R_{2}^{(5)}[h_3] &= -\frac{1}{3}-\frac{x^2}{3}+\frac{x^4}{3}+x^5+\frac{2 x^6}{3}+\frac{2 x^7}{3}+\frac{x^8}{3}-\frac{x^{10}}{3}-x^{11}-\frac{2 x^{12}}{3}+O\left(x^{13}\right)
\end{align*}
For $N=3$ and coefficient of $1$, we have
\begin{align*}
    L_{3}[1] &= 1-x^4-x^5-x^6-x^7-x^8+x^{11}+x^{12}+2 x^{13}+2 x^{14}+2 x^{15}+O\left(x^{16}\right) \\
    R_{3}^{(0)}[1] &= 1 + +O\left(x^{16}\right) \\
    R_{3}^{(1)}[1] &= 1-x^4-x^5-x^6-x^7-x^8-x^9-x^{10}-x^{11}-x^{12}-x^{13}-x^{14}-x^{15}+O\left(x^{16}\right) \\
    R_{3}^{(2)}[1] &= 1-x^4-x^5-x^6-x^7-x^8+x^{11}+x^{12}+2 x^{13}+2 x^{14}+3 x^{15}+O\left(x^{16}\right) \\
    R_{3}^{(3)}[1] &= 1-x^4-x^5-x^6-x^7-x^8+x^{11}+x^{12}+2 x^{13}+2 x^{14}+2 x^{15}+O\left(x^{16}\right) \\
    R_{3}^{(4)}[1] &= 1-x^4-x^5-x^6-x^7-x^8+x^{11}+x^{12}+2 x^{13}+2 x^{14}+2 x^{15}+O\left(x^{16}\right)
\end{align*}
For $N=3$ and coefficient of $h_1$, we have
\begin{align*}
    L_{3}[h_1] &= -x^3+x^7+x^8+x^9+x^{10}+x^{11}-x^{14}+O\left(x^{15}\right) \\
    R_{3}^{(0)}[h_1] &= O\left(x^{15}\right) \\
    R_{3}^{(1)}[h_1] &= -x^3+O\left(x^{15}\right) \\
    R_{3}^{(2)}[h_1] &= -x^3+x^7+x^8+x^9+x^{10}+x^{11}+x^{12}+x^{13}+x^{14}+O\left(x^{15}\right) \\
    R_{3}^{(3)}[h_1] &= -x^3+x^7+x^8+x^9+x^{10}+x^{11}-x^{14}+O\left(x^{15}\right) \\
    R_{3}^{(4)}[h_1] &= -x^3+x^7+x^8+x^9+x^{10}+x^{11}-x^{14}+O\left(x^{15}\right)
\end{align*}
For $N=3$ and coefficient of $h_1^2$, we have
\begin{align*}
    L_{3}[h_1^2] &= -\frac{x^2}{2}-\frac{x^3}{2}+\frac{x^5}{2}+x^6+x^7+x^8+\frac{x^9}{2}-\frac{x^{11}}{2}-x^{12}-\frac{3 x^{13}}{2}+O\left(x^{14}\right) \\
    R_{3}^{(0)}[h_1^2] &= O\left(x^{14}\right) \\
    R_{3}^{(1)}[h_1^2] &= -\frac{x^2}{2}-\frac{x^3}{2}+O\left(x^{14}\right) \\
    R_{3}^{(2)}[h_1^2] &= -\frac{x^2}{2}-\frac{x^3}{2}+\frac{x^5}{2}+x^6+x^7+x^8+x^9+x^{10}+x^{11}+x^{12}+x^{13}+O\left(x^{14}\right) \\
    R_{3}^{(3)}[h_1^2] &= -\frac{x^2}{2}-\frac{x^3}{2}+\frac{x^5}{2}+x^6+x^7+x^8+\frac{x^9}{2}-\frac{x^{11}}{2}-x^{12}-\frac{3 x^{13}}{2}+O\left(x^{14}\right) \\
    R_{3}^{(4)}[h_1^2] &= -\frac{x^2}{2}-\frac{x^3}{2}+\frac{x^5}{2}+x^6+x^7+x^8+\frac{x^9}{2}-\frac{x^{11}}{2}-x^{12}-\frac{3 x^{13}}{2}+O\left(x^{14}\right)
\end{align*}
For $N=3$ and coefficient of $h_2$, we have
\begin{align*}
    L_{3}[h_2] &= -\frac{x^2}{2}-\frac{x^3}{2}+\frac{x^5}{2}+x^7+x^8+\frac{x^9}{2}+x^{10}+\frac{x^{11}}{2}-\frac{x^{13}}{2}+O\left(x^{14}\right) \\
    R_{3}^{(0)}[h_2] &= O\left(x^{14}\right) \\
    R_{3}^{(1)}[h_2] &= -\frac{x^2}{2}-\frac{x^3}{2}+O\left(x^{14}\right) \\
    R_{3}^{(2)}[h_2] &= -\frac{x^2}{2}-\frac{x^3}{2}+\frac{x^5}{2}+x^7+x^8+x^9+x^{10}+x^{11}+x^{12}+x^{13}+O\left(x^{14}\right) \\
    R_{3}^{(3)}[h_2] &= -\frac{x^2}{2}-\frac{x^3}{2}+\frac{x^5}{2}+x^7+x^8+\frac{x^9}{2}+x^{10}+\frac{x^{11}}{2}-\frac{x^{13}}{2}+O\left(x^{14}\right) \\
    R_{3}^{(4)}[h_2] &= -\frac{x^2}{2}-\frac{x^3}{2}+\frac{x^5}{2}+x^7+x^8+\frac{x^9}{2}+x^{10}+\frac{x^{11}}{2}-\frac{x^{13}}{2}+O\left(x^{14}\right)
\end{align*}
For $N=3$ and coefficient of $h_1 h_2$, we have
\begin{align*}
    L_{3}[h_1 h_2] &= -\frac{x}{2}+x^3+\frac{x^4}{2}+\frac{x^5}{2}-\frac{x^7}{2}-x^8-\frac{x^9}{2}-x^{10}-x^{11}-\frac{x^{12}}{2}+O\left(x^{13}\right) \\
    R_{3}^{(0)}[h_1 h_2] &= O\left(x^{13}\right) \\
    R_{3}^{(1)}[h_1 h_2] &= -\frac{x}{2}+\frac{x^3}{2}+O\left(x^{13}\right) \\
    R_{3}^{(2)}[h_1 h_2] &= -\frac{x}{2}+x^3+\frac{x^4}{2}+\frac{x^5}{2}+\frac{x^6}{2}+O\left(x^{13}\right) \\
    R_{3}^{(3)}[h_1 h_2] &= -\frac{x}{2}+x^3+\frac{x^4}{2}+\frac{x^5}{2}-\frac{x^7}{2}-x^8-\frac{x^9}{2}-\frac{3 x^{10}}{2}-\frac{3 x^{11}}{2}-\frac{3 x^{12}}{2}+O\left(x^{13}\right) \\
    R_{3}^{(4)}[h_1 h_2] &= -\frac{x}{2}+x^3+\frac{x^4}{2}+\frac{x^5}{2}-\frac{x^7}{2}-x^8-\frac{x^9}{2}-x^{10}-x^{11}-\frac{x^{12}}{2}+O\left(x^{13}\right)
\end{align*}
For $N=3$ and coefficient of $h_3$, we have
\begin{align*}
    L_{3}[h_3] &= -\frac{x}{3}-\frac{x^2}{3}+\frac{x^5}{3}+x^7+x^8+\frac{x^9}{3}+x^{10}+\frac{x^{11}}{3}-\frac{x^{12}}{3}+O\left(x^{13}\right) \\
    R_{3}^{(0)}[h_3] &= O\left(x^{13}\right) \\
    R_{3}^{(1)}[h_3] &= -\frac{x}{3}-\frac{x^2}{3}-\frac{x^3}{3}+O\left(x^{13}\right) \\
    R_{3}^{(2)}[h_3] &= -\frac{x}{3}-\frac{x^2}{3}+\frac{x^5}{3}+\frac{x^6}{3}+x^7+x^8+x^9+x^{10}+x^{11}+x^{12}+O\left(x^{13}\right) \\
    R_{3}^{(3)}[h_3] &= -\frac{x}{3}-\frac{x^2}{3}+\frac{x^5}{3}+x^7+x^8+\frac{x^9}{3}+\frac{2 x^{10}}{3}+\frac{x^{11}}{3}-\frac{x^{12}}{3}+O\left(x^{13}\right) \\
    R_{3}^{(4)}[h_3] &= -\frac{x}{3}-\frac{x^2}{3}+\frac{x^5}{3}+x^7+x^8+\frac{x^9}{3}+x^{10}+\frac{x^{11}}{3}-\frac{x^{12}}{3}+O\left(x^{13}\right)
\end{align*}
For $N=0$ and coefficient of $1$, we have
\begin{align*}
    L_{0}[1] &= 1-x-x^2+x^5+x^7-x^{12}-x^{15}+O\left(x^{16}\right) \\
    R_{0}^{(0)}[1] &= 1+O\left(x^{16}\right) \\
    R_{0}^{(1)}[1] &= 1-x-x^2-x^3-x^4-x^5-x^6-x^7-x^8-x^9-x^{10}-x^{11}-x^{12}-x^{13}-x^{14}-x^{15}+O\left(x^{16}\right) \\
    R_{0}^{(2)}[1] &= 1-x-x^2+x^5+x^6+2 x^7+2 x^8+3 x^9+3 x^{10}+4 x^{11}+4 x^{12}+5 x^{13}+5 x^{14}+6 x^{15}+O\left(x^{16}\right) \\
    R_{0}^{(3)}[1] &= 1-x-x^2+x^5+x^7-x^{10}-x^{11}-3 x^{12}-3 x^{13}-5 x^{14}-6 x^{15}+O\left(x^{16}\right) \\
    R_{0}^{(4)}[1] &= 1-x-x^2+x^5+x^7-x^{12}+O\left(x^{16}\right) \\
    R_{0}^{(5)}[1] &= 1-x-x^2+x^5+x^7-x^{12}-x^{15}+O\left(x^{16}\right)
\end{align*}
For $N=0$ and coefficient of $h_1$, we have
\begin{align*}
    L_{0}[h_1] &= -1+x+x^2-x^5-x^7+x^{12}+O\left(x^{15}\right) \\
    R_{0}^{(0)}[h_1] &= O\left(x^{15}\right) \\
    R_{0}^{(1)}[h_1] &= -1 + O\left(x^{15}\right) \\
    R_{0}^{(2)}[h_1] &= -1+x+x^2+x^3+x^4+x^5+x^6+x^7+x^8+x^9+x^{10}+x^{11}+x^{12}+x^{13}+x^{14}+O\left(x^{15}\right) \\
    R_{0}^{(3)}[h_1] &= -1+x+x^2-x^5-x^6-2 x^7-2 x^8-3 x^9-3 x^{10}-4 x^{11}-4 x^{12}-5 x^{13}-5 x^{14}+O\left(x^{15}\right) \\
    R_{0}^{(4)}[h_1] &= -1+x+x^2-x^5-x^7+x^{10}+x^{11}+3 x^{12}+3 x^{13}+5 x^{14}+O\left(x^{15}\right) \\
    R_{0}^{(5)}[h_1] &= -1+x+x^2-x^5-x^7+x^{12}+O\left(x^{15}\right)
\end{align*}
For $N=0$ and coefficient of $h_1^2$, we have
\begin{align*}
    L_{0}[h_1^2] &= 0 \\
    R_{0}^{(0)}[h_1^2] &= O\left(x^{14}\right) \\
    R_{0}^{(1)}[h_1^2] &= -\frac{1}{2 x}-\frac{1}{2}+O\left(x^{14}\right) \\
    R_{0}^{(2)}[h_1^2] &= \frac{1}{2}+x+x^2+x^3+x^4+x^5+x^6+x^7+x^8+x^9+x^{10}+x^{11}+x^{12}+x^{13}+O\left(x^{14}\right) \\
    R_{0}^{(3)}[h_1^2] &= -\frac{x^2}{2}-x^3-\frac{3 x^4}{2}-2 x^5-\frac{5 x^6}{2}-3 x^7-\frac{7 x^8}{2}-4 x^9-\frac{9 x^{10}}{2}-5 x^{11}-\frac{11 x^{12}}{2}-6 x^{13}+O\left(x^{14}\right) \\
    R_{0}^{(4)}[h_1^2] &= \frac{x^5}{2}+x^6+\frac{3 x^7}{2}+\frac{5 x^8}{2}+\frac{7 x^9}{2}+\frac{9 x^{10}}{2}+6 x^{11}+\frac{15 x^{12}}{2}+9 x^{13}+O\left(x^{14}\right) \\
    R_{0}^{(5)}[h_1^2] &= -\frac{x^9}{2}-x^{10}-\frac{3 x^{11}}{2}-\frac{5 x^{12}}{2}-4 x^{13}+O\left(x^{14}\right) \\
    R_{0}^{(6)}[h_1^2] &= O\left(x^{14}\right)
\end{align*}
For $N=0$ and coefficient of $h_2$, we have
\begin{align*}
    L_{0}[h_2] &= -1+x+x^2-x^5-x^7+x^{12}+O\left(x^{14}\right) \\
    R_{0}^{(0)}[h_2] &= O\left(x^{14}\right) \\
    R_{0}^{(1)}[h_2] &= -\frac{1}{2 x}-\frac{1}{2}+O\left(x^{14}\right) \\
    R_{0}^{(2)}[h_2] &= -\frac{1}{2}+x+x^2+x^3+x^4+x^5+x^6+x^7+x^8+x^9+x^{10}+x^{11}+x^{12}+x^{13}+O\left(x^{14}\right) \\
    R_{0}^{(3)}[h_2] &= -1+x+\frac{x^2}{2}-\frac{x^4}{2}-x^5-\frac{3 x^6}{2}-2 x^7-\frac{5 x^8}{2}-3 x^9-\frac{7 x^{10}}{2}-4 x^{11}-\frac{9 x^{12}}{2}-5 x^{13}+O\left(x^{14}\right) \\
    R_{0}^{(4)}[h_2] &= -1+x+x^2-\frac{x^5}{2}-\frac{x^7}{2}+\frac{x^8}{2}+\frac{x^9}{2}+\frac{3 x^{10}}{2}+2 x^{11}+\frac{7 x^{12}}{2}+4 x^{13}+O\left(x^{14}\right) \\
    R_{0}^{(5)}[h_2] &= -1+x+x^2-x^5-x^7-\frac{x^9}{2}-\frac{x^{11}}{2}+\frac{x^{12}}{2}-x^{13}+O\left(x^{14}\right) \\
    R_{0}^{(6)}[h_2] &= -1+x+x^2-x^5-x^7+x^{12}+O\left(x^{14}\right)
\end{align*}
For $N=-1$ and coefficient of $1$, we have
\begin{align*}
    L_{-1}[1] &= 0 \\
    R_{-1}^{(0)}[1] &= 1+O\left(x^{16}\right) \\
    R_{-1}^{(1)}[1] &= -x-x^2-x^3-x^4-x^5-x^6-x^7-x^8-x^9-x^{10}-x^{11}-x^{12}-x^{13}-x^{14}+O\left(x^{16}\right) \\
    R_{-1}^{(2)}[1] &= x^3+x^4+2 x^5+2 x^6+3 x^7+3 x^8+4 x^9+4 x^{10}+5 x^{11}+5 x^{12}+6 x^{13}+O\left(x^{16}\right) \\
    R_{-1}^{(3)}[1] &= -x^6-x^7-2 x^8-3 x^9-4 x^{10}-5 x^{11}-7 x^{12}+O\left(x^{16}\right) \\
    R_{-1}^{(4)}[1] &= x^{10}+x^{11}+O\left(x^{16}\right) \\
    R_{-1}^{(5)}[1] &= O\left(x^{16}\right)
\end{align*}
For $N=-1$ and coefficient of $h_1$, we have
\begin{align*}
    L_{-1}[h_1] &= 0 \\
    R_{-1}^{(0)}[h_1] &= O\left(x^{15}\right) \\
    R_{-1}^{(1)}[h_1] &= -\frac{1}{x}+O\left(x^{15}\right) \\
    R_{-1}^{(2)}[h_1] &= 1+x+x^2+x^3+x^4+x^5+x^6+x^7+x^8+x^9+x^{10}+x^{11}+x^{12}+O\left(x^{15}\right) \\
    R_{-1}^{(3)}[h_1] &= -x^2-x^3-2 x^4-2 x^5-3 x^6-3 x^7-4 x^8-4 x^9-5 x^{10}-5 x^{11}+O\left(x^{15}\right) \\
    R_{-1}^{(4)}[h_1] &= x^5+x^6+2 x^7+3 x^8+4 x^9+5 x^{10}+O\left(x^{15}\right) \\
    R_{-1}^{(5)}[h_1] &= -x^9+O\left(x^{15}\right) \\
    R_{-1}^{(6)}[h_1] &= O\left(x^{15}\right)
\end{align*}
For $N=-1$ and coefficient of $h_1^2$, we have
\begin{align*}
    L_{-1}[h_1^2] &= 0 \\
    R_{-1}^{(0)}[h_1^2] &= O\left(x^{14}\right) \\
    R_{-1}^{(1)}[h_1^2] &= -\frac{1}{2 x^2}-\frac{1}{2 x}+O\left(x^{14}\right) \\
    R_{-1}^{(2)}[h_1^2] &= \frac{1}{2 x^3}+\frac{1}{2 x^2}+\frac{1}{2 x}+1+x+x^2+x^3+x^4+x^5+x^6+x^7+x^8+x^9+x^{10}+x^{11}+O\left(x^{14}\right) \\
    R_{-1}^{(3)}[h_1^2] &= -\frac{1}{2 x^2}-\frac{1}{x}-1-\frac{3 x}{2}-2 x^2-\frac{5 x^3}{2}-3 x^4-\frac{7 x^5}{2}-4 x^6-\frac{9 x^7}{2}-5 x^8-\frac{11 x^9}{2}-6 x^{10}+O\left(x^{14}\right) \\
    R_{-1}^{(4)}[h_1^2] &= \frac{1}{2}+x+\frac{3 x^2}{2}+2 x^3+3 x^4+4 x^5+5 x^6+\frac{13 x^7}{2}+8 x^8+\frac{19 x^9}{2}+O\left(x^{14}\right) \\
    R_{-1}^{(5)}[h_1^2] &= -\frac{x^3}{2}-x^4-\frac{3 x^5}{2}-\frac{5 x^6}{2}-\frac{7 x^7}{2}-5 x^8+O\left(x^{14}\right) \\
    R_{-1}^{(6)}[h_1^2] &= \frac{x^7}{2}+O\left(x^{14}\right)
\end{align*}
For $N=-1$ and coefficient of $h_2$, we have
\begin{align*}
    L_{-1}[h_2] &= 0 \\
    R_{-1}^{(0)}[h_2] &= O\left(x^{14}\right) \\
    R_{-1}^{(1)}[h_2] &= -\frac{1}{2 x^2}-\frac{1}{2 x}+O\left(x^{14}\right) \\
    R_{-1}^{(2)}[h_2] &= \frac{1}{2 x^3}-\frac{1}{2 x^2}+\frac{1}{2 x}+1+x+x^2+x^3+x^4+x^5+x^6+x^7+x^8+x^9+x^{10}+x^{11}+O\left(x^{14}\right) \\
    R_{-1}^{(3)}[h_2] &= -\frac{1}{2 x^2}-\frac{x}{2}-x^2-\frac{3 x^3}{2}-2 x^4-\frac{5 x^5}{2}-3 x^6-\frac{7 x^7}{2}-4 x^8-\frac{9 x^9}{2}-5 x^{10}+O\left(x^{14}\right) \\
    R_{-1}^{(4)}[h_2] &= \frac{1}{2}+\frac{x^2}{2}+x^4+x^5+2 x^6+\frac{5 x^7}{2}+4 x^8+\frac{9 x^9}{2}+O\left(x^{14}\right) \\
    R_{-1}^{(5)}[h_2] &= -\frac{x^3}{2}-\frac{x^5}{2}-\frac{x^6}{2}-\frac{x^7}{2}-x^8+O\left(x^{14}\right) \\
    R_{-1}^{(6)}[h_2] &= \frac{x^7}{2}+O\left(x^{14}\right)
\end{align*}

\bibliographystyle{JHEP}

\bibliography{index}

\end{document}